\documentclass[aps,prb,twocolumn,floatfix,showpacs]{revtex4-1}
\usepackage{graphicx}
\usepackage{amsmath,relsize,}
\usepackage{subfigure}
\begin{document}

\title{Point Defects in Twisted Bilayer Graphene: A Density Functional Theory Study}

\author{Kanchan Ulman}
\author{Shobhana Narasimhan}
\affiliation{Theoretical Sciences Unit, Jawaharlal Nehru Centre for Advanced Scientific Research, Jakkur, Bangalore 560064, India}

\begin{abstract}
We have used \textit{ab initio} density functional theory, incorporating van der Waals corrections, to study twisted bilayer graphene (TBLG) where Stone-Wales defects or monovacancies are introduced in one of the layers. We compare these results to those for defects in single layer graphene or Bernal stacked graphene. The energetics of defect formation is not very sensitive to the stacking of the layers or the specific site at which the defect is created, suggesting a weak interlayer coupling. However signatures of the interlayer coupling are manifested clearly in the electronic band structure. For the ``$\gamma\gamma$" Stone Wales defect in TBLG, we observe two Dirac cones that are shifted in both momentum space and energy. This up/down shift in energy results from the combined effect of a charge transfer between the two graphene layers, and a chemical interaction between the layers,  which mimics the effects of a transverse electric field. Charge density plots show that states near the Dirac points have significant admixture between the two layers. For Stone Wales defects at other sites in TBLG, this basic structure is modified by the creation of mini gaps at energy crossings. For a monovacancy, the Dirac cone of the pristine layer is shifted up in energy by $\sim0.25$ eV due to a combination of the requirements of the equilibration of Fermi energy between the
two layers with different numbers of electrons, charge transfer, and chemical interactions. Both kinds of defects increase the density of states at the Fermi level. The monovacancy also results in spin polarization, with magnetic moments on the defect of 1.2 -- 1.8 $\mu_B$. 

\end{abstract}


\maketitle

\section{Introduction}
The mechanical and electronic properties of graphene make it an attractive candidate for use in the electronics industry, and indeed
it has even been speculated that it could one day replace silicon as the primary material used.\cite{Geim-NatMat2007,CastroNeto-RMP2009} In addition, it exhibits several exotic properties, such as a room
temperature quantum Hall effect.\cite{Geim-NatMat2007} Many of the interesting electronic properties of single-layer graphene arise from the presence in the
band structure of Dirac cones at the K points in the Brillouin zone that touch at the Fermi level $E_F$, as a result of which electrons behave
like massless Dirac fermions.\cite{Novoselov-Nature2005}

When graphite or multilayer graphene is assembled by arranging single graphene layers in the conventional Bernal stacking, also referred to as AB-stacking, where successive layers are displaced by both a vertical and a lateral shift, many of the unique properties of single-layer graphene, such as the Dirac cones,
disappear. However, in recent years there has been the appealing discovery that if instead a few layers of graphene are grown so that successive
layers are aligned with a twist with respect to each other, there then appears to be an effective electronic decoupling between layers, so that
the Dirac cones are maintained, at least for large angles of twist.\cite{Santos-PRL2007, Shallcross-PRL2008, Luican-PRL2011} 
Twisted bilayer graphene has also been shown to have other intriguing properties, e.g., for small angles of twist, there is a reduction in the Fermi velocity, and a localization of electrons, and one can bring van
Hove singularities in the electronic density of states very close to the Fermi level.\cite{Li-NatPhys2010} This in turn raises the possibility of seeing other interesting
phenomena such as superconductivity. Other fascinating results on TBLG include the demonstration of Hofstadter's Butterfly in the energy spectrum in a magnetic field,\cite{Moon-PRB2012} and neutrino-like oscillations as a result of coupling of the Dirac cones of the two rotated layers.\cite{Chou-NL2013}
Graphene layers with a twist can now be grown by a variety of methods, which allow one to access a range of twist angles. \cite{Haas-PRL2008, Meng-PRB2012, Rong-PRB1993}
However, the nature and extent of the interlayer coupling, and the consequences thereof, continue to be debated.\cite{OhtaPRL2012}

The question then arises: in what way are these properties altered upon the introduction of defects in TBLG?
There has always been a lot of interest in defects in graphene and other low-dimensional carbon materials.\cite{Hashimoto-Nature2004} Much of this interest stems from concerns
about how the presence of such defects will affect the functioning of these materials in applications. Defects such as monovacancies, divacancies,
interstitials and Stone-Wales defects\cite{SW-CPL1986} are known to affect the mechanical, electronic and magnetic properties of carbon materials. There have,
for example, been several studies to check to what extent the presence of defects affects electrical conductance.\cite{Banhart-ACSNano2011, Zhan-AdvMat2012, Pisani-NJP2008} 
However, it has also been appreciated that it might be possible to purposely design defective structures with a view toward creating certain functionalities.\cite{Zhan-AdvMat2012,Chen-NatPhys2011}
Among the possible advantages that the presence of defects in graphene can confer, it has been shown that they can  induce magnetic moments,\cite{Lehtinen-PRL2003, Lehtinen-PRL2004, Yazyev-PRB2007, Yazyev-RepProgPhys2010, Ma-NJP2007, Nanda-NJP2012} and improve gas sensing properties.\cite{Zhang-Nanotech2009} 
It has also been suggested that the introduction of defects can be used for band-gap engineering.\cite{Zhou-NatMat2007,Zhan-AdvMat2012,Peng-NanoLett2008} 
Defects can be deliberately created by, e.g., irradiation with electrons\cite{He-APL2011, Liu-IEEE2011, Teweldebrhan-APL-94-2009, Teweldebrhan-APL-95-2009} 
or ions,\cite{Mathew-JAP2011,Mathew-Carbon2011,Bai-NatNanotec2010,Ugeda-PRL2010} 
oxidation,\cite{ Stankovich-Nature2006} hydrogenation,\cite{Ryu-NanoLett2008, Elias-Science2009, Luo-APL2010, Ni-NanoLett2010} or fluorination.\cite{Nair-Small2010, Jeon-ACSNano2011}

In this theoretical study we perform density functional theory (DFT) calculations to study the properties of two kinds of point defects: Stone-Wales defects (SW) and monovacancies in twisted bilayer graphene. We are particularly interested in the question of how the electronic band structure in the neighborhood of $E_F$ ( i.e., the Dirac cones) are affected by the presence of defects.

\section{Computational Details}
Our spin-polarized DFT calculations were performed using the PWscf package of the Quantum ESPRESSO distribution.\cite{PWSCF-JPCM2009} 
Ultrasoft pseudopotentials\cite{Vanderbilt-PRB1990} were used to describe the interactions between the ion cores and valence electrons. 
A plane wave basis set was used, with kinetic energy cut-offs of $40$ Ry and $480$ Ry for the wavefunctions and charge densities, respectively. 
We have considered two different levels of approximation for the exchange-correlation interactions: (i) the local density approximation (LDA) in the Perdew-Zunger form, \cite{PZ-PRB1981} and (ii) the generalized gradient approximation (GGA) in the Perdew-Burke-Ernzerhof (PBE) form.\cite{PBE-PRL1996} Many of the previous DFT studies on TBLG made use of the LDA. However, given that 
the (weak) coupling between the two graphene layers is important in TBLG, and that there has been much debate about its nature and extent, we feel that it is desirable to have an accurate treatment
of these interlayer interactions, where van der Waals interactions (which are absent in conventional DFT calculations) may be expected to play a crucial role. 
For this reason, along with the PBE exchange-correlation, we also incorporate the van der Waals interactions as a semi-empirical correction, using the ``DFT-D2" treatment suggested by Grimme.\cite{Grimme-JComChem2004,Grimme-JComChem2006} Of the three kinds of theoretical treatments used in this paper (LDA, PBE and DFT-D2), we believe that those results using DFT-D2 should be regarded as being the most reliable, and most of our results have been obtained using this approach.
However, we also present some results with the other two treatments for purposes of comparison; we believe that this is of interest since several authors still continue to use these other functionals,
especially LDA, in order to treat such systems, and it is therefore worth examining their reliability.

System geometries are described in Section~\ref{SysGeom} below. 
Along the $z$ direction (normal to the graphene sheets) periodic images were separated by a vacuum spacing of about 13 \AA. 
The Brillouin zone was sampled with a Monkhorst-Pack mesh of $(21\times21\times1)$ k-points for the primitive unit cell of graphene, 
and proportionately equivalent meshes for larger supercells of graphene systems considered in this paper, when performing the self-consistent-field calculations. However, when performing
the post-processing calculations, when an extremely accurate k-point sampling was needed in order to obtain a precise computation of the Fermi energy, much finer grids were used, and the grid
spacing was further reduced in regions of the Brillouin zone (near the Dirac points) where it was found necessary to have a particularly dense sampling; the weights assigned to k-points were appropriately adjusted. Convergence was aided by making use of the Marzari-Vanderbilt smearing scheme, \cite{MV-PRL1999} with a small smearing width of 0.001 Ry. 
All the structures were relaxed using Hellmann-Feynman forces \cite{Feynman-PR1939} until the force on each atom was less than 0.001 Ry/bohr along all the three directions.
Simulated scanning tunneling microscopy (STM) images were obtained using the Tersoff-Hamann theory \cite{Tersoff-PRL1998} for STM images, as
incorporated in the ``pp.x" post-processing tool supplied with Quantum-ESPRESSO.

\begin{figure}[t!]
\centering
\includegraphics[scale=0.35]{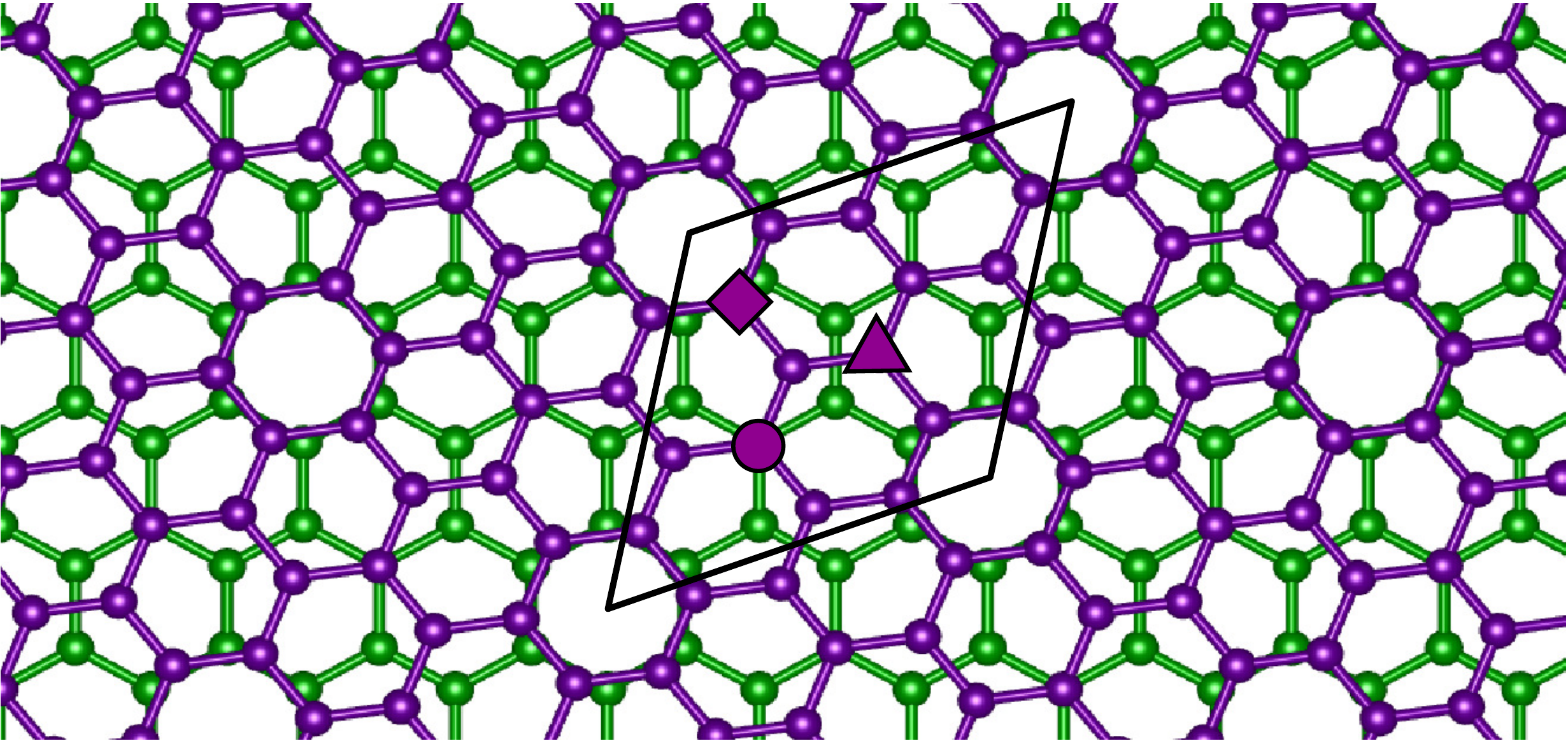}
\caption{(color online) A bilayer of graphene with rotational stacking fault (TBLG) with a rotation angle of $38.213^{\circ}$. 
The upper layer is shown in violet and lower layer in green, the 28-atom primitive unit cell ($S_1$) is also shown. Examples of three symmetry-inequivalent types of sites in the upper layer are shown with different
shapes: $\alpha$-site (circle), $\beta$-site (triangle) and  $\gamma$-site (square).}
\label{Fig:tBLG} 
\end{figure}

\section{Twisted Bilayer Graphene}
Modeling twisted bilayer graphene, i.e., a bilayer of graphene with a rotational stacking fault, requires the construction of a commensurate 
supercell with the necessary finiteness required for practical computation. A number of such commensurate supercells 
are possible depending on the angle of rotation between the two graphene layers.~\cite{Shallcross-PRB2008} 
The derivation of the commensuration condition and the construction of the commensurate unit cell for an arbitrary set of angles between the two graphene layers 
is described below. Though this commensuration condition and the resulting Brillouin zone have been described earlier in Ref.~\onlinecite{Shallcross-PRB2008}, 
we briefly summarize them here, in order to make
clear the notation used in this paper, and to enable the reader to follow the analysis of results that is
presented later in Section \ref{SW-ElProp}.

\subsection{Geometric Properties of Twisted Bilayer Graphene:}\label{App-Model-TBLG}

A TBLG system is created by placing two layers of graphene on top of each other, one of which is ``unrotated" (U), and the other ``rotated" (R).
In order to describe the TBLG system, one must find a commensurate unit cell for both the layers.

Let $\textbf{a}_1^{\rm U}$ and $\textbf{a}_2^{\rm U}$ be the primitive lattice vectors of the (unrotated) graphene honeycomb lattice. 
A supercell for the unrotated layer can be constructed by the supercell lattice vectors: 
$\textbf{A}_1^{\rm U}=m_1\textbf{a}_1^{\rm U}+m_2\textbf{a}_2^{\rm U}$  and $\textbf{A}_2^{\rm U}=-m_2\textbf{a}_1^{\rm U}+(m_1+m_2)\textbf{a}_2^{\rm U}$, 
where $m_1$ and $m_2$ are integers. The primitive lattice vectors of the second graphene layer, which has been rotated by an angle $\theta$, are given by:
$\textbf{a}_1^{\rm R}=R(\theta)\textbf{a}_1^{\rm U}$ and $\textbf{a}_2^{\rm R}=R(\theta)\textbf{a}_2^{\rm U}$, where $R(\theta)$ is the rotation matrix.
For such a rotated lattice, a supercell is given by the lattice vectors: 
$\textbf{A}_1^{\rm R}=n_1\textbf{a}_1^{\rm R}+n_2\textbf{a}_2^{\rm R}$  and $\textbf{A}_2^{\rm R}=-n_2\textbf{a}_1^{\rm R}+(n_1+n_2)\textbf{a}_2^{\rm R}$, 
where $n_1$ and $n_2$ are integers.

The condition for commensuration is given by:
\begin{equation}
n_1\textbf{a}_1^{\rm R}+n_2\textbf{a}_2^{\rm R} = m_1\textbf{a}_1^{\rm U}+m_2\textbf{a}_2^{\rm U}
\end{equation}

For a standard choice of graphene primitive lattice vectors $\textbf{a}_1^{\rm U} = a(1,0)$  and $\textbf{a}_2^{\rm U} = a(1/2,\sqrt{3}/2)$, where $a$ is the lattice constant, the above commensuration relation becomes:

\begin{equation}\label{Diphant}
\Bigg(\begin{array}{c} n_1\\ n_2 \end{array}\Bigg) = \Bigg(\begin{array}{cc} {\rm cos}\theta+\frac{1}{\sqrt{3}}{\rm sin}\theta & +\frac{2}{\sqrt{3}}{\rm sin}\theta \\ -\frac{2}{\sqrt{3}}{\rm sin}\theta & {\rm cos}\theta-\frac{1}{\sqrt{3}}{\rm sin}\theta \end{array}\Bigg) \Bigg(\begin{array}{c} m_1\\ m_2 \end{array}\Bigg)
\end{equation}

This maps one integer pair ($n_1,n_2$) to another ($m_1,m_2$), with the constraint that $m_1^2+m_2^2+m_1m_2 = n_1^2+n_2^2+n_1n_2$. As pointed out in Ref.~\onlinecite{Shallcross-PRB2008},
this is a Diophantine problem, which has to be solved  to find out integer pairs ($n_1,n_2$) and ($m_1,m_2$) that satisfy Eq.~(\ref{Diphant}). 
For integer pairs satisfying the above commensuration condition, the twist angle $\theta$ is given by: 
\begin{equation}\label{eqn-theta}
\theta ={\rm cos}^{-1}\Bigg(  \frac{2m_1n_1 + m_1n_2 + m_2n_1 + 2m_2n_2}{2(m_1^2+m_2^2+m_1m_2)} \Bigg)
\end{equation}

The primitive  lattice vectors of the twisted bilayer, $\textbf{A}_1^{\rm B}$ and $\textbf{A}_2^{\rm B}$,  are related to the unrotated primitive lattice vectors of the single layer, $\textbf{a}_1^{\rm U}$ and $\textbf{a}_2^{\rm U}$, by:
\begin{equation}
\textbf{A}_1^{\rm B} = m_1\textbf{a}_1^{\rm U}+m_2\textbf{a}_2^{\rm U} ,
\end{equation}
and 
\begin{equation}
\textbf{A}_2^{\rm B} = -m_2\textbf{a}_1^{\rm U}+(m_1+m_2)\textbf{a}_2^{\rm U} .
\end{equation}

\begin{figure}
\centering
\includegraphics[scale=0.35]{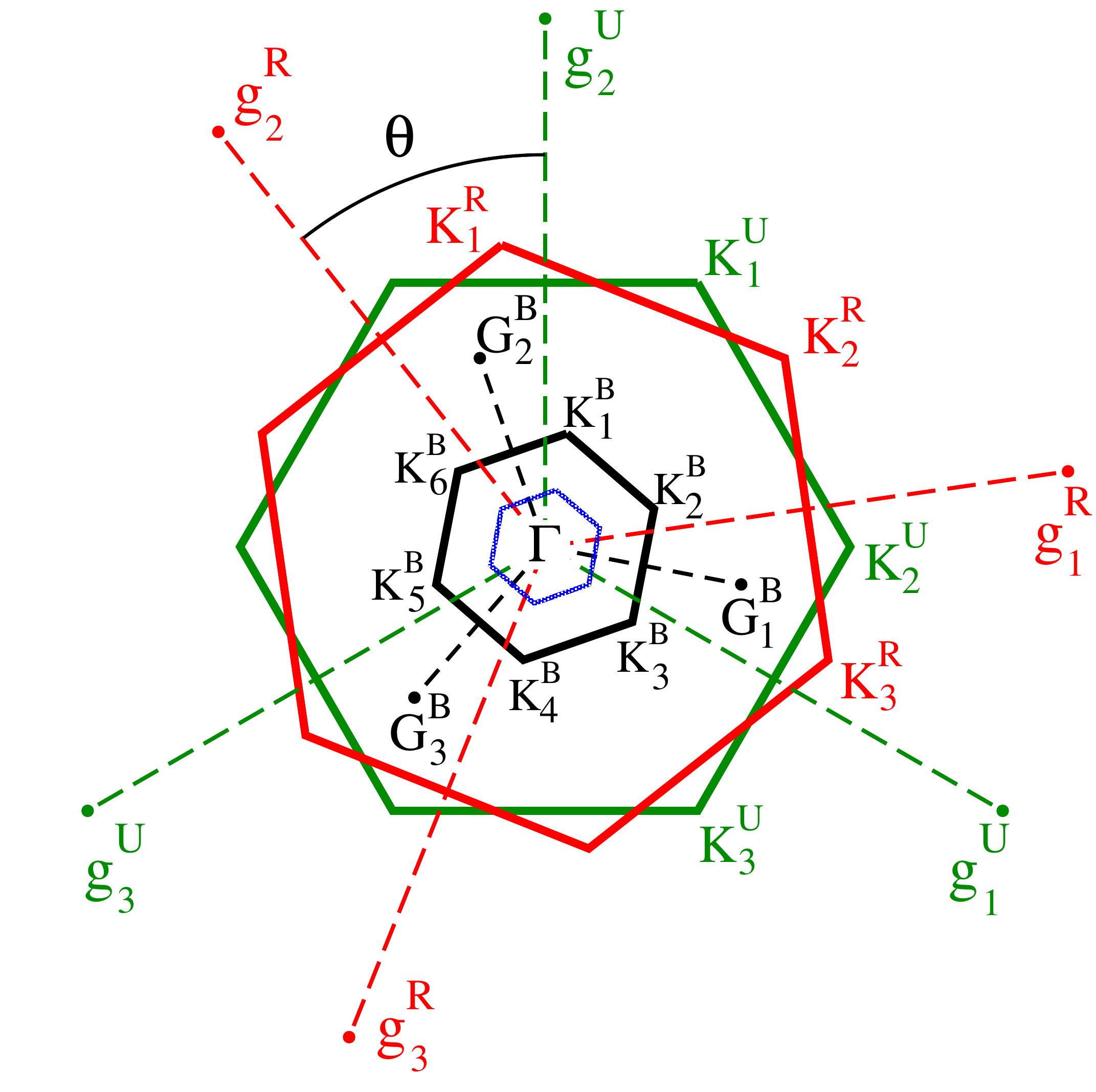}
\caption{(color online) The First Brillouin Zone (FBZ) corresponding to the the unrotated graphene layer (green), rotated graphene layer (red) and TBLG (for $S_1$ in black and $S_2$ in blue). 
The Dirac points K$_1$, K$_2$ and K$_3$ for the unrotated (superscript U), rotated (superscript R) and TBLG-$S_1$ (superscript B) are also shown.}
\label{Fig:TBLG-FBZ} 
\end{figure}
The smallest possible such unit cell corresponds to $m_1 = 2$, $m_2 = 1$; by Eqs.~(\ref{Diphant}) and (\ref{eqn-theta}), this gives $n_1 = 3$, $n_2 = -1$, and $\theta = 38.213^\circ$.
This is the angle of twist used for the twisted bilayer graphene considered in our study, its primitive unit cell contains 14 atoms in each layer. The unit cell vectors for this primitive
unit cell $S_1$ are
$\textbf{A}_1^{\rm B}=2\textbf{a}_1^{\rm U}+\textbf{a}_2^{\rm U}$ and $\textbf{A}_2^{\rm B}=-\textbf{a}_1^{\rm U}+3\textbf{a}_2^{\rm U}$. 

In Fig.~\ref{Fig:tBLG} we have shown the structure of TBLG with this twist angle, with the primitive unit cell $S_1$ of the twisted bilayer indicated by the black rhombus. 
Note that for this choice of twist angle, the upper (rotated) layer contains three symmetry-inequivalent types of carbon atoms, 
which we label $\alpha$, $\beta$ and $\gamma$ (see Fig.~\ref{Fig:tBLG}).

However, most of our calculations have been performed using a larger unit cell $S_2$, which has lattice vectors that are twice as long: $\textbf{A}_1^{\rm B}=4\textbf{a}_1^{\rm U}+ 2 \textbf{a}_2^{\rm U}$ and $\textbf{A}_2^{\rm B}=- 2 \textbf{a}_1^{\rm U}+6 \textbf{a}_2^{\rm U}$. This unit cell contains 56 atoms in each layer, i.e., 112 atoms in the bilayer unit cell; this corresponds to the choice $m_1 = 4$, $m_2 = 2$.

It is important to also have a knowledge of the structure of the reciprocal space of TBLG, in order to gain an  understanding of its electronic properties.
If $\textbf{g}_1^{\rm U}$ and $\textbf{g}_2^{\rm U}$ are the reciprocal lattice vectors corresponding to the unrotated layer, and $\textbf{g}_1^{\rm R}$ and $\textbf{g}_2^{\rm R}$ are the reciprocal lattice vectors corresponding to the rotated layer, the reciprocal lattice vectors of the TBLG are given by:

\smaller
\begin{equation}
\textbf{G}_1^{\rm B} = \Big\{ \frac{m_1+m_2}{m_1^2+m_2^2+m_1m_2} \Big\} \textbf{g}_1^{\rm U} + \Big\{ \frac{m_2}{m_1^2+m_2^2+m_1m_2} \Big\} \textbf{g}_2^{\rm U}
\end{equation}
\normalsize
\[
\textbf{G}_1^{\rm B} = \left\{
  \begin{array}{l l l}
    \frac{3}{7}  \textbf{g}_1^{\rm U} + \frac{1}{7} \textbf{g}_2^{\rm U},     & \quad \quad \text{for $S_1$} & \quad \text{(6a)}\\ \\
    \frac{3}{14}  \textbf{g}_1^{\rm U} + \frac{1}{14} \textbf{g}_2^{\rm U},   & \quad \quad \text{for $S_2$} & \quad \text{(6b)}  
  \end{array} \right.
\]

and 
\smaller
\begin{equation}
\textbf{G}_2^{\rm B} = -\Big\{ \frac{m_2}{m_1^2+m_2^2+m_1m_2} \Big\} \textbf{g}_1^{\rm U}   + \Big\{ \frac{m_1}{m_1^2+m_2^2+m_1m_2} \Big\} \textbf{g}_2^{\rm U}
\end{equation}
\normalsize
\[
\textbf{G}_2^{\rm B} = \left\{
  \begin{array}{l l l}
    -\frac{1}{7}  \textbf{g}_1^{\rm U} +  \frac{2}{7} \textbf{g}_2^{\rm U},     & \quad \quad \text{for $S_1$} & \quad \text{(7a)}\\ \\
    -\frac{1}{14}  \textbf{g}_1^{\rm U} +  \frac{2}{14} \textbf{g}_2^{\rm U},   & \quad \quad \text{for $S_2$} & \quad \text{(7b)}  
  \end{array} \right.
\]

\noindent

Thus,
\begin{equation}
\textbf{g}_1^{\rm U} = m_1\textbf{G}_1^{\rm B} - m_2\textbf{G}_2^{\rm B}
\end{equation}
\[
\textbf{g}_1^{\rm U} = \left\{
  \begin{array}{l l l}
    2\textbf{G}_1^{\rm B} - \textbf{G}_2^{\rm B},     & \quad \quad \text{for $S_1$} & \quad \text{(8a)}\\ \\
    4\textbf{G}_1^{\rm B} - 2\textbf{G}_2^{\rm B},    & \quad \quad \text{for $S_2$} & \quad \text{(8b)}  
  \end{array} \right.
\]

and
\begin{equation}
\textbf{g}_2^{\rm U} = m_2\textbf{G}_1^{\rm B} + (m_1+m_2)\textbf{G}_2^{\rm B}
\end{equation}
\[
\textbf{g}_2^{\rm U} = \left\{
  \begin{array}{l l l}
    \textbf{G}_1^{\rm B} + 3\textbf{G}_2^{\rm B},     & \quad \quad \text{for $S_1$} & \quad \text{(9a)}\\ \\
    2\textbf{G}_1^{\rm B} + 6\textbf{G}_2^{\rm B},    & \quad \quad \text{for $S_2$} & \quad \text{(9b)}  
  \end{array} \right.
\]

The reciprocal lattice and the first Brillouin zone for the unrotated and rotated layers, and the twisted bilayer graphene, are shown in Fig.~\ref{Fig:TBLG-FBZ}. 
The first Brillouin zone for the TBLG (shown by the black hexagon in Fig.~\ref{Fig:TBLG-FBZ}) corresponds to a twist angle of $38.213^{\circ}$, 
for the smallest commensurate supercell $S_1$. A similar figure can be obtained when $S_2$ is used; this is provided in the Supplementary material for easy reference.
Note that by the translational symmetry of the lattice, $\textbf{K}_1^{\rm B}$, $\textbf{K}_3^{\rm B}$ and $\textbf{K}_5^{\rm B}$ are identical; 
similarly, $\textbf{K}_2^{\rm B}$, $\textbf{K}_4^{\rm B}$ and $\textbf{K}_6^{\rm B}$ are identical. 
If, in addition, one has inversion symmetry or time-reversal symmetry, then all six $\textbf{K}$ points are identical. 
However, these symmetries are valid at the corners of the Brillouin zone only, and need not necessarily hold in the interior of the Brillouin zone, 
i.e., one need not have three-fold or six-fold symmetry always present. In particular, the introduction of defects may lower the symmetry.
The $\textbf{K}_1^{\rm U}$ and $\textbf{K}_2^{\rm U}$ points in the Brillouin zone of the unrotated layer can be folded onto points in the first Brillouin zone of the TBLG lattice:

\begin{equation}
\textbf{K}_1^{\rm U} = (m_1-m_2)\textbf{K}_1^{\rm B} + m_2(\textbf{G}_1^{\rm B} + \textbf{G}_2^{\rm B})
\end{equation}
\[
\textbf{K}_1^{\rm U} = \left\{
  \begin{array}{l l l}
    \textbf{K}_1^{\rm B} + (\textbf{G}_1^{\rm B} + \textbf{G}_2^{\rm B}),   & \quad \quad \text{for $S_1$} & \quad \text{(10a)}\\ \\
    \textbf{K}_2^{\rm B} + (2\textbf{G}_1^{\rm B} + 3\textbf{G}_2^{\rm B}), & \quad \quad \text{for $S_2$} & \quad \text{(10b)}
  \end{array} \right.
\]

and

\begin{equation}
\textbf{K}_2^{\rm U} = (m_1-m_2)\textbf{K}_2^{\rm B} + m_2\textbf{G}_1^{\rm B}
\end{equation}
\[
\textbf{K}_2^{\rm U} = \left\{
  \begin{array}{l l l}
    \textbf{K}_2^{\rm B} +  \textbf{G}_1^{\rm B}, & \quad \quad \text{for $S_1$} & \quad \text{(11a)}\\ \\
    \textbf{K}_1^{\rm B} + 3\textbf{G}_1^{\rm B}, & \quad \quad \text{for $S_2$} & \quad \text{(11b)}
  \end{array} \right.
\]

Similarly for the rotated layer,

\begin{equation}
\textbf{K}_1^{\rm R} = (n_1-n_2)\textbf{K}_1^{\rm B} + n_2(\textbf{G}_1^{\rm B} + \textbf{G}_2^{\rm B})
\end{equation}
\[
\textbf{K}_1^{\rm R} = \left\{
  \begin{array}{l l l}
    \textbf{K}_1^{\rm B} + \textbf{G}_2^{\rm B},  & \quad \quad \text{for $S_1$} & \quad \text{(12a)}\\ \\
    \textbf{K}_2^{\rm B} + 3\textbf{G}_2^{\rm B}, & \quad \quad \text{for $S_2$} & \quad \text{(12b)}
  \end{array} \right.
\]

and

\begin{equation}
\textbf{K}_2^{\rm R} = (n_1-n_2)\textbf{K}_2^{\rm B} + n_2\textbf{G}_1^{\rm B}
\end{equation}
\[
\textbf{K}_2^{\rm R} = \left\{
  \begin{array}{l l l}
    \textbf{K}_2^{\rm B} + (\textbf{G}_1^{\rm B}+\textbf{G}_2^{\rm B}),   & \quad \quad \text{for $S_1$} & \quad \text{(13a)}\\ \\
    \textbf{K}_1^{\rm B} + (3\textbf{G}_1^{\rm B}+2\textbf{G}_2^{\rm B}), & \quad \quad \text{for $S_2$} & \quad \text{(13b)}
  \end{array} \right.
\]

Thus, \textbf{K}$_1^{\rm U}$ folds onto $\Gamma$, \textbf{K}$_1^{\rm B}$ or \textbf{K}$_2^{\rm B}$, depending on whether ($m_1-m_2$) is $3n$, $3n+1$ or $3n+2$, respectively. 
Similarly, for the rotated layer, \textbf{K}$_1^{\rm R}$ folds onto $\Gamma$, \textbf{K}$_1^{\rm B}$ or \textbf{K}$_2^{\rm B}$, 
depending on whether ($n_1-n_2$) is $3n$, $3n+1$ or $3n+2$, respectively. 
For TBLG with the supercell $S_1$, we therefore have the following mappings: \textbf{K}$_1^{\rm U} $ $\rightarrow$ \textbf{K}$_1^{\rm B}$ and 
\textbf{K}$_2^{\rm U} $ $\rightarrow$ \textbf{K}$_2^{\rm B}$. 
Similarly, when performing calculations with the supercell $S_2$, we have: \textbf{K}$_1^{\rm U} $ $\rightarrow$ \textbf{K}$_2^{\rm B}$ and 
\textbf{K}$_2^{\rm U} $ $\rightarrow$ \textbf{K}$_1^{\rm B}$. 
 
This tells us that when we use $S_1$ or $S_2$ to analyze the band-structure of the TBLG, the Dirac cones will continue to appear at the $\textbf{K}$ points of the Brillouin zone of the supercell.
Thus, these are the points whose vicinity we will focus on, to see whether the introduction of defects opens up a band gap and/or alters the linear dispersion relation.

\subsection{Systems Studied}\label{SysGeom}

We consider two types of point defects in TBLG, viz., the Stone-Wales defect and the monovacancy. 
We primarily work with the supercell $S_2$ which is $2\times2$ times as large as the supercell $S_1$. 
For both the Stone-Wales defect and the monovacancy, we perform calculations considering all possible inequivalent sites of one defect in the $S_2$ supercell. 
This corresponds to one defect in a supercell of 56 carbon atoms per layer, i.e., a defect density of 1.8\%. 
For the monovacancy, we have in addition considered defect densities of 7.2\% (corresponding to one vacancy in the $S_1$ supercell with 14 carbon atoms in each layer) and 0.8\% (corresponding to one defect in supercell $S_3$, which is $3\times3$ times as large as $S_1$, and contains 126 carbon atoms in each layer).

\begin{figure*}[t!]
\centering
\subfigure{\label{fig: SLG}\includegraphics[width=0.225\textwidth]{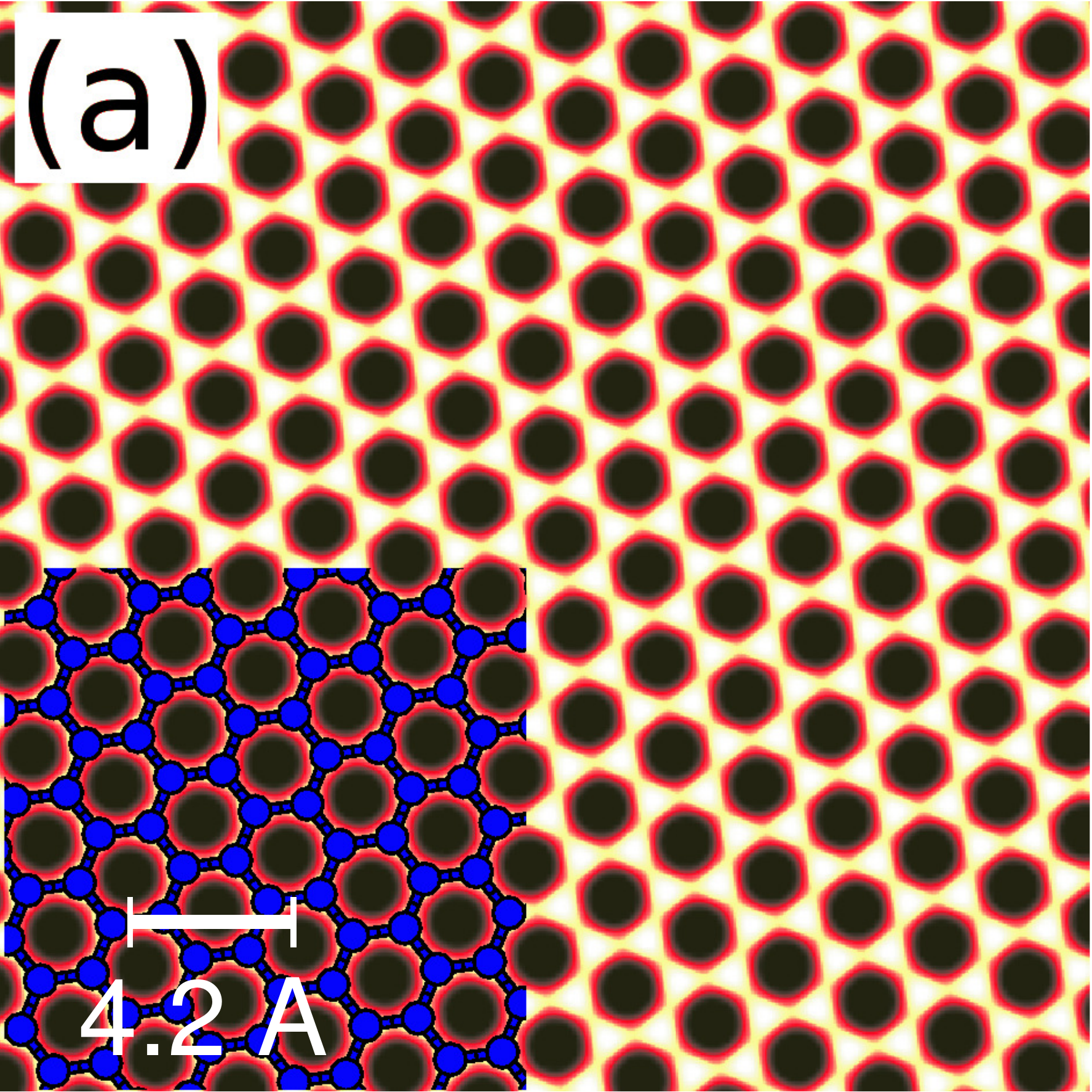}}
\quad
\subfigure{\label{fig: AB-BLG}\includegraphics[width=0.225\textwidth]{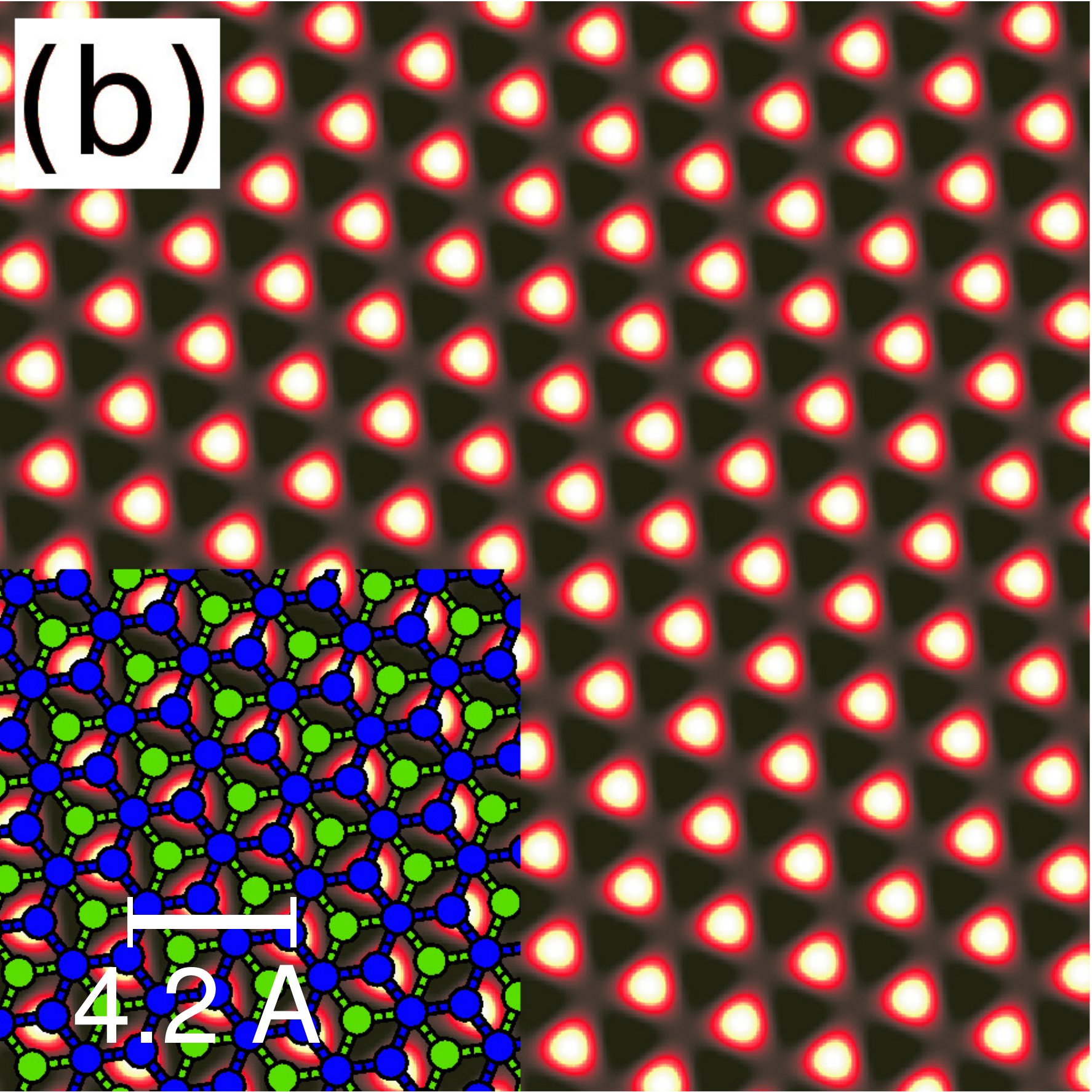}}
\quad
\subfigure{\label{fig: TBLG}\includegraphics[width=0.225\textwidth]{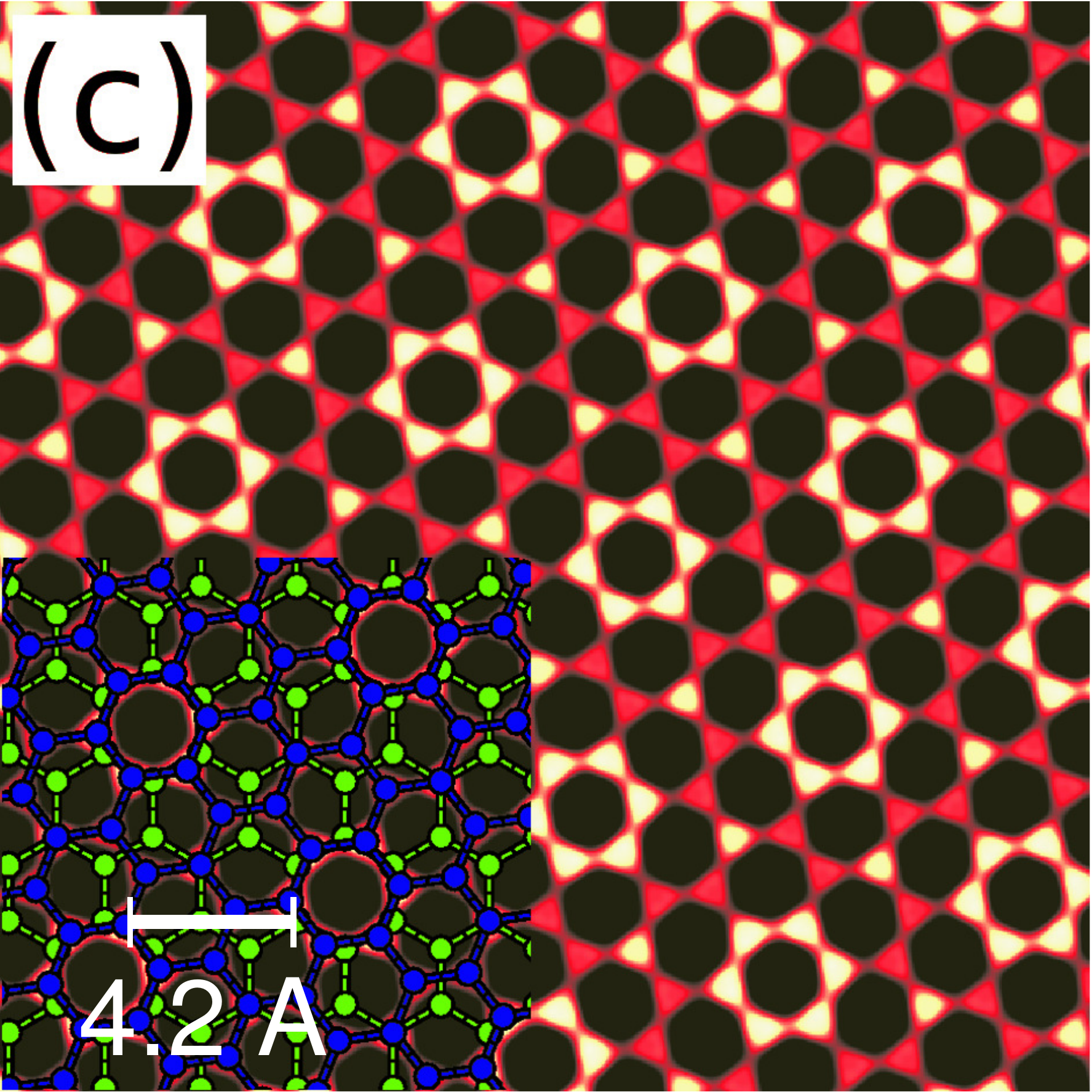}}
\caption{\label{Prist-STM} (color online) Simulated STM images for (a) Single layer of graphene (SLG), (b) AB-stacked BLG, and (c) Twisted-BLG. 
The insets show an overlay of the atomic positions of the system. The blue circles are the carbon atoms of the upper layer of 
graphene, and the green circles are the carbon atoms of the lower layer.}
\end{figure*}

For comparison of various properties of the above-mentioned defects in TBLG, we also consider the same defects in a single layer of graphene (SLG), 
as well as in an AB-stacked bilayer graphene (we refer to this as AB-BLG). 
To facilitate comparison, i.e., maintain the same defect density and system symmetry, we use a similar $S_2$ supercell also for our calculations on SLG and AB-BLG.

\section{Results and Discussion}
\begingroup
\begin{table}[h!]
\begin{center}
\begin{tabular}{ l  c  c  c  }
\multicolumn{4}{l}{ AB stacked BLG} \\ 
\hline 
\hline
Method & in-plane C-C   & Interlayer        & Exfoliation                      \\
(XC)     & bond length    & distance          & Energy       \\
           & $a_0$ (\AA)    & $d$ (\AA)        & $E_{\rm{exf}}$ (meV/atom) \\ 
\hline
LDA       &               1.41                      &               3.34               &             12.4                 \\
PBE       &               1.42                      &               4.15               &              0.8                 \\
DFT-D2  &               1.42                      &               3.20               &             26.1                 \\
Expt.     &               1.42                      &             3.35 [Ref.~\onlinecite{Bernal-RSC}]                    &       43 [Ref.~\onlinecite{Girifalco-JCP1956}]                     \\
             &                                         &                                  &         35$\pm$10 [Ref.~\onlinecite{Benedict-CPL1998}]   \\
              &                                         &                                  &                31$\pm$2 [Ref.~\onlinecite{Liu-PRB2012}] \\
\hline
\\
\multicolumn{4}{l}{ Twisted BLG} \\ 
\hline 
Method & in-plane C-C   & Interlayer        & Exfoliation                      \\
(XC)     & bond length    & distance          & Energy       \\
           & $a_0$ (\AA)    & $d$ (\AA)        & $E_{\rm{exf}}$ (meV) \\ 
\hline
\hline
LDA       &               1.41                      &               3.40               &             10.8                \\
PBE       &               1.42                      &               4.35               &              0.8                \\
DFT-D2    &               1.42                     &               3.30               &             24.0                \\
\hline
\end{tabular}
\caption{Results for structure and energetics of AB-stacked bilayer graphene and TBLG, as obtained with different approaches. $a_0$ is the in-plane C-C bond length,
$d$ is the interlayer distance, and $E_{\rm{exf}}$ is the exfoliation energy. Note that the experimental values are for graphite, not for bilayer graphene.}
\label{tab-AB-TBLG}
\end{center}
\end{table}
\endgroup

\subsection{Pristine Graphene}

First, as benchmarks, we perform calculations on defect-free bilayer graphene, with both Bernal Stacking
and twisted stacking.
Our results for the energetics and geometry, for AB-BLG and TBLG, with both LDA and DFT-D2 approaches, as well as with the PBE alone, are shown in Table~\ref{tab-AB-TBLG}. 
While both LDA and DFT-D2 give values of the interlayer distance $d$ that are in reasonably good agreement with experiment, use of the PBE alone leads to a value that is far too large. 
However, the DFT-D2 leads to a result for the exfoliation energy $E_{\rm{exf}}$, the energy required for exfoliating a graphene layer from the bilayer system, 
that is closer to experimental estimates than the LDA result; once again the PBE result is completely erroneous. 
All of these results are in agreement with previous theoretical work.\cite{Haas-PRL2008, Latil-PRB2007, Mao-Nanotech2008} 
Also note that both LDA and DFT-D2 show that AB-BLG is lower in total energy than TBLG by about 2 meV/atom; however PBE alone leads to TBLG being more
stable by the very small amount of 0.04 meV per atom. 
On going from AB-BLG to TBLG, there is a very slight increase in $d$, which causes a small decrease in $E_{\rm exf}$; however, these changes are very small.

\begin{figure}[tbp]
\centering
\subfigure{\includegraphics[width=5cm]{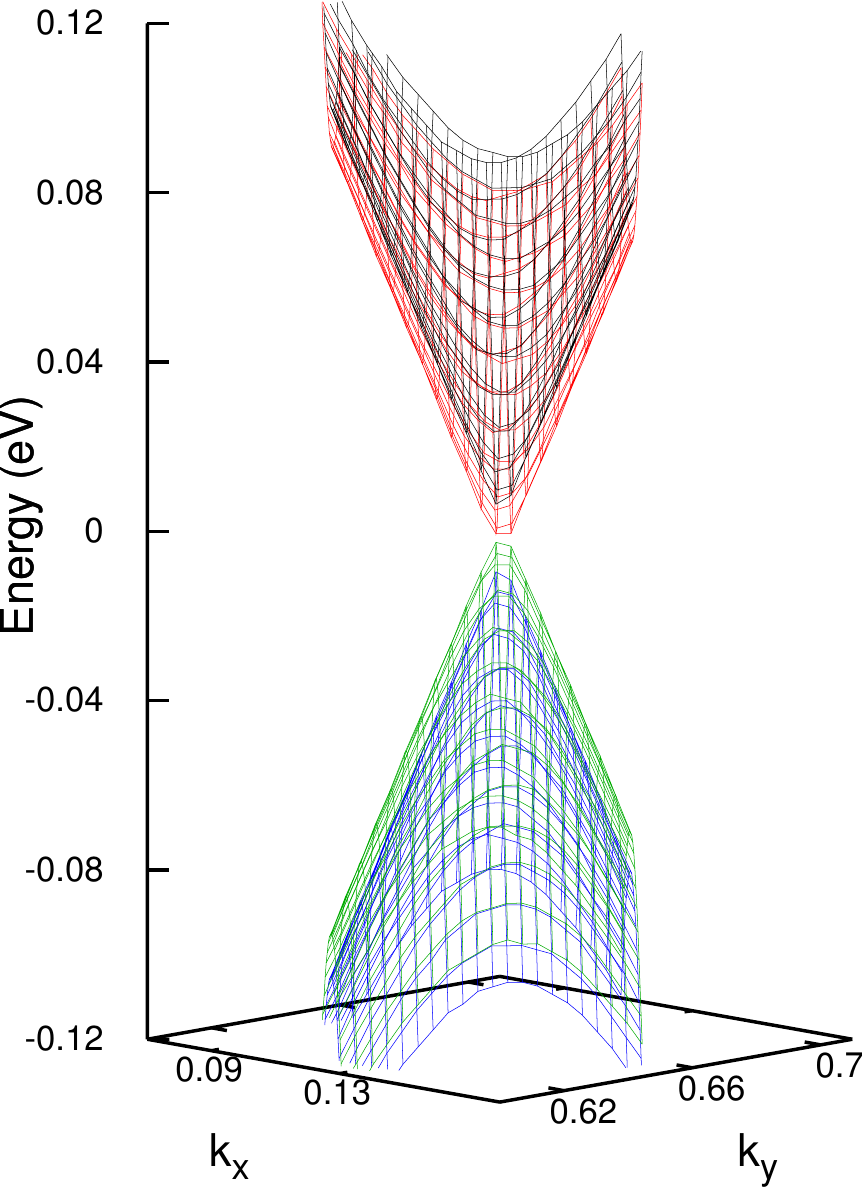}}
\caption{ Band structure of pristine TBLG in the vicinity of \textbf{K}$_1^{\rm B}$. Four bands are shown: two valence bands (blue and green online), and two conduction bands (red and black online).
The Fermi energy $E_F$ is at 0.}
\label{Fig:BS-TBLG}
\end{figure}

In Fig.~\ref{Fig:BS-TBLG} we plot the electronic band structure, as obtained using DFT-D2 for pristine (defect-free) twisted bilayer graphene, using the $S_2$ supercell, in the vicinity of the \textbf{K}$_1^{\rm B}$ point. In agreement with previous authors,\cite{Shallcross-PRB2008} we obtain a conical dispersion, with the Dirac point lying exactly at \textbf{K}$_1^{\rm B}$, and the Dirac crossing energy $E_D$=$E_F$. We have plotted the energy surfaces of the two topmost valence bands, we call these VB2 (blue in Fig.~\ref{Fig:BS-TBLG}), VB1 (green), and the two lowest-lying conduction bands, CB1 (red) and CB2 (black). Note that VB1 and VB2 are almost, but not quite, degenerate, being shifted in energy by $\sim$9 meV; similarly CB1 and CB2 are displaced in energy by $\sim$9 meV. Note also that there is no band gap between VB1 and CB1. These results are similar to those obtained by previous authors who have performed calculations on TBLG.\cite{Shallcross-PRB2008} Fig.~\ref{Fig:BS-TBLG} is useful in that it serves as a baseline to compare our other results to further below, when we will see how its features are modified upon the introduction of defects.

Figs.~2 (a)--(c) show simulated STM images of SLG, AB-BLG and TBLG; the input local densities of states for these calculations were
obtained from DFT-D2 calculations. For a single layer of graphene, the STM image clearly shows the honeycomb like arrangement of carbon atoms. 
In the case of AB-stacked bilayer graphene, the image reflects the three-fold symmetry of Bernal stacking. 
These images are in excellent agreement with earlier reported experimental STM images. \cite{Rutter-NatPhys2011} 
In the case of TBLG, at first sight, the STM image might appear rather similar to that of the single layer of graphene, 
however closer examination of the STM image brings out the effect of the twisted lower layer. 
The relatively brighter spots in this image correspond to the positions where the atoms of both the layers lie directly atop each other, thus enhancing the STM signal.

\subsection{Stone-Wales Defect}
A Stone-Wales (SW) defect is formed by rotating a single carbon-carbon bond in the graphene sheet by $90^{\circ}$, 
resulting in a structure with a pair each of seven-membered and five-membered rings. 
This is known to be one of the most common defects in graphene-related systems and carbon nanotubes.\cite{SW-CPL1986} 
Since it has been reported from calculations on SLG \cite{Ma-PRB2009}  that an out-of-plane distortion of the carbon atoms in the vicinity of the SW defect further stabilizes this defect, we explicitly check if permitting such distortions stabilizes the SW defect in TBLG. 

\subsubsection{Structural properties}
\begingroup
\squeezetable
\begin{table}[!b]
\begin{center}
\begin{tabular}{  l  c  c  c  c  c  c}
\hline 
\hline 
\multicolumn{7}{l}{Stone-Wales defect: Formation energies (eV)} \\ 
\hline 
\hline
Method/XC & SLG & AB-BLG    & \multicolumn{4}{c}{TBLG}                                            \\
          &                      & $\alpha\beta$-SW  & $\alpha\beta$-SW & $\beta\beta$-SW & $\beta\gamma$-SW & $\gamma\gamma$-SW \\
\hline
LDA       & 5.25              & 5.39           & 5.36         & 5.38          & 5.35           & 5.36            \\
          & (5.47)            & (5.47)         & (5.45)        & (5.47)        & (5.45)         & (5.46)          \\
PBE       & 5.14              & 5.15           & 5.16          & 5.16          & 5.15           & 5.16           \\
          & (5.35)            & (5.38)          & (5.35)         & (5.35)         & (5.35)          & (5.35)           \\
DFT-D2    & 5.09              & 5.27           & 5.24          & 5.27         & 5.23           & 5.24            \\
          & (5.32)            & (5.31)          & (5.30)         & (5.33)         & (5.29)          & (5.31)           \\
Expt.     &  6.02\cite{Zhou-APL2007} &                   &                  &                 &                  &                \\
\hline
\hline
\end{tabular}
\caption{Stone Wales defect formation energies for SLG, AB-BLG, and TBLG,  with  sine-like distortions in the defective layer. 
The values in  parentheses are the formation energies  when there is no distortion of the defective layer. }
\label{tab-SW-TBLG}
\end{center}
\end{table}
\endgroup

\begin{figure}[]
\centering
\subfigure[\ ]{\includegraphics[width=0.2\textwidth]{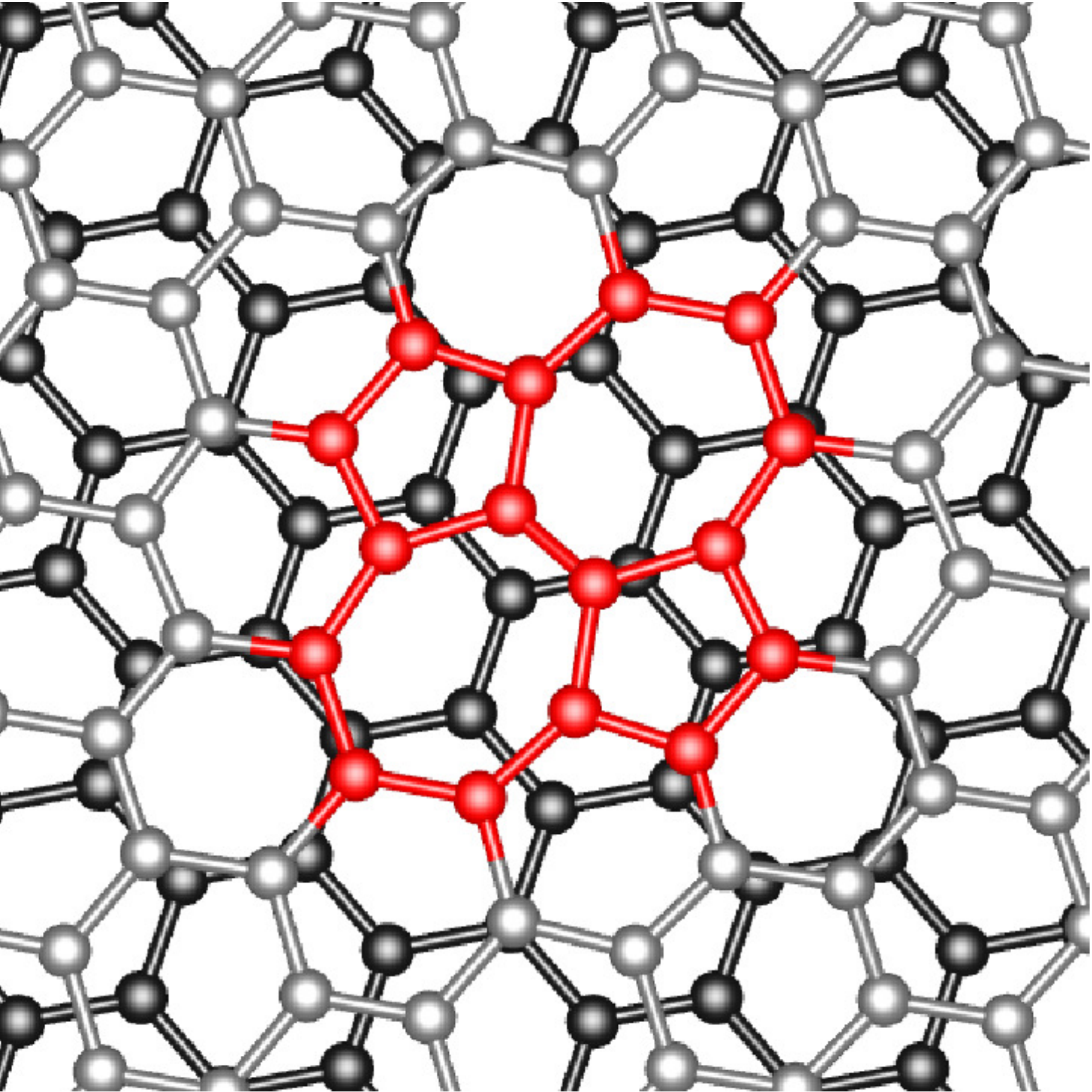}}
\quad
\subfigure[\ ]{\includegraphics[width=0.2\textwidth]{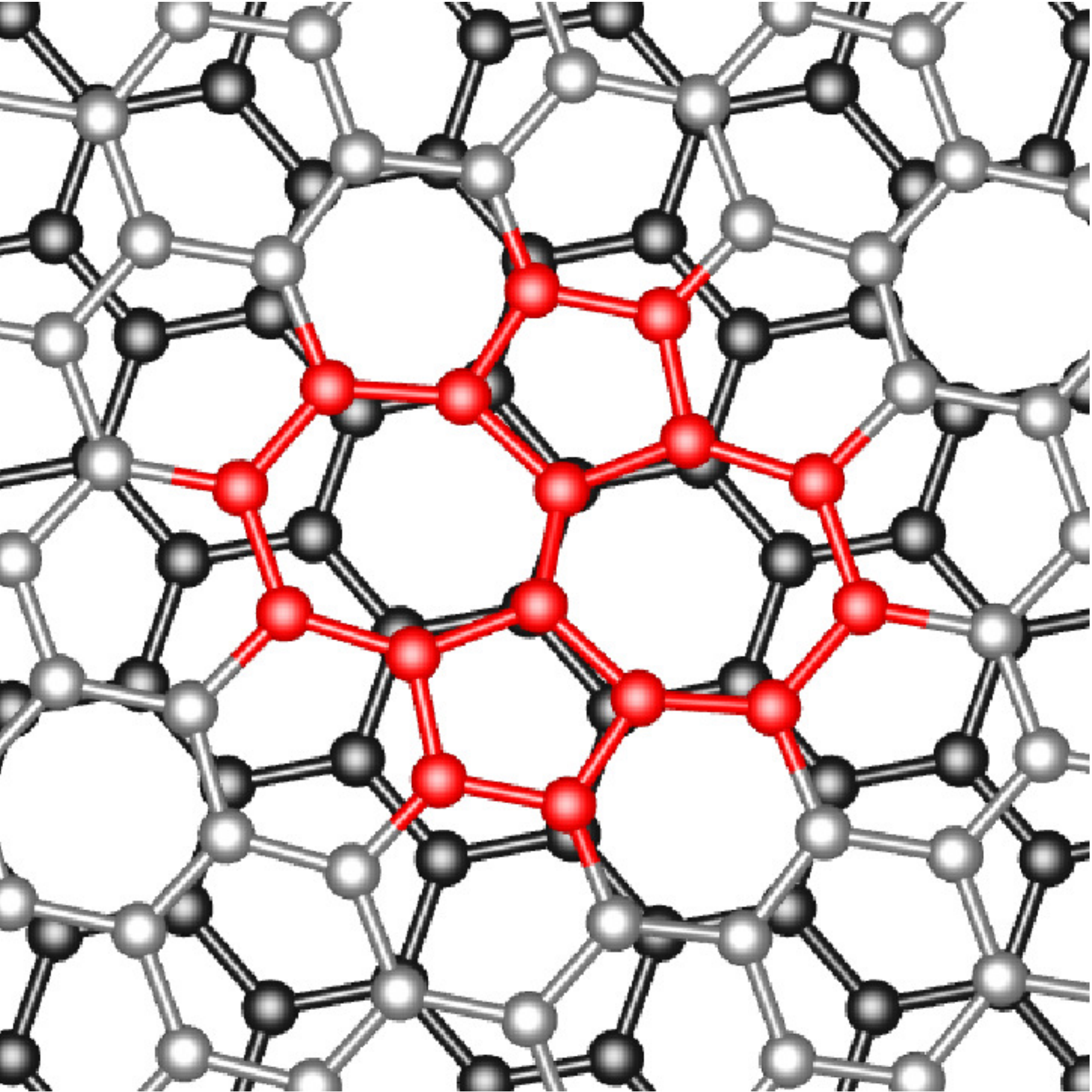}}
\quad
\subfigure[\ ]{\includegraphics[width=0.2\textwidth]{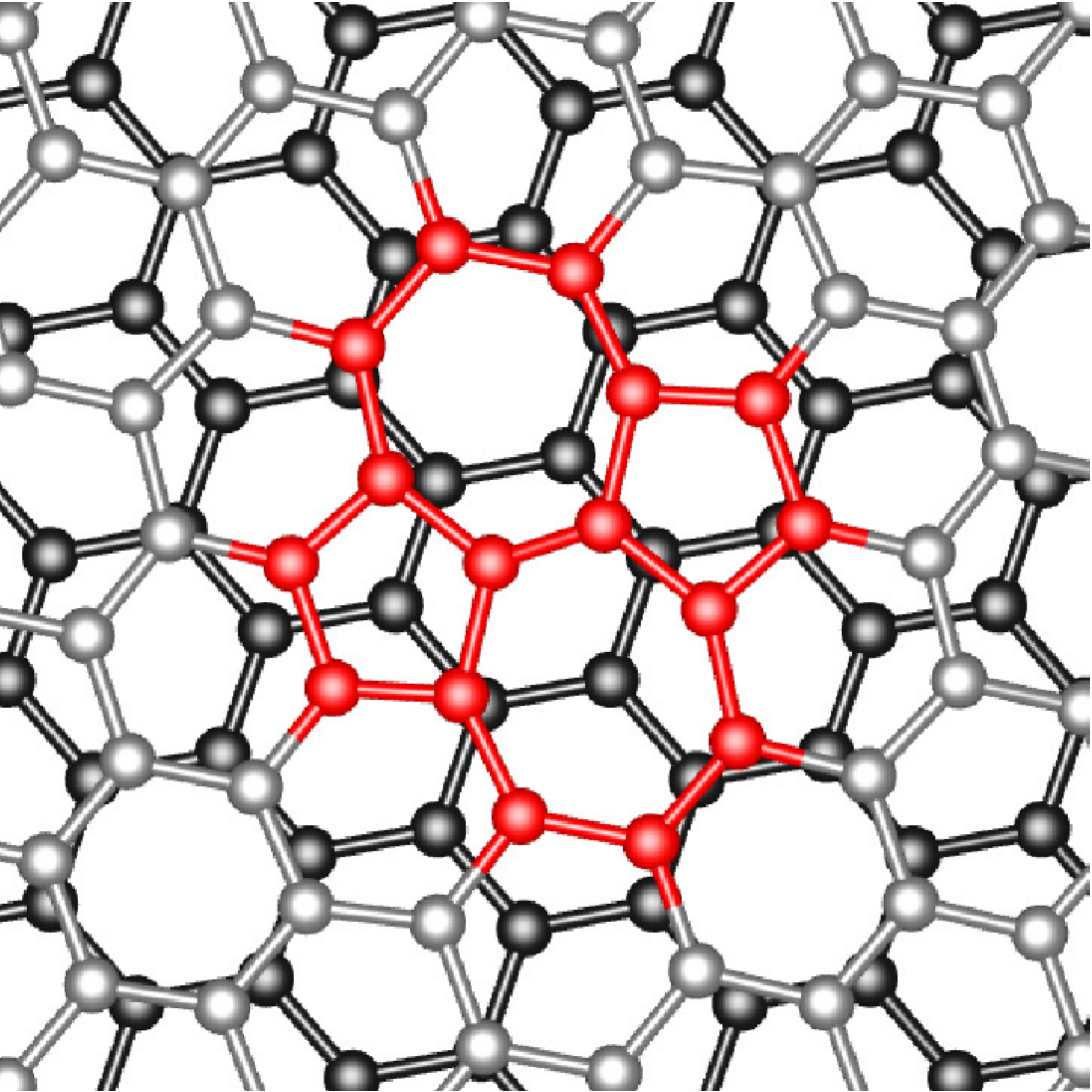}}
\quad
\subfigure[\ ]{\includegraphics[width=0.2\textwidth]{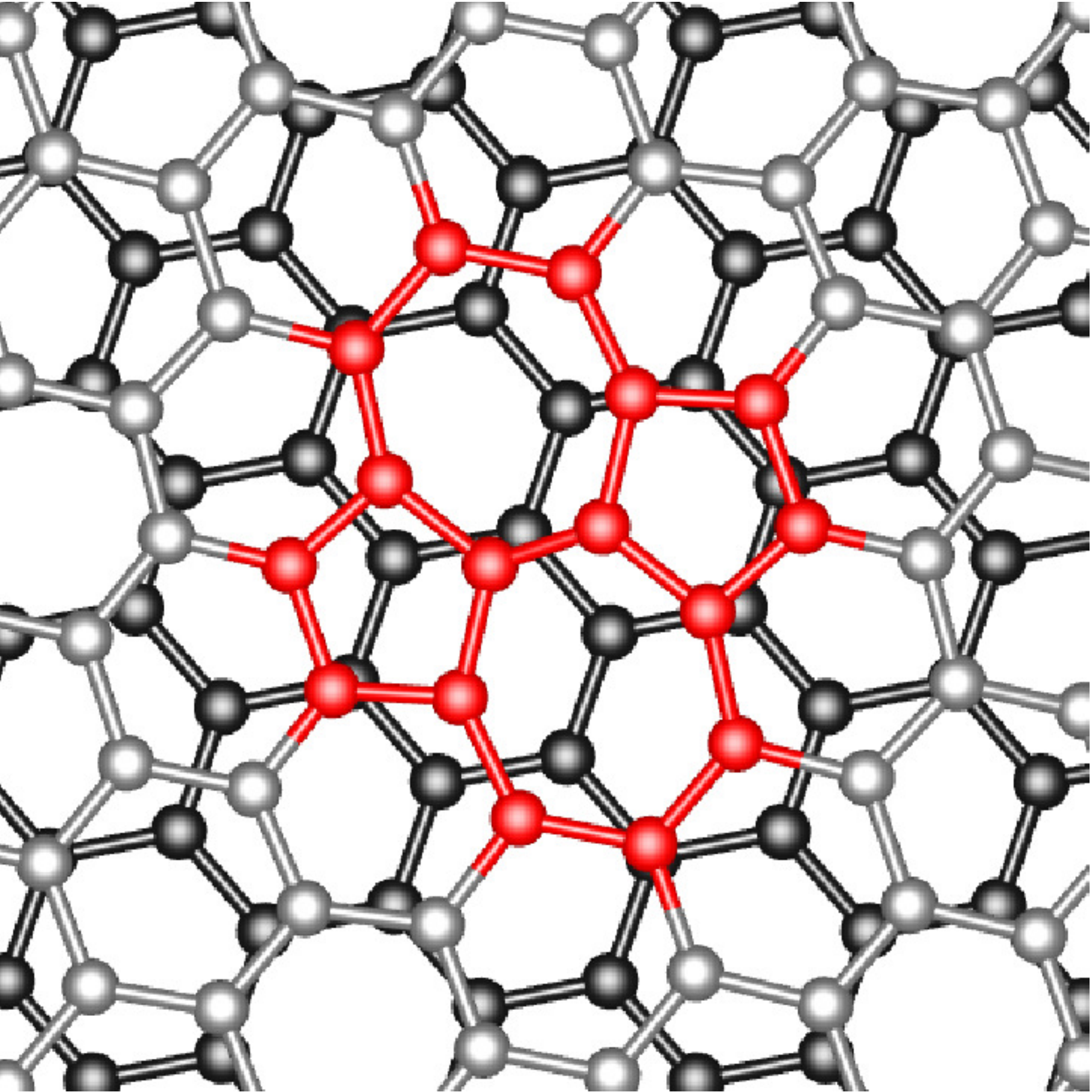}}
\caption{(Color online) Relaxed geometries for (a) $\alpha\beta$-SW, (b) $\beta\beta$-SW, (c) $\beta\gamma$-SW, and $\gamma\gamma$-SW defects in TBLG. Lower layer atoms are black, and the upper defective layer atoms are light gray. The C atoms near the SW defect are colored red for visual ease. }
\label{tBLG-SW}
\end{figure}

We create the SW defect in one of the layers of the TBLG, and compare the structural and electronic properties with those of SLG and AB-BLG with SW defects.
Unlike SLG or AB-BLG, where only one type of site for a SW defect is possible, in TBLG, there can be many inequivalent defect-sites, with the number depending on the angle of twist between the two layers. 
For the case considered by us, where the two layers are rotated by a relative angle of 38.213$^{\circ}$, 
there can be four inequivalent geometries (see Fig.~\ref{tBLG-SW}), corresponding to four different choices of the rotated bond; we name these  $\alpha$$\beta$-SW, 
$\beta$$\gamma$-SW, $\beta$$\beta$-SW, and $\gamma$$\gamma$-SW, with the labeling convention denoting which pair of adjacent atoms 
is connected by the rotated bond (see Fig.~\ref{Fig:tBLG}). 

\begin{figure}[]
\centering
\subfigure{\includegraphics[width=4.5cm]{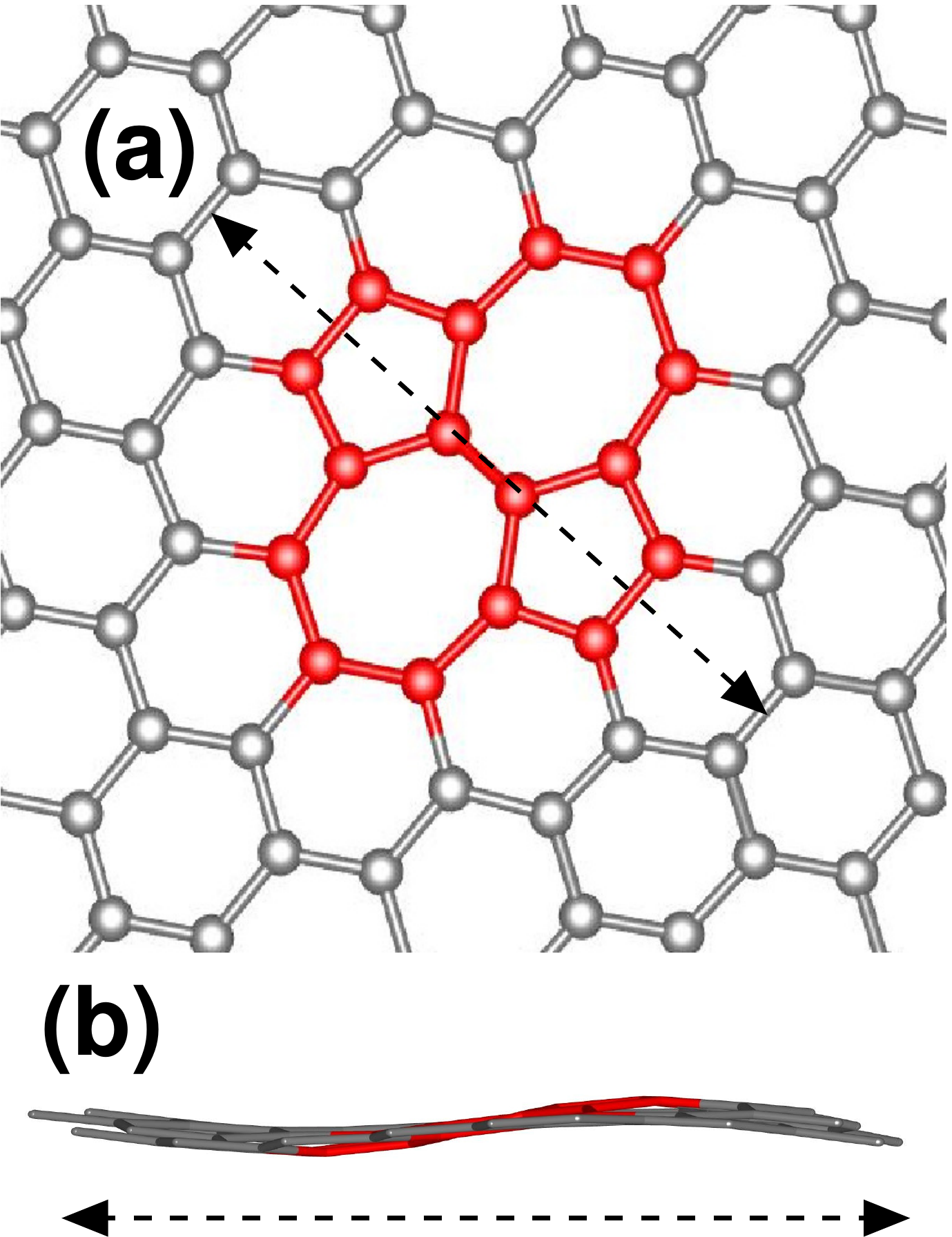}}
\quad\quad
\subfigure{\includegraphics[width=8.5cm]{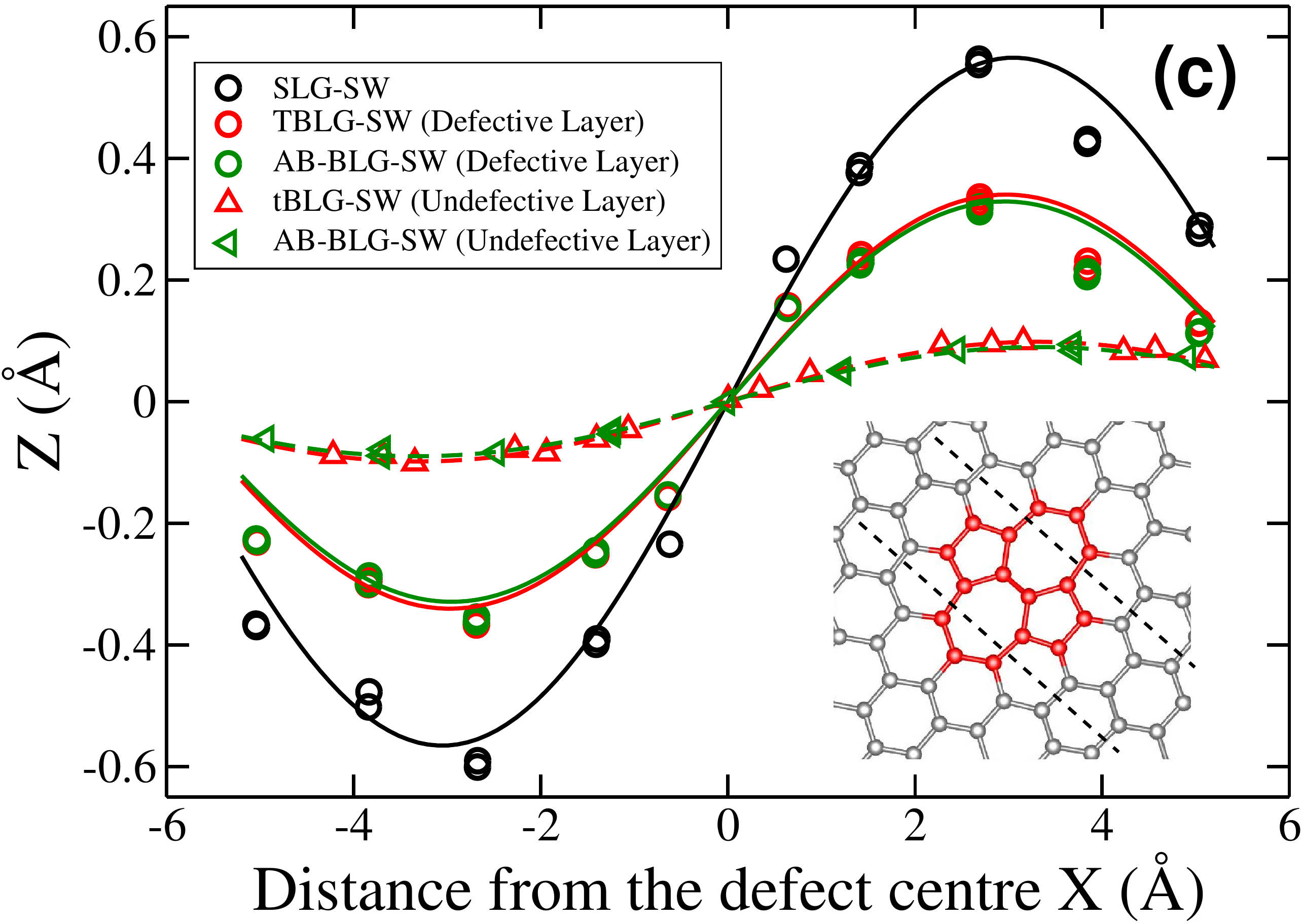}}
\caption{ Structure for SW defect in SLG, (a) the top-view and (b) the side-view showing the distortion of the defective graphene sheet. 
The side view in (b) is taken by viewing perpendicular to the dashed line in (a). 
(c) The distortion pattern for the SW defects in SLG, TBLG and AB-BLG. The zero of the distortion amplitude is taken at the center of the SW defect. 
The dashed line in the inset encloses the atoms over which the deviations from the zero are measured. }
\label{Fig:SW-Structures}
\end{figure}

For SLG, we find that a sinusoidal distortion pattern about the defect center 
results in the most stable geometry of the SW defect [see Figs.~\ref{Fig:SW-Structures}(a) and (b)], in agreement with earlier reports. \cite{Ma-PRB2009} 
The amplitude of distortion, $\rm Z$, (the difference between the largest upward displacement and the largest downward displacement) 
is as large as 1.2 \AA\ [see Fig.~\ref{Fig:SW-Structures}(c)] for a SW defect in SLG within a supercell of size $S_2$, 
which is in good agreement with earlier values of about 1 \AA\ for supercells of similar size. 
In the case of the bilayers, we find that SW defects in both AB-BLG and TBLG are stabilized by a similar sinusoidal distortion pattern, however the amplitude of the distortion in the defective layer is reduced 
to about 0.7 \AA; the presence of the second layer inhibits the distortion. 
In our study, we have also allowed the undefective layer to relax, it exhibits a smaller distortion amplitude of $\sim$0.2 \AA. 
Fig.~\ref{Fig:SW-Structures}(c) shows the comparison of the net distortion of the graphene layers for the cases of single layer, AB-stacked bilayer, and twisted bilayer of graphene. 
Both the twisted and the AB-stacked bilayers show a very similar extent of distortion. 

The energy of formation of the SW defect, as obtained with the various approaches, is given in Table~\ref{tab-SW-TBLG}.  The values listed in parentheses in Table~\ref{tab-SW-TBLG} give the formation energy for the SW defect when no out-of-plane distortion is permitted. Firstly for SLG, by comparing the results when this distortion is permitted with those
when it is suppressed, we find that the sinusoidal distortion leads to a stabilization of about 220 meV/defect; this is in good agreement with earlier results.\cite{Ma-PRB2009}

In the case of bilayers, the distortion results in a stabilization of the SW defect, as compared to the undistorted case, 
by an energy of about 40 meV in AB-BLG, and about 60 meV in TBLG. For TBLG, we find that the defect formation energy depends only very slightly on the position of the defect, this is because the interaction between the layers is small. Similarly, the energy to form a SW defect in TBLG is very similar to that in AB-stacked BLG. The small differences in defect formation energy (of the order of tens of meV) between the different kinds of Stone-Wales defects in TBLG are due to the differences in interlayer coupling when the SW defect is
situated at different positions in the graphene sheet. Note that these differences disappear when the calculations are performed with the PBE alone, since in this case the interlayer coupling
is described very poorly and becomes negligible. From the DFT-D2 calculation, we see that it costs an additional 140--180 meV to create a SW defect in TBLG than in SLG; the majority of this increase comes
because the energy-lowering due to distortion along the $z$-direction is hindered by the presence of the second layer. The same effect is reflected in our LDA results; once again, it is absent
in the PBE results.

\subsubsection{Band Structure}\label{SW-ElProp}
The band structure of AB-stacked bilayer graphene is characterized by parabolic bands that touch near the $\textbf{K}$ points in the Brillouin zone, at the Fermi level $E_F$. In contrast,
for twisted bilayer graphene, there are Dirac cones, i.e., linear bands that touch at the $\textbf{K}$ points, at $E_F$ (recall Fig.~\ref{Fig:BS-TBLG} above). 
Given the great interest in linear dispersion, as well as the potential importance of the band-gap engineering of graphene, we wish to know 
whether, upon the introduction of defects such as Stone-Wales defects and monovacancies, (i) the Dirac cones are maintained (ii) if not, in what way are they altered, i.e., are they shifted in energy (iii)  whether gaps open up, and if so, at what energies. As we will show below, clear signatures of the importance of interlayer coupling emerge when looking at the band structure, as opposed to the energetics of defect formation.

\begin{figure}[t!]
\centering
\hspace{1.3cm}
\subfigure{\label{BZ-TBLGonly}\includegraphics[width=0.4\textwidth]{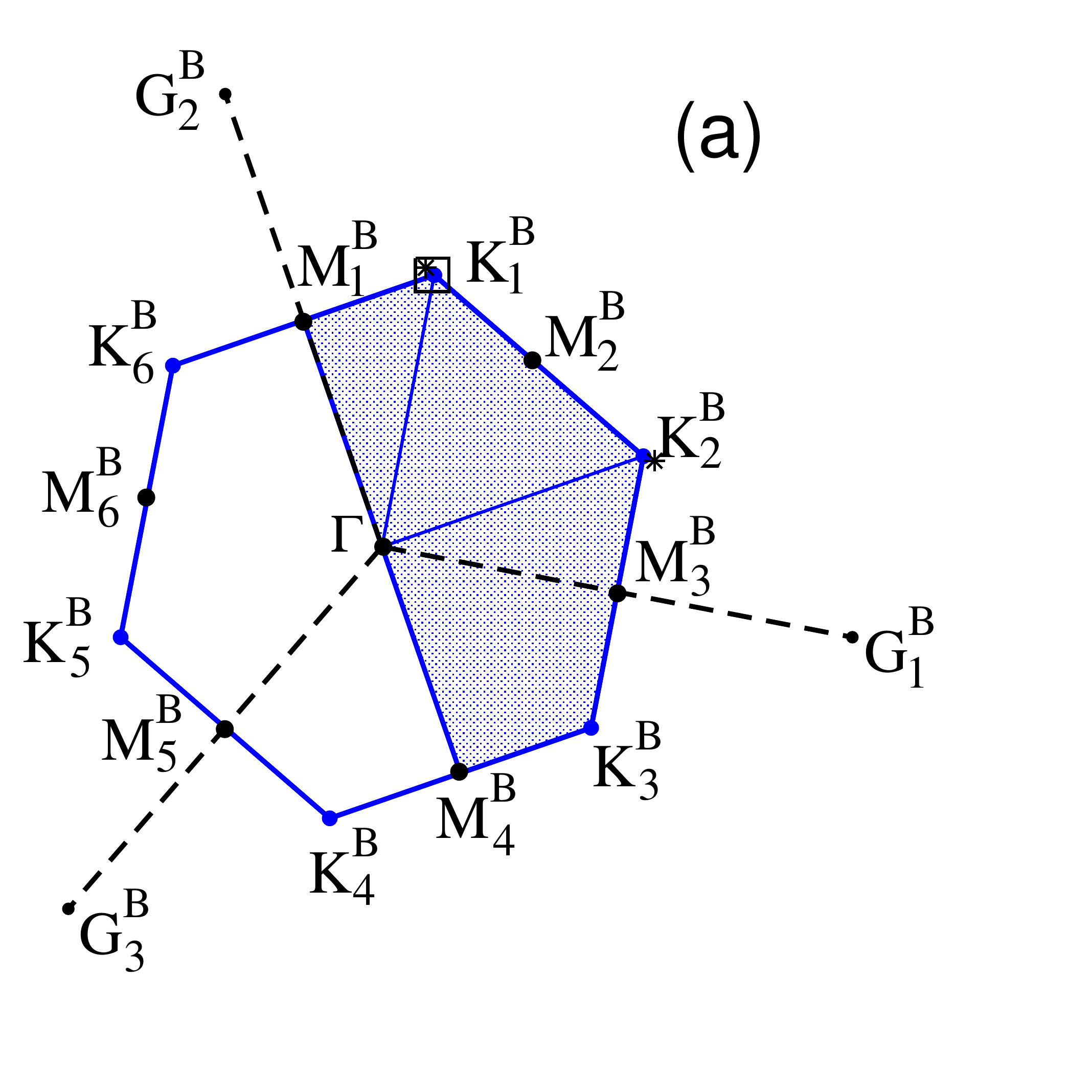}}
\quad
\subfigure{\label{BZ-w-shift}\includegraphics[width=0.4\textwidth]{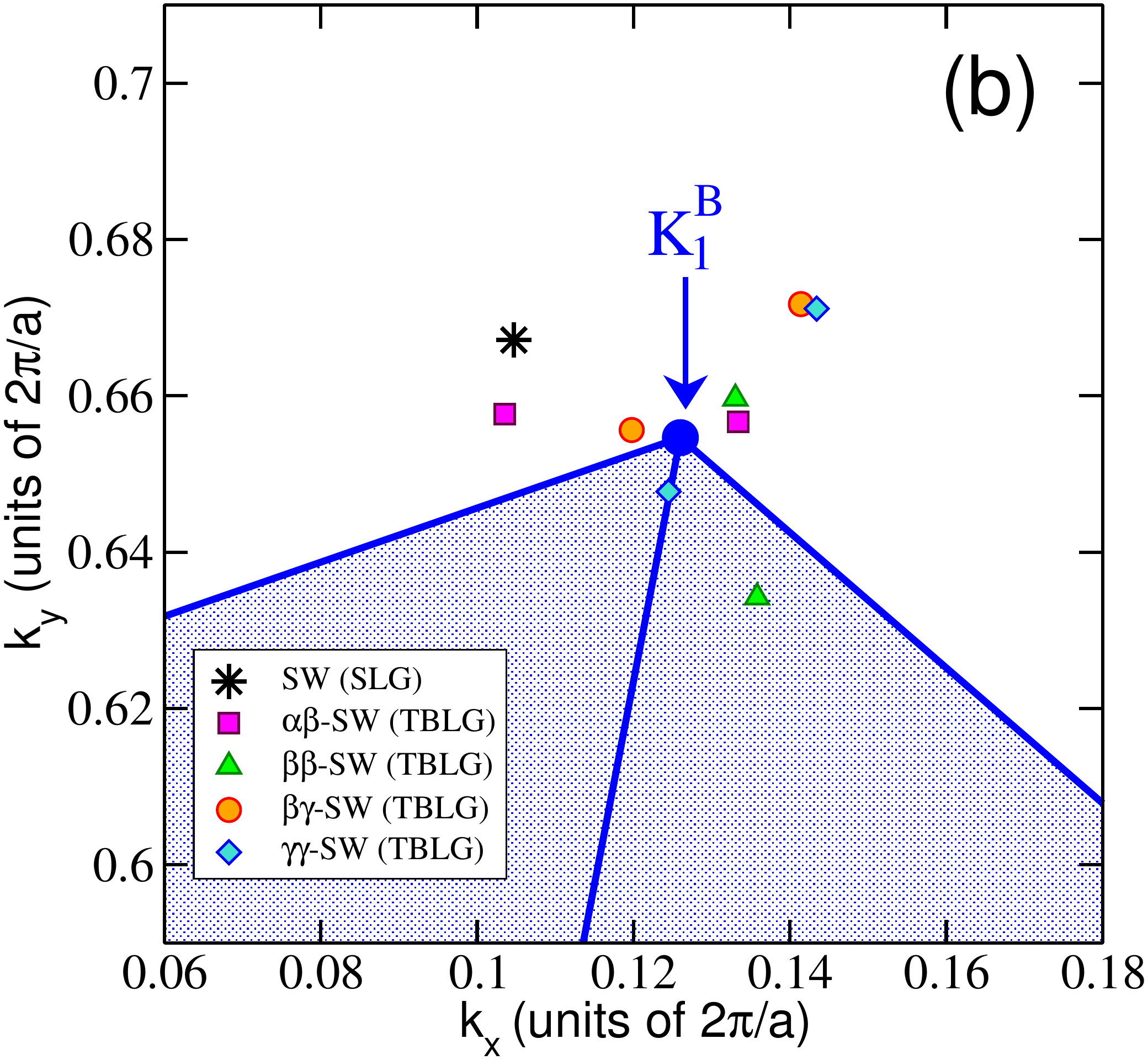}}
\quad
\caption{(Color online) (a) The First Brillouin Zone for the $S_2$ supercell of TBLG, and (b) the zoomed view of the region marked by the square in (a) showing the position of the Dirac point $E_D$ for SLG with SW defect (asterix) and Dirac points $E_D^1$ and $E_D^2$ for the four types of SW defect in TBLG.  }
\label{Fig:BZ-S2}
\end{figure}

\begin{figure}[]
\centering
\subfigure[\ SW (SLG)]{\label{SWBS}\includegraphics[width=4cm]{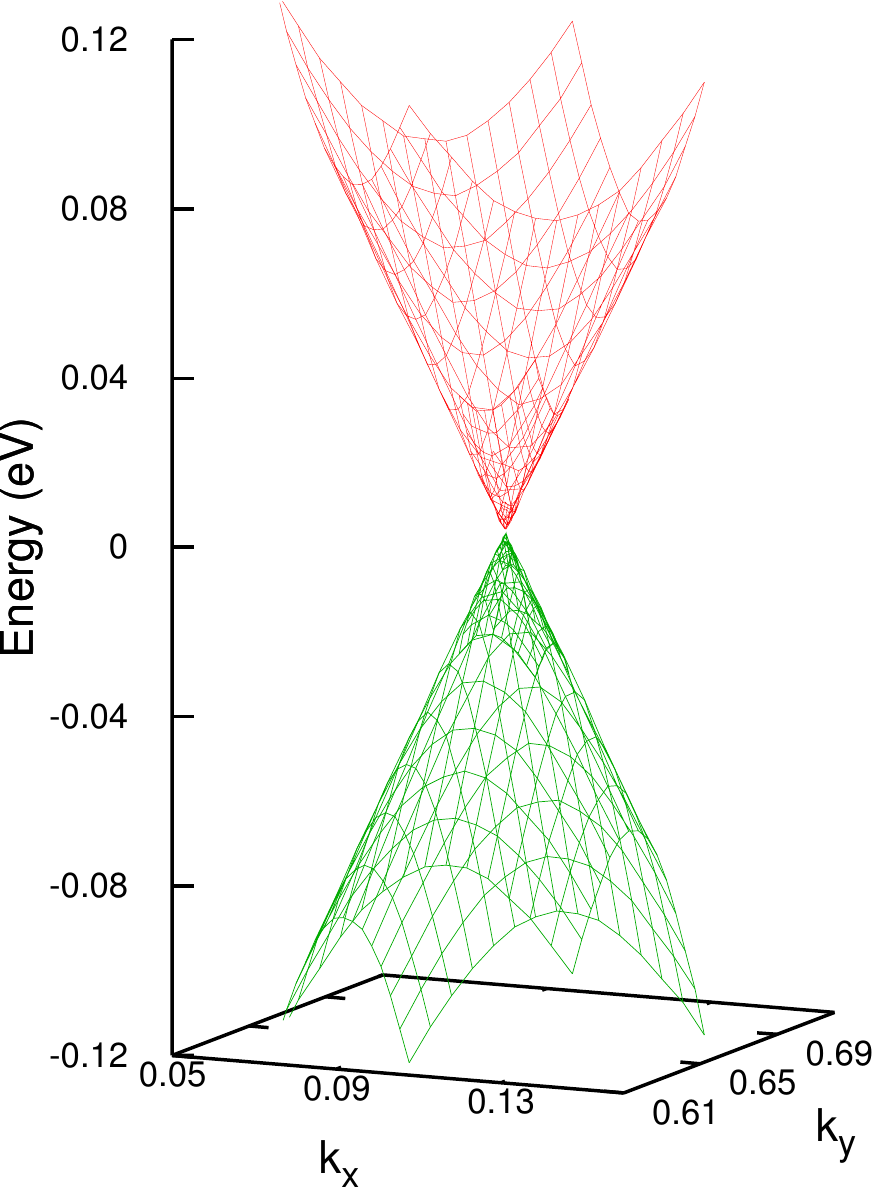}}
\quad
\subfigure[\ SW (AB-stacked BLG)]{\label{ABBLGBS}\includegraphics[width=4cm]{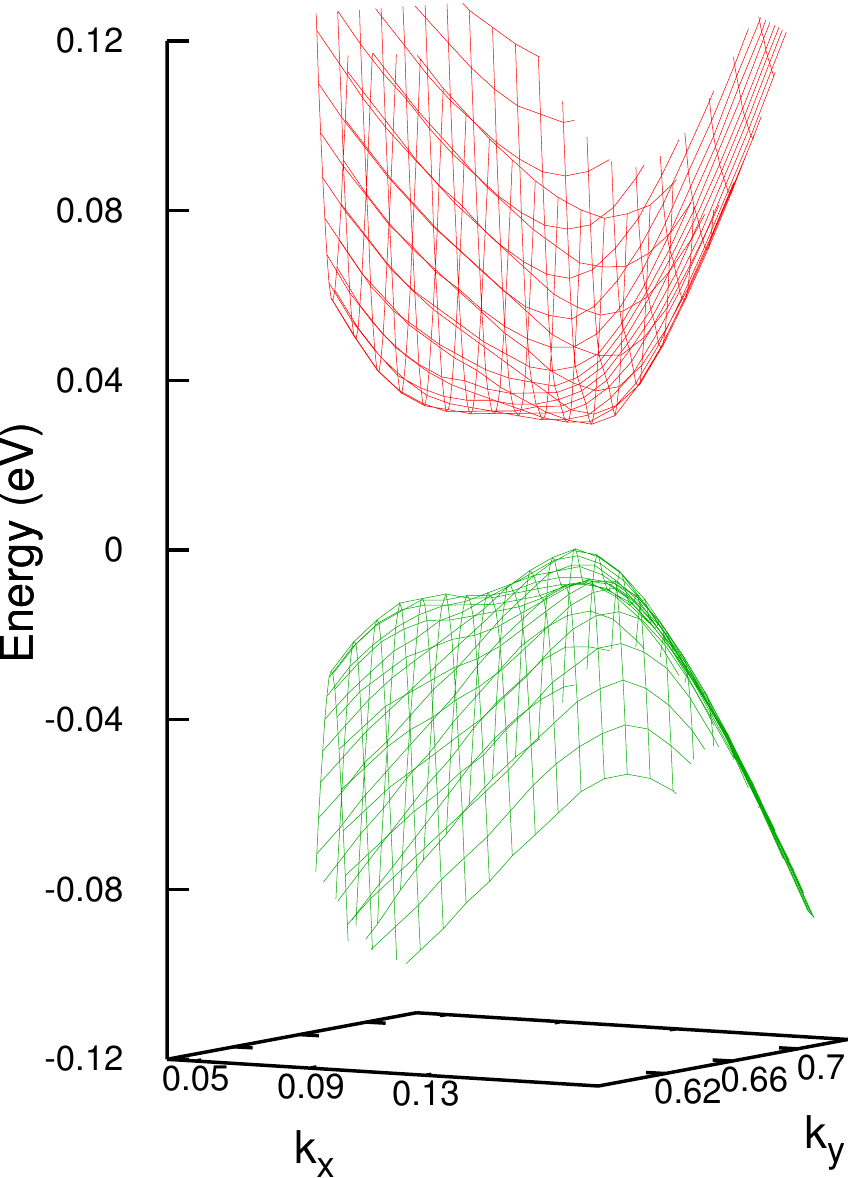}}
\quad
\subfigure[\ $\gamma$$\gamma$-SW (TBLG)]{\label{TBLGCC-BS}\includegraphics[width=4cm]{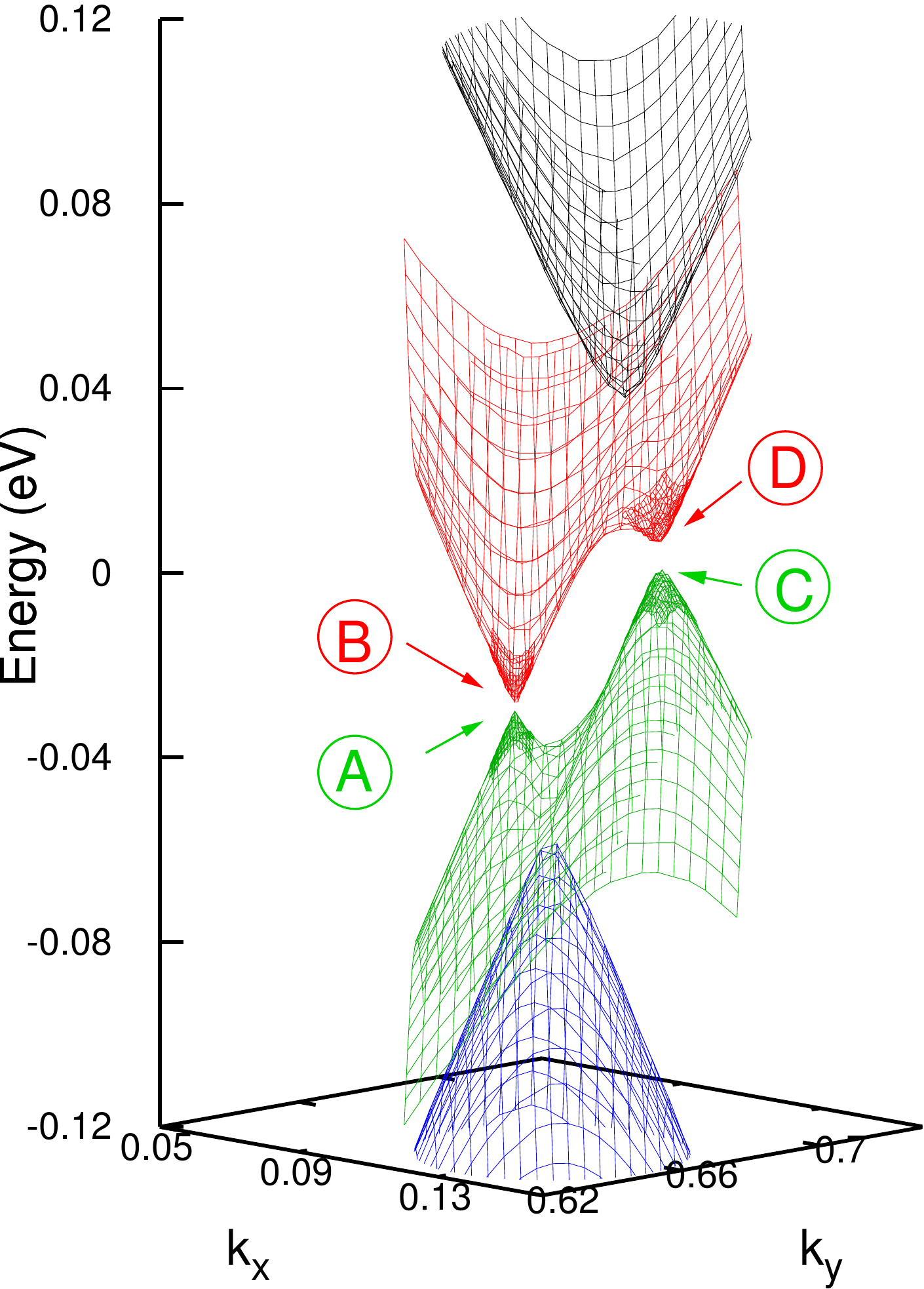}}
\quad
\subfigure[\ $\alpha$$\beta$-SW (TBLG)]{\label{TBLGAB-BS}\includegraphics[width=4cm]{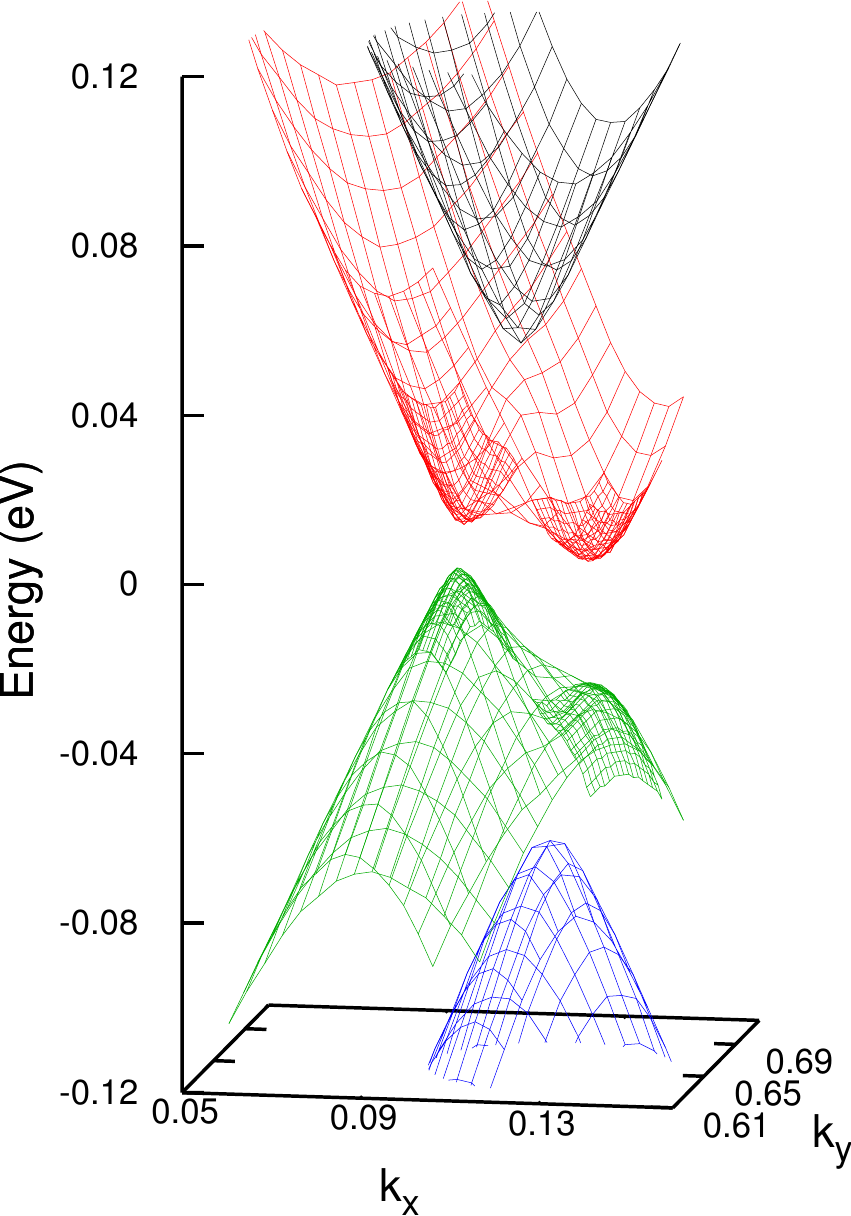}}
\quad
\subfigure[\ $\beta$$\beta$-SW (TBLG)]{\label{TBLG-BBBS}\includegraphics[width=4cm]{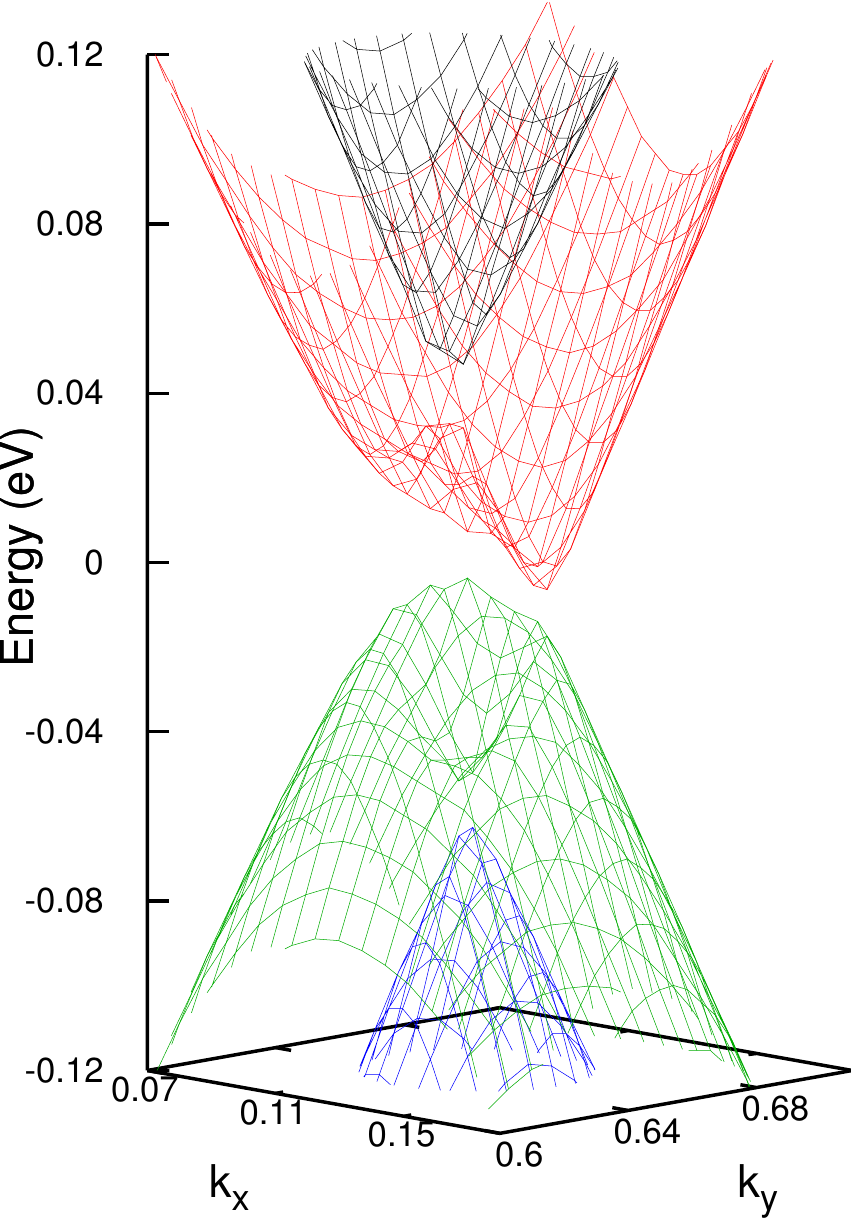}}
\quad
\subfigure[\ $\beta$$\gamma$-SW (TBLG)]{\label{TBLGBC-BS}\includegraphics[width=4cm]{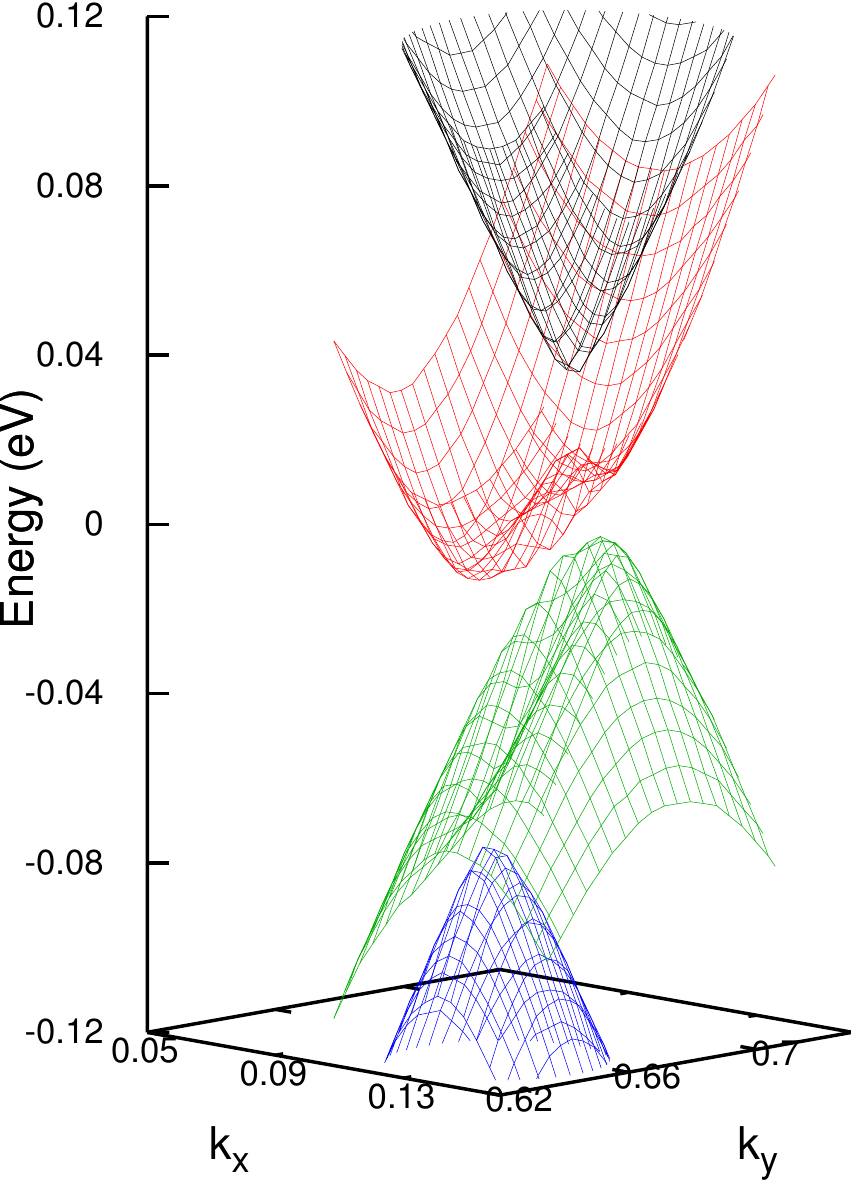}}
\caption{ (a) 3$d$ band structure in the vicinity of the Dirac point for SW defect in $S_2$ supercell of SLG, and (b) 3$d$ band structure in the vicinity of \textbf{K}$_1^{\rm B}$  for SW defect in $S_2$ supercell of AB-stacked BLG. Panels (c), (d), (e), and (f) show the 3$d$ band structure for $\gamma$$\gamma$-SW, $\alpha$$\beta$-SW, $\beta$$\gamma$-SW, and $\beta$$\beta$-SW, respectively in the vicinity of \textbf{K}$_1^{\rm B}$. The degenerate Dirac bands for the TBLG split into a complex band-structure in addition to the shift from \textbf{K}$_1^{\rm B}$. The valence band maxima (VBM) and the conduction band minima (CBM) are shown in panel (b) of Fig.~\ref{Fig:BZ-S2}, for all the cases of SW defect in TBLG. }
\label{Fig:SLG-AB_BLG-TBLG-SW-Bands}
\end{figure}

We will examine the electronic band structure of SLG, AB-BLG and TBLG, containing a single SW defect in the $S_2$ supercell. 
No perceptible spin-polarization was found in any of the cases, and we therefore present non-spin-polarized results.
Our structures break the reflection symmetry about the two axes passing through the SW defect. As a result, the irreducible Brillouin zone (IBZ) is half of the first Brillouin zone [shown by the shaded region in the TBLG Brillouin zone in Fig.~\ref{BZ-TBLGonly}]. Let us consider the high-symmetry points that lie in this IBZ. As mentioned in Section III above, by the translational symmetry of the lattice, \textbf{K}$_1^{\rm B}$ is identical to \textbf{K}$_3^{\rm B}$; further, by making use of time-reversal symmetry, these are also identical to  \textbf{K}$_2^{\rm B}$. Thus, electronic eigenvalues should be identical at these three \textbf{K} points. On the other hand, no symmetries relate the three points \textbf{M}$_1^{\rm B}$, \textbf{M}$_2^{\rm B}$ and \textbf{M}$_3^{\rm B}$. However, as we will see below, it is not particularly useful, and perhaps even misleading, to plot the band structure for these systems by taking  cuts along high-symmetry directions of the Brillouin zone, as is frequently done. Instead, we plot the surfaces of energy dispersion $E(k_x, k_y)$, in the vicinity of the point \textbf{K}$_1^{\rm B}$, which is the region of especial interest. These results, for SW defects in SLG and AB-BLG, and the four distinct types of SW defects in TBLG, are plotted in Fig.~\ref{Fig:SLG-AB_BLG-TBLG-SW-Bands}.

Before we proceed to the band structure of TBLG, it is of interest to first consider the band structure of SLG with a SW defect, as it will help us interpret our results for TBLG with  SW defects. The band-structure of SW defects in SLG has been the subject of a debate in the literature. While some authors reported that the presence of the SW defect opened up a gap,\cite{Peng-NanoLett2008,Popov-Carbon2009} other authors have shown that this phenomenon depends on the size of the unit cell used;\cite{PRB2010-Sharmila} in addition, there is a shift of the Dirac point away from the K point. The cause of these shifts in the Dirac point has been attributed to electron-phonon coupling;\cite{Pisana-NatMat2007} note that the  90$^\circ$ bond rotation involved in the formation of SW defects can be viewed as a 
linear combination of $\Gamma$-point phonon modes.\cite{PRB2010-Sharmila} In Fig.~\ref{SWBS} we see that the Dirac cone is preserved even after introducing the SW defect, but the Dirac point has been shifted in k-space [see also Fig.~\ref{BZ-w-shift}, where the position of the shifted Dirac point is indicated by the asterisk]. Note however that the Dirac crossing energy $E_D$ remains at the Fermi energy $E_F$.

Next, we consider the band structure of a SW defect in AB-BLG [see Fig.~\ref{ABBLGBS}]. The higher valence band VB1, and the lower conduction band CB1, have been plotted; VB2/CB2 lie
considerably lower/higher in energy and are therefore not seen in this figure. A band gap of $\sim$34 meV has opened up. The bands are very flat in the vicinity
of \textbf{K}$_1^{\rm B}$, and we therefore do not define a Dirac point. 

Finally, we consider the band structure of the four kinds of SW defects in TBLG; these results, in the vicinity of \textbf{K}$_1^{\rm B}$, are shown in Figs.~\ref{TBLGCC-BS}--\ref{TBLGBC-BS}.
It is interesting to compare these figures with Fig.~\ref{Fig:BS-TBLG}, which shows the band
structure of TBLG in the absence of the SW defect. We see that the introduction of the SW defect has had a significant impact on the band structure of TBLG. Moreover, the four figures ~\ref{TBLGCC-BS}--\ref{TBLGBC-BS} show perceptible differences. This is because, as can be seen in Fig.~\ref{tBLG-SW}, the registry between the atoms of the two layers is quite different for the four types of SW defects, as a result
of which the interlayer coupling is quite different in the four cases. 

Let us first consider Fig.~\ref{TBLGCC-BS}, which is the easiest of the four to understand in terms of how it arises from two twisted bilayers, one of which contains a SW defect. The Dirac points of the two layers are shifted from each other in k-space, \textit{even when we consider the ``small" Brillouin zone that is commensurate with both layers} [See the blue hexagon in Fig.~\ref{Fig:TBLG-FBZ}, also shown in Fig.~\ref{Fig:BZ-S2}]. We wish to emphasize that this is different from the usual case of pristine TBLG with, e.g., the $S_2$ supercell, where the Dirac points corresponding to the upper and lower layers are shifted in extended k-space, but fold back to the  \textbf{K}$_1^{\rm B}$ point when considering the ``small" Brillouin zone corresponding to the commensurate supercell. For conceptual purposes, it is easy
to think of each of these two Dirac points as arising from the Dirac points of the two individual layers, one unrotated and pristine, and the other rotated and containing a SW defect (note, however, that we will see further below that this picture is oversimplified, and needs to be further qualified). For TBLG with a $\gamma\gamma$ SW defect, the Dirac point that arises primarily from the unrotated and undefective layer lies very close to the \textbf{K}$_1^{\rm B}$ point [see the position of the blue diamond that lies within the first BZ in Fig.~\ref{Fig:BZ-S2}(b)]. The second Dirac point, which arises primarily from the rotated defective layer, is shifted further away from \textbf{K}$_1^{\rm B}$ [see the position of the second blue diamond in Fig.~\ref{Fig:BZ-S2}(b)], as a result of the SW defect, as was seen, e.g., in Fig.~\ref{SWBS}. Very small minigaps open up at the two Dirac points due to interactions between the two layers. The two Dirac crossing energies $E_D^1$ and $E_D^2$ are slightly shifted with respect to each other, with one lying slightly below the Fermi energy $E_F$, and the other lying slightly above it. At energies further away from $E_F$, the two Dirac cones intersect, and the resulting avoided crossings result in the opening up of band gaps, as a result of which VB1 has a skewed ``M" shape, CB1 has a skewed ``W" shape, and the bands VB2 and CB2 directly below and above these have conical shapes. The saddle points in VB1 and CB1 that form where the two cones fuse will give rise to van Hove singularities in the electronic density of states.

\begin{figure}[]
\centering
\subfigure[\ ]{\includegraphics[width=0.22\textwidth]{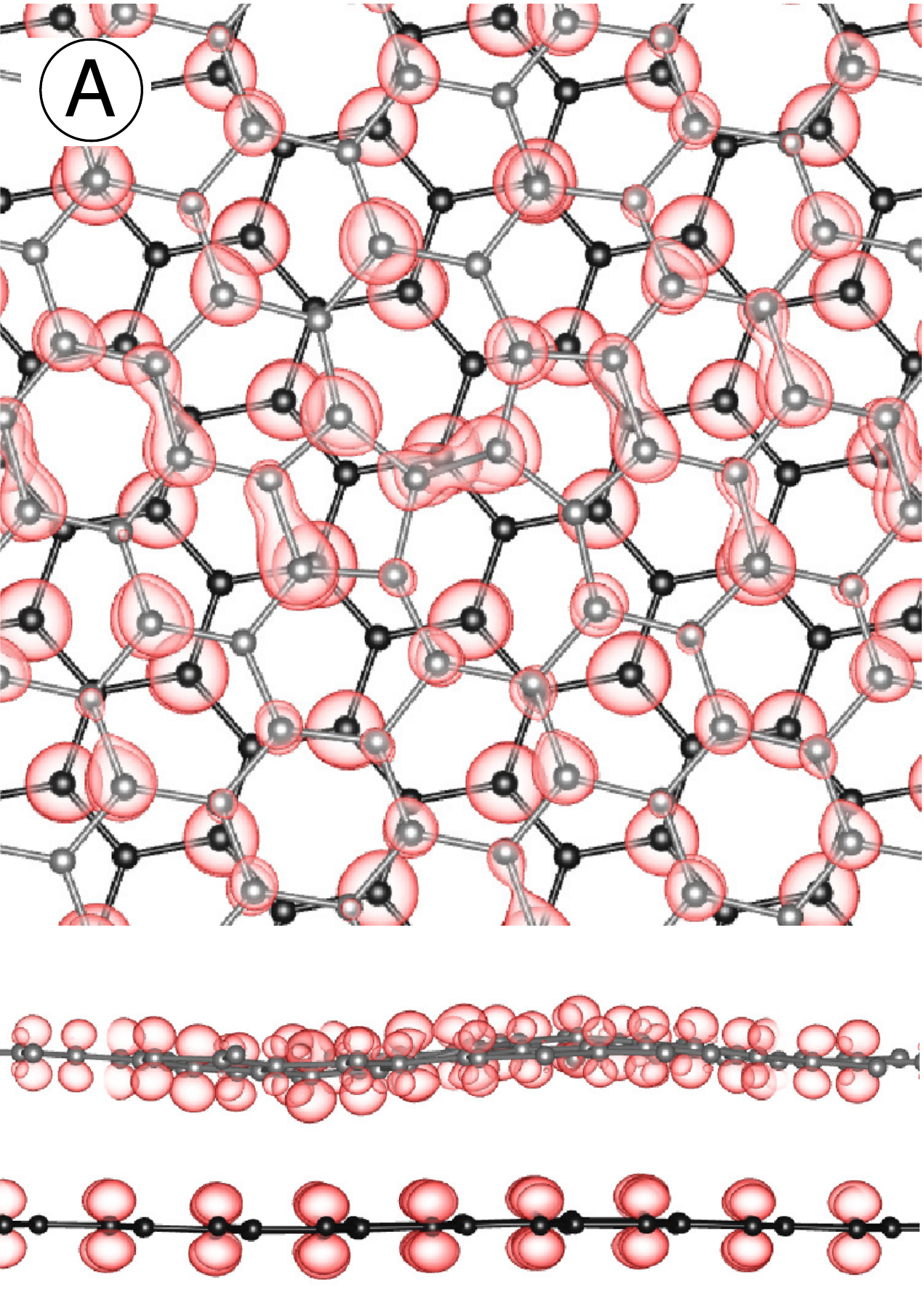}}
\quad
\subfigure[\ ]{\includegraphics[width=0.22\textwidth]{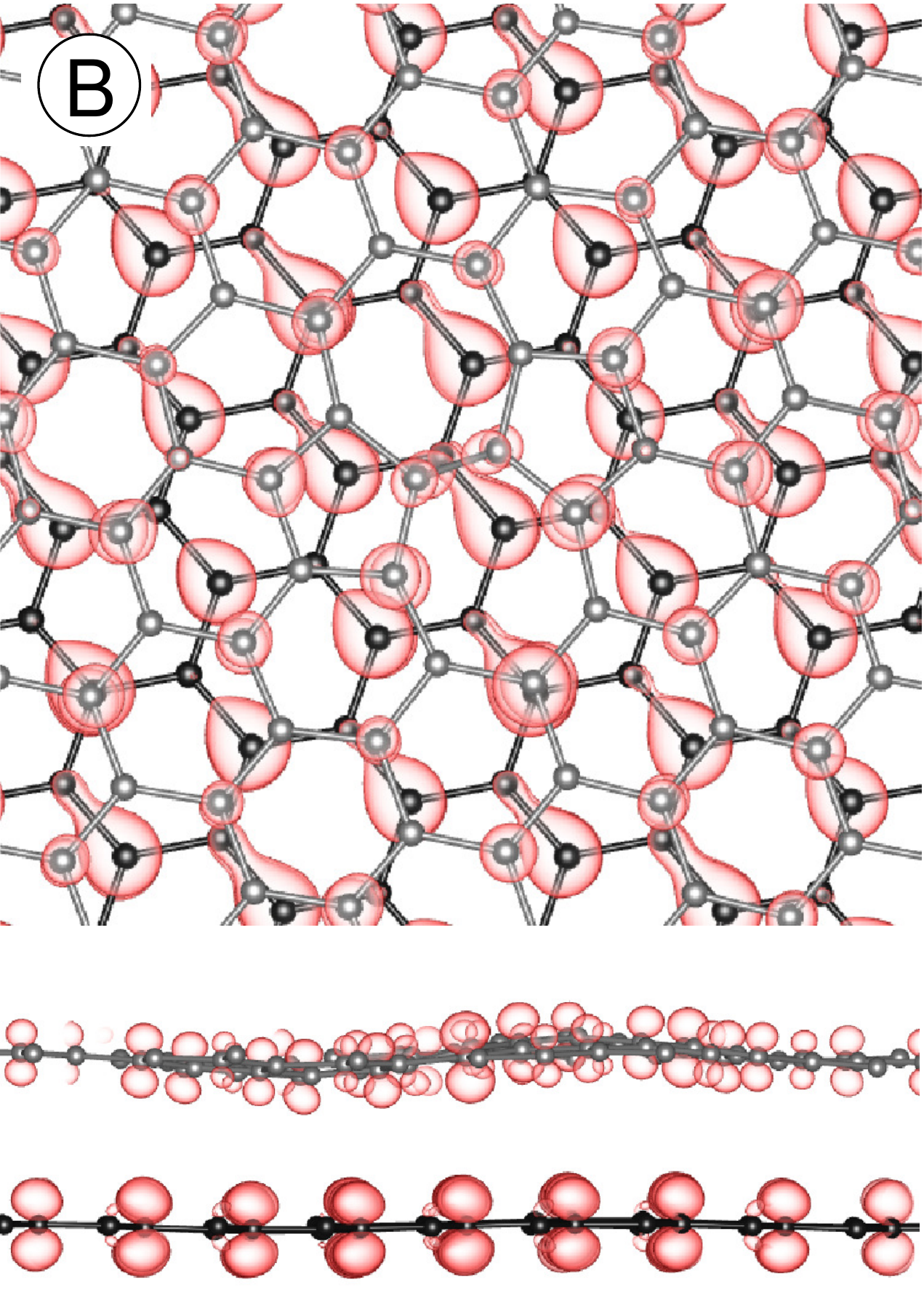}}
\quad
\subfigure[\ ]{\includegraphics[width=0.22\textwidth]{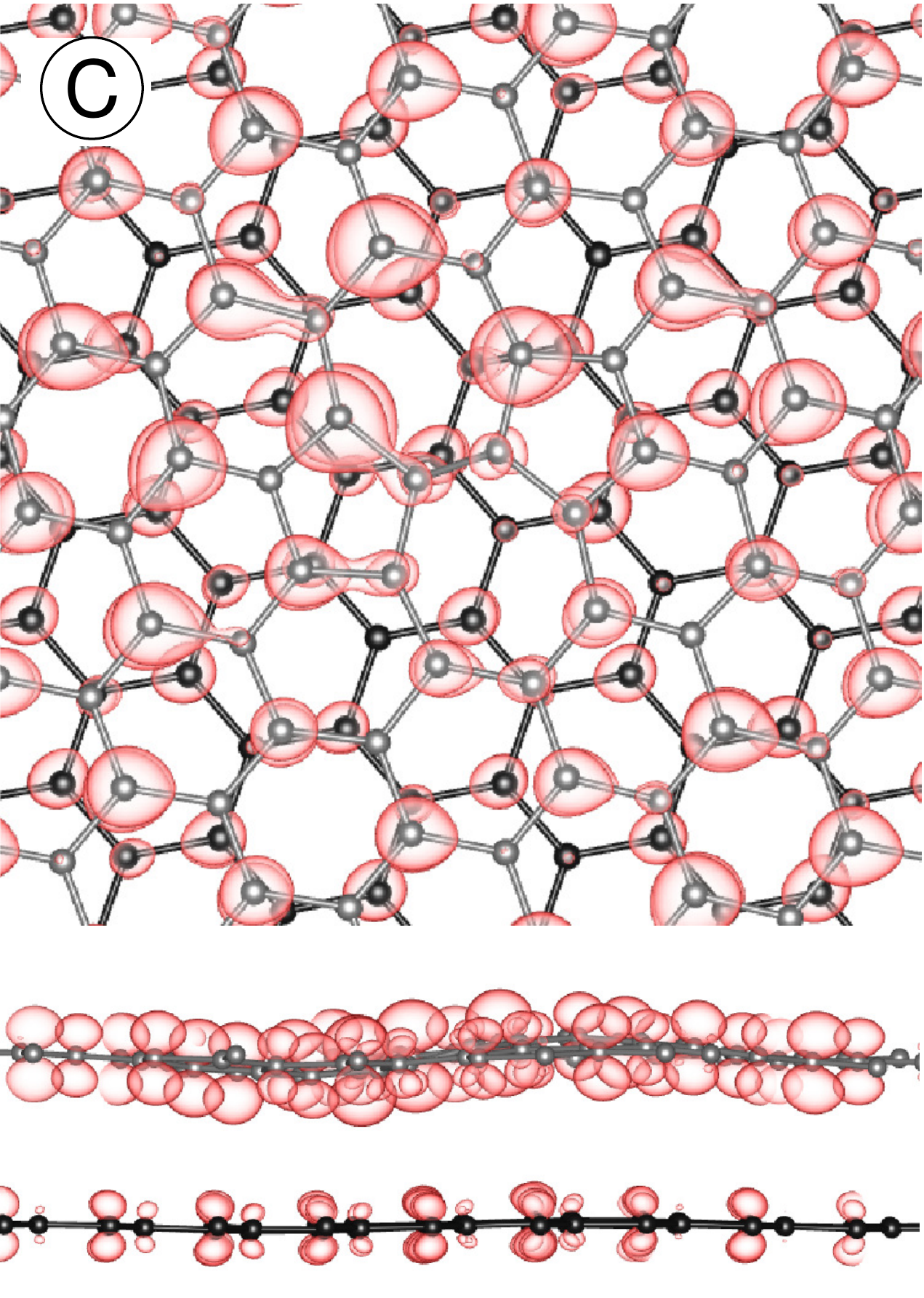}}
\quad
\subfigure[\ ]{\includegraphics[width=0.22\textwidth]{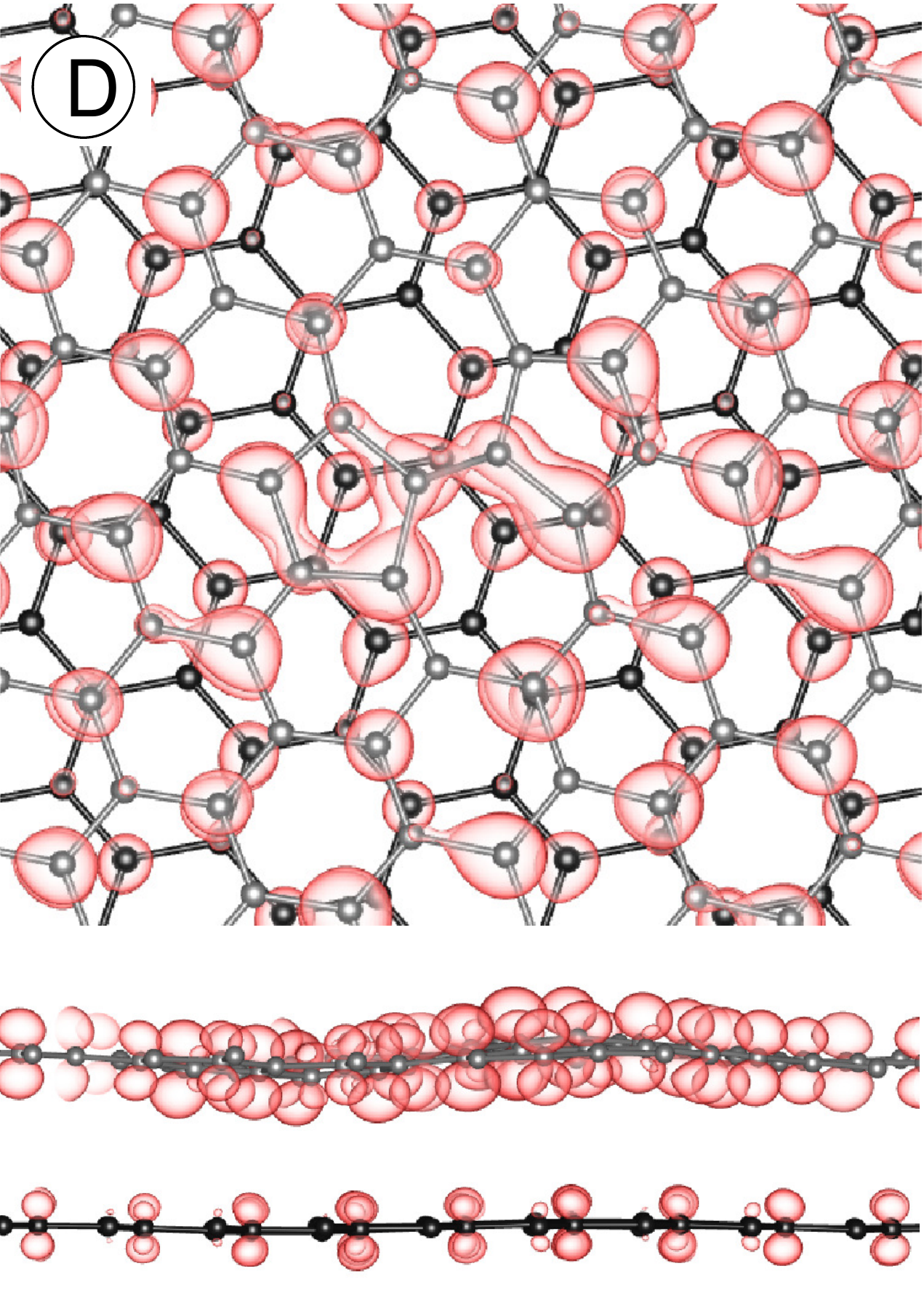}}
\caption{ The distribution of charge density corresponding to the four states that lie just above and below the Dirac points for the case of $\gamma\gamma$-SW defect in TBLG (corresponding to the labels (A), (B), (C), and (D) shown in Fig.~\ref{Fig:SLG-AB_BLG-TBLG-SW-Bands}(f)), shown by the red lobes. The top panel shows the top view, where the upper layer is shown in gray and the lower layer in black. The bottom panels show the side view.}
\label{SWCC-DPcharge}
\end{figure}

The same kind of interpretation applies to the other three types of SW
defects ($\alpha\beta$, $\beta\beta$ and $\beta\gamma$) shown in Figs.~\ref{Fig:SLG-AB_BLG-TBLG-SW-Bands}(d)--(f), though in these cases the underlying double cone structure is more distorted since the gaps that open up are larger, because of a greater effect of
interlayer coupling. We point out that one effect of the introduction of the SW defects is an increase in the density of states at the Fermi energy, since the Dirac points $E_D^1$ and $E_D^2$ have now been shifted slightly  below  and above $E_F$. 

As already noted, in Fig.~\ref{Fig:BZ-S2}(b) we have  marked the positions in k-space of the Dirac points for the four kinds of SW defects in our TBLG. The Dirac points have been defined as the points in k-space corresponding to the maximum of VB1 and the minimum of CB1. Of these, one (arising primarily from the pristine layer) lies close to \textbf{K}$_1^{\rm B}$, and one (arising primarily from the layer with the Stone-Wales defect) further away. The direction of the shift of the latter in k-space is determined by the orientation of the SW defect in real space. 

\begin{figure}[t!]
\centering
\includegraphics[scale=0.35]{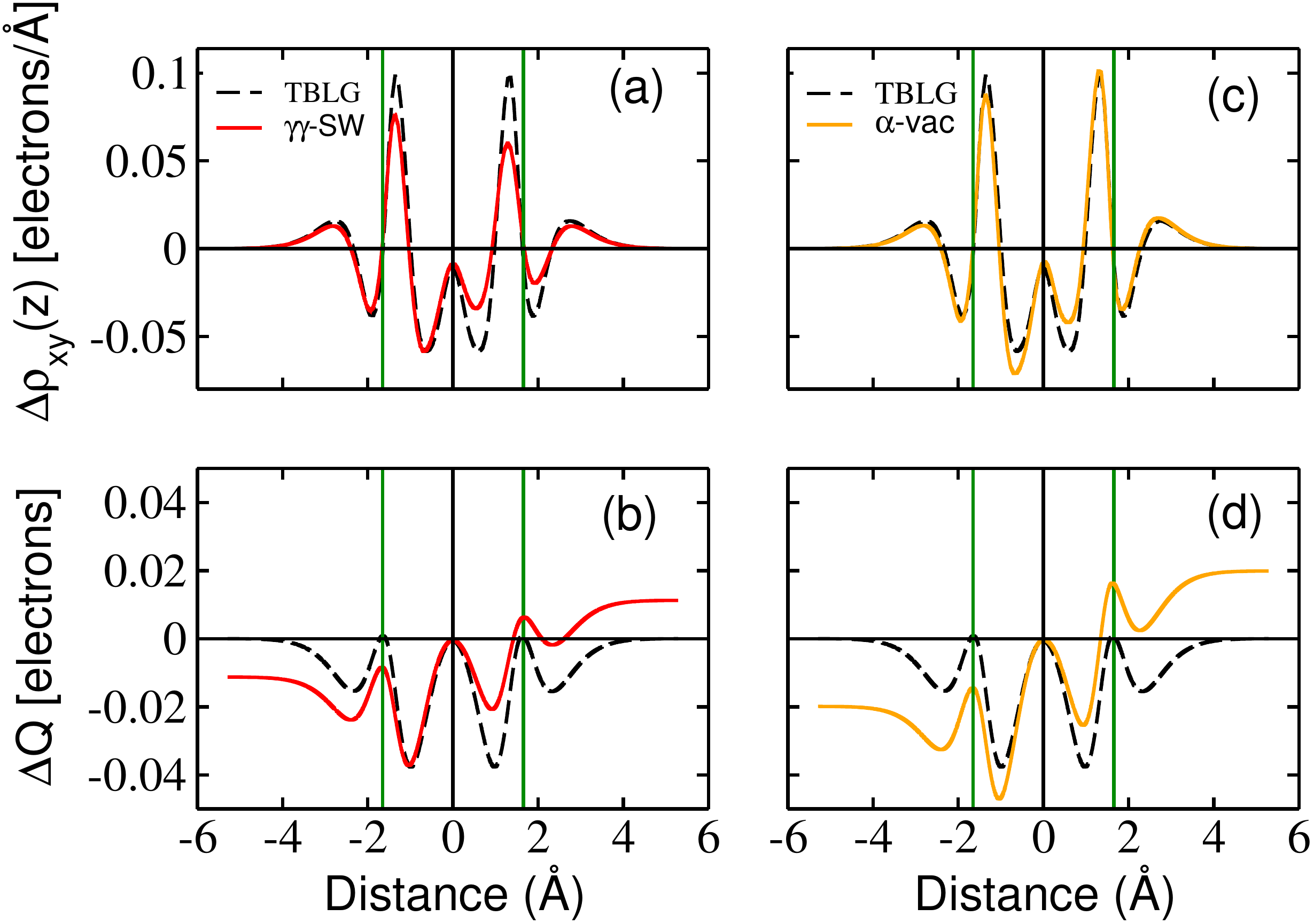}
\caption{(color online) Planar integral of the change in electronic charge density $\Delta\rho_{xy}(z)$ for (a) SW defect (solid red line) and (c) monovacancy (solid orange line) in TBLG. The black dashed line in (a) and (b) shows the value of $\Delta\rho_{xy}(z)$ for pristine TBLG. The zero of the abscissa marks the midpoint between the two layers of TBLG. The net electronic charge lost or gained by the individual layers, $\Delta Q$, is found by integrating $\Delta\rho_{xy}$ outwards from the midpoint of the layers. The cumulative value of $\Delta Q(z)$ for (c) SW defect and (d) monovacancy in TBLG is shown. The black dashed line in (c) and (d) shows the value of $\Delta Q$ for pristine TBLG.}
\label{CT-TBLG-SW-VacA} 
\end{figure}

The skewed shape of the band structure arising from double Dirac cones [see Fig.~\ref{Fig:SLG-AB_BLG-TBLG-SW-Bands}], where VB1 looks like a tilted ``M", and CB1 like a tilted ``W", is reminiscent of the effects of a transverse electric field on
the band structure of twisted bilayer graphene.\cite{Santos-PRL2007,Li-NatPhys2010} Note that even quite small electric fields (arising, e.g., due to a very minute charge transfer) can be expected to have a big effect on the positioning of the
Dirac points with respect to $E_F$, because of the very low electronic density of states in this region. We next check what the nature of the charge transfer (between the two graphene layers) and charge
redistribution (in the vicinity of each graphene layer) is when two graphene sheets are brought together to form our TBLG systems. In Fig.~\ref{CT-TBLG-SW-VacA} (a) we have plotted our results for $\Delta\rho_{xy}(z)$, the
planar integral of the change in electronic charge density of TBLG with a $\gamma\gamma$ SW defect, referenced to the sum of the charge densities of an isolated pristine graphene layer and an
isolated graphene layer with a SW defect. Note that the charge densities of the latter two systems are computed making use of their geometries in the combined system. For comparison, we also
plot the same results for TBLG without a SW defect. One can clearly see that while the curve for TBLG with a SW defect (red solid line) is asymmetric about the midpoint, that for TBLG with no defect
(black dashed line) is symmetric. By integrating $\Delta\rho_{xy}(z)$ outward from the midpoint, i.e., $z=0$, we get the results shown in Fig.~\ref{CT-TBLG-SW-VacA} (b). This shows clearly that upon bringing the two layers
together, there is a depletion of electronic charge from the pristine layer, and an accumulation of charge in the defective layer; the net charge transferred is 0.0106 electrons.
It is therefore quite tempting to apply a model where there is a uniform positive charge (depletion of electrons) on the pristine layer and uniform negative charge (accumulation of electrons) on the defective layer, and approximate its effects as those of an electric field due to a parallel plate capacitor. This would yield an electric field of strength $1.3 \times 10^8$ NC$^{-1}$. We see from Fig.~\ref{Fig:SLG-AB_BLG-TBLG-SW-Bands}(c) that in our case, $E_D$ shifts by $\sim$0.02 eV with respect to $E_F$. These results appear to be consistent with those of a previous study of undefective TBLG in an electric field, where the authors found that an electric field of $8.95 \times 10^8$ NC$^{-1}$ resulted in  the Dirac points getting shifted with respect to $E_F$ by $\sim$0.1 eV. However, while these results might seem initially encouraging,
the actual situation is not so simple, since in our case, the shift in energy of the two Dirac points is opposite in sign to that predicted by this simple model of charge transfer, whether interpreted in terms of the direction of the electric
field, or whether interpreted (equivalently) in terms of how $E_F$ and $E_D$ should shift with respect to one another when electrons are removed or added.

A similar situation, in which the relative shift between $E_F$ and $E_D$ is, in some cases, in the direction opposite to that expected from a naive model of charge transfer, has been observed by previous authors who studied the electronic structure
of graphene on metals.\cite{Giovannetti-PRL2008, Khomyakov-PRB2009,Gong-JAP2010, Pi-PRB2009} The authors of these previous studies attributed this to the presence of chemical interactions between the graphene and metal surface, which cannot be accounted for by the
simple picture of charge transfer. For our systems, we find unmistakable signatures of the interaction between the two layers; it is clear that a scenario in which one Dirac cone is seen as arising entirely from the pristine layer,
and the other as arising entirely from the defective layer, is not correct. As evidence of this, in Fig.~\ref{SWCC-DPcharge}, we have plotted the charge densities of four states very near the Dirac points, one each just above
and below $E_D^1$, and one each just above and below $E_D^2$, for the case of the $\gamma\gamma$ Stone-Wales defect in TBLG. In all four cases, it is obvious that the states have
significant charge density on both graphene layers, even though the two states near $E_D^1$ have a greater charge density on the pristine (lower) layer, and those near $E_D^2$ have a greater charge
density on the defective (upper) layer.

We have also computed the band-structure of the $\gamma\gamma$ SW defect in TBLG using LDA.  Upon comparing with the results obtained using DFT-D2, (see Fig.~\ref{Fig:SLG-AB_BLG-TBLG-SW-Bands}) we find almost no change
in the vicinity of $E_F$, though there are some slight differences at energies away from $E_F$.

\subsection{Monovacancy Defect}

A monovacancy defect can be created in a graphene sheet by removing one of the carbon atoms. 
In our example of TBLG, there are three inequivalent choices for the site at which this can be done; 
we term the resulting monovacancies $\alpha$, $\beta$ and $\gamma$ (see Fig.~\ref{Fig:tBLG}). 
We consider one defect per $S_2$ supercell (corresponding to a defect density of 1.8\%) for all these three types of monovacancies.  
For the case of the $\alpha$-monovacancy, we have also considered different supercells $S_1$ and $S_3$, corresponding to defect densities of 7.2\% and 0.8\%, respectively.
This allows us to examine the role of defect-defect interactions and strain relaxation in the defective systems.

To create these vacancies one needs to supply energy to the system; this is known as the monovacancy formation energy. 
The formation energy of a monovacancy is defined as:\cite{Ranber-JPCM2009} 

\begin{equation}
\Delta{E}_{\rm vac} = {E}^{\rm vac}_{{S_i} } - \left(\frac{N-1}{N}\right) {E}^{\rm novac}_{{S_i} }
\end{equation}

\noindent where ${E}^{\rm novac}_{{S_i} }$ is the total energy of the pristine system containing $N$ atoms in the supercell $S_i$,  and  ${E}^{\rm vac}_{{S_i} }$ is the total energy of the supercell $S_i$ that contains a single monovacancy and is comprised of $N-1$ atoms. Note that the system with the monovacancy is found to possess a magnetic moment, and all these total energies are therefore obtained from spin-polarized DFT calculations. 
 
 Our results for the monovacancy formation energy  for various single-layer and bilayer graphene systems are shown in Table~\ref{tab-Vac-TBLG}. As in the case of the SW defect, we see that the different atomic arrangements in the different types of monovacancies are not significantly reflected in the energetics of defect formation: the values of $\Delta E_{\rm vac}$ for the three types of monovacancy for TBLG are very close to one another, and also to the values for SLG and AB-BLG.

\begin{figure*}[t!]
\centering
\subfigure[]{\label{tBLG-VacA}\includegraphics[width=0.25\textwidth]{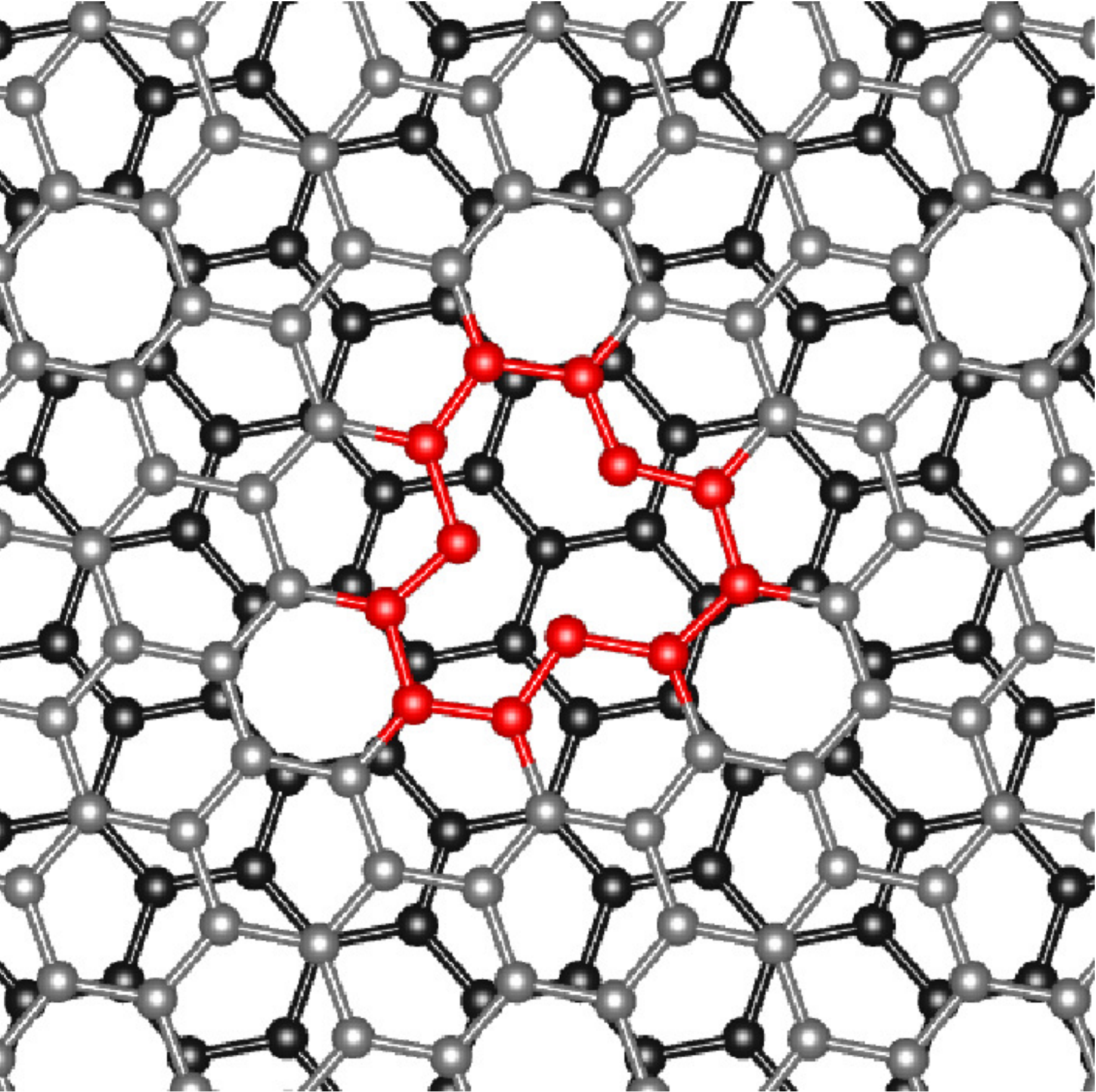}}
\quad
\subfigure[]{\label{tBLG-VacB}\includegraphics[width=0.25\textwidth]{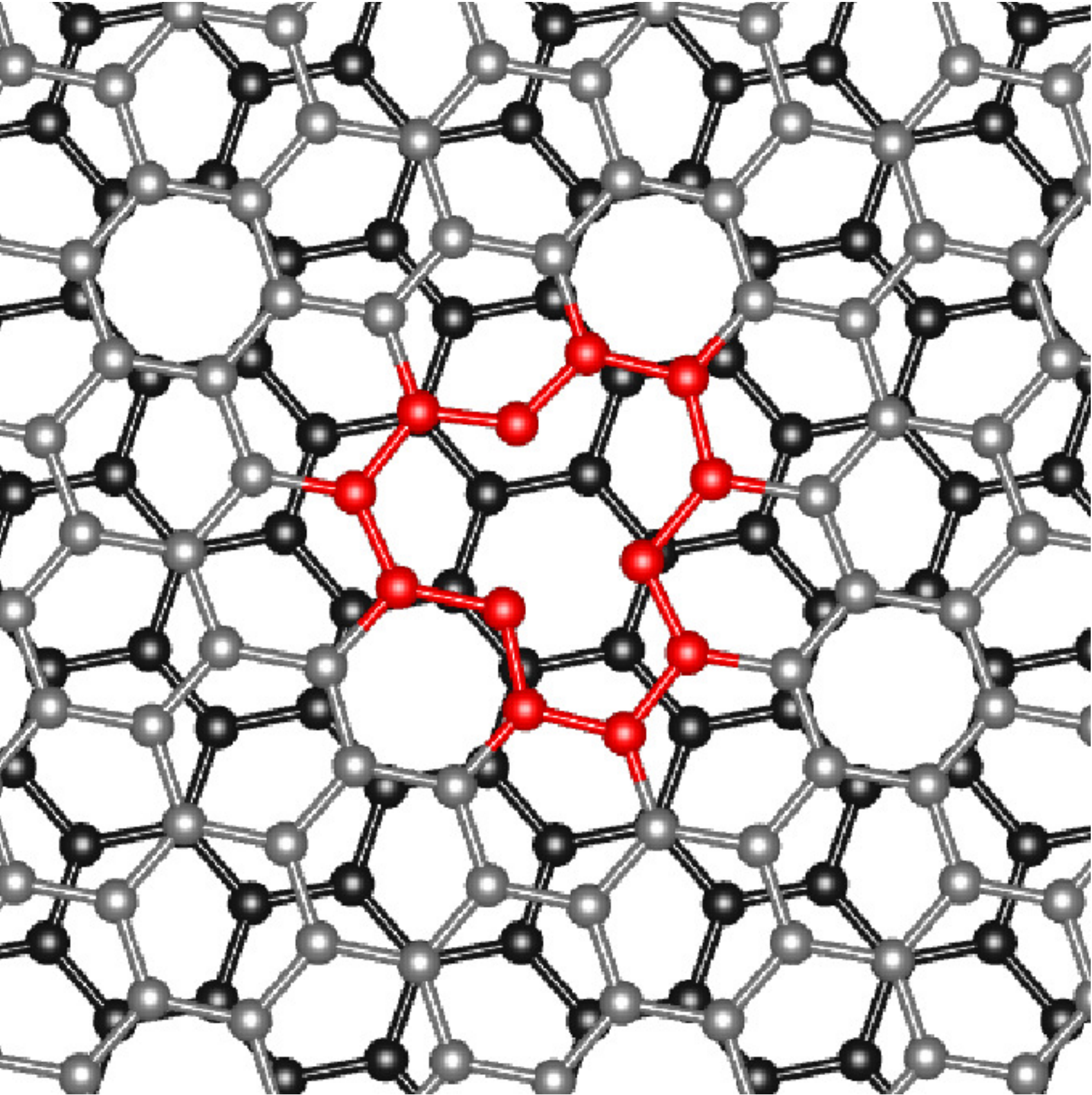}}
\quad
\subfigure[]{\label{tBLG-VacC}\includegraphics[width=0.25\textwidth]{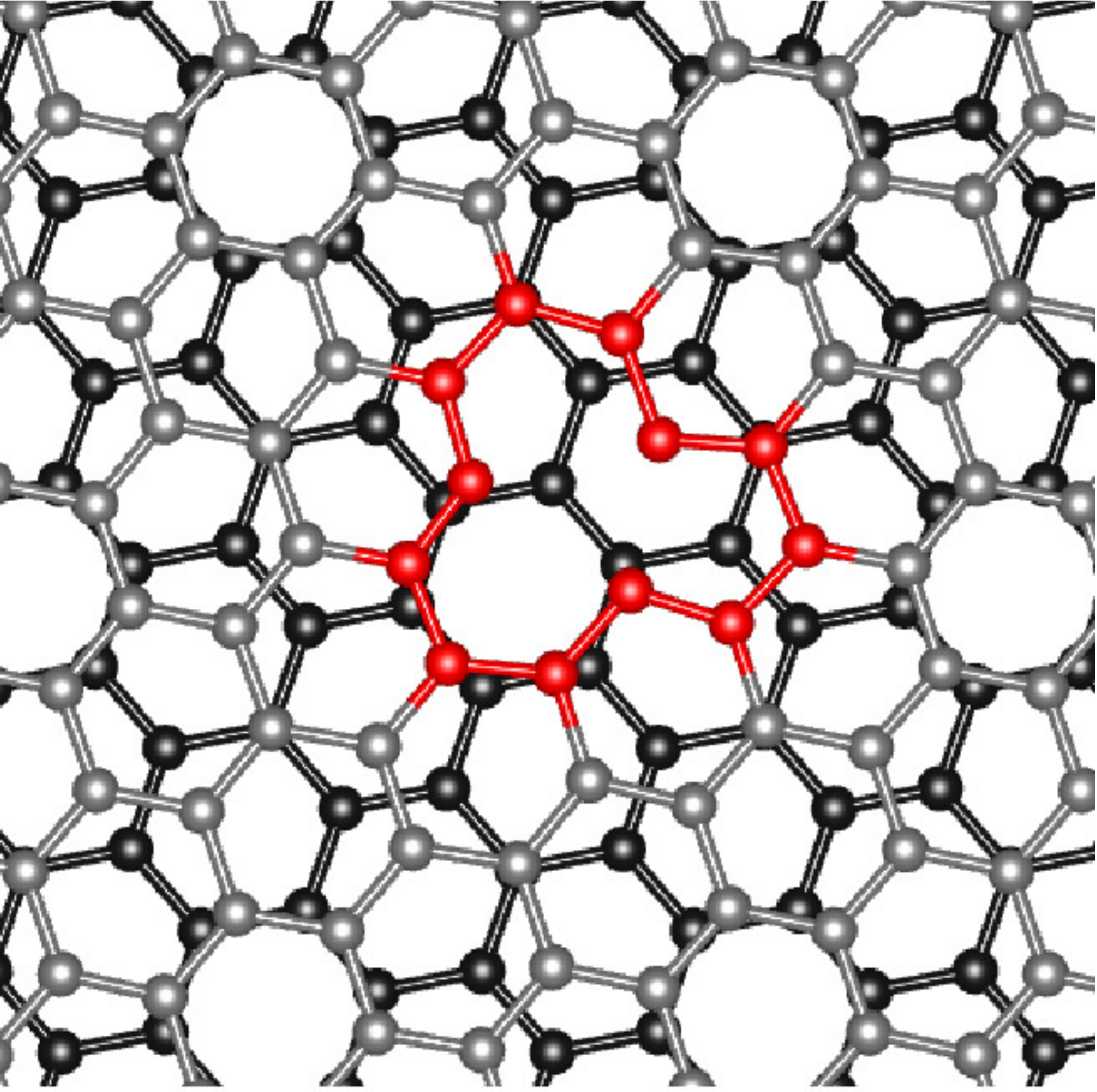}}
\quad
\subfigure[]{\label{tBLG-VacA-STM-Top}\includegraphics[width=0.25\textwidth]{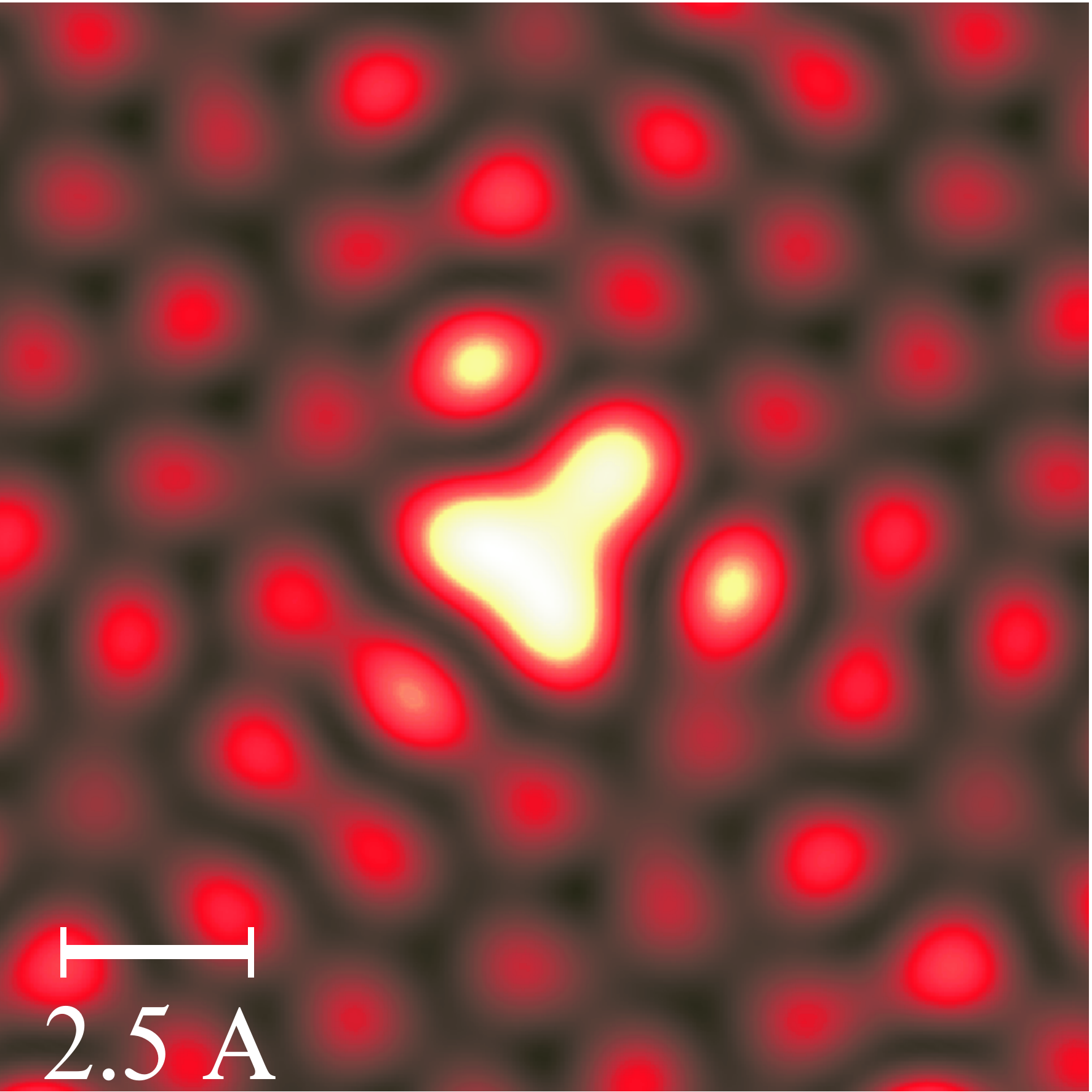}}
\quad
\subfigure[]{\label{tBLG-VacB-STM-Top}\includegraphics[width=0.25\textwidth]{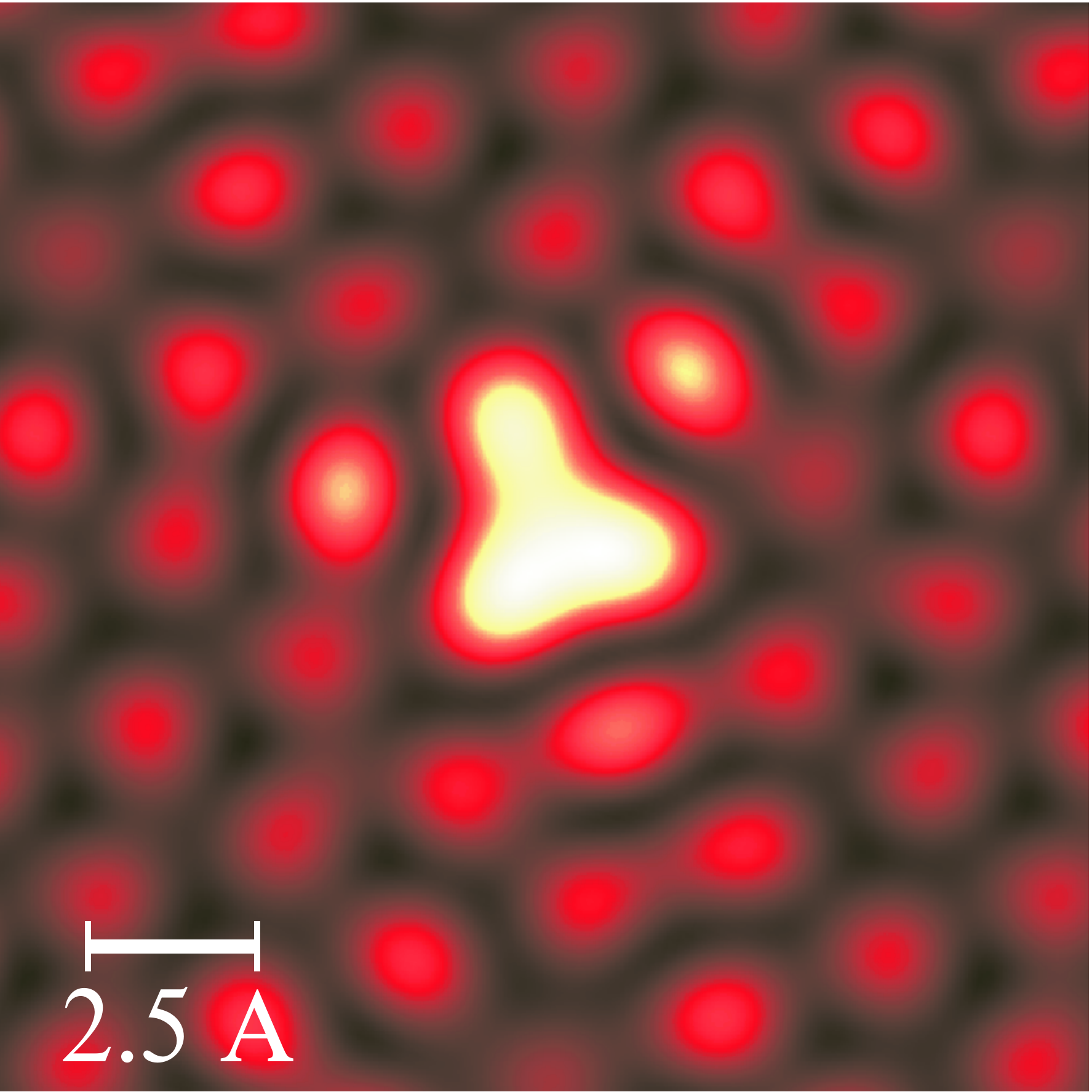}}
\quad
\subfigure[]{\label{tBLG-VacC-STM-Top}\includegraphics[width=0.25\textwidth]{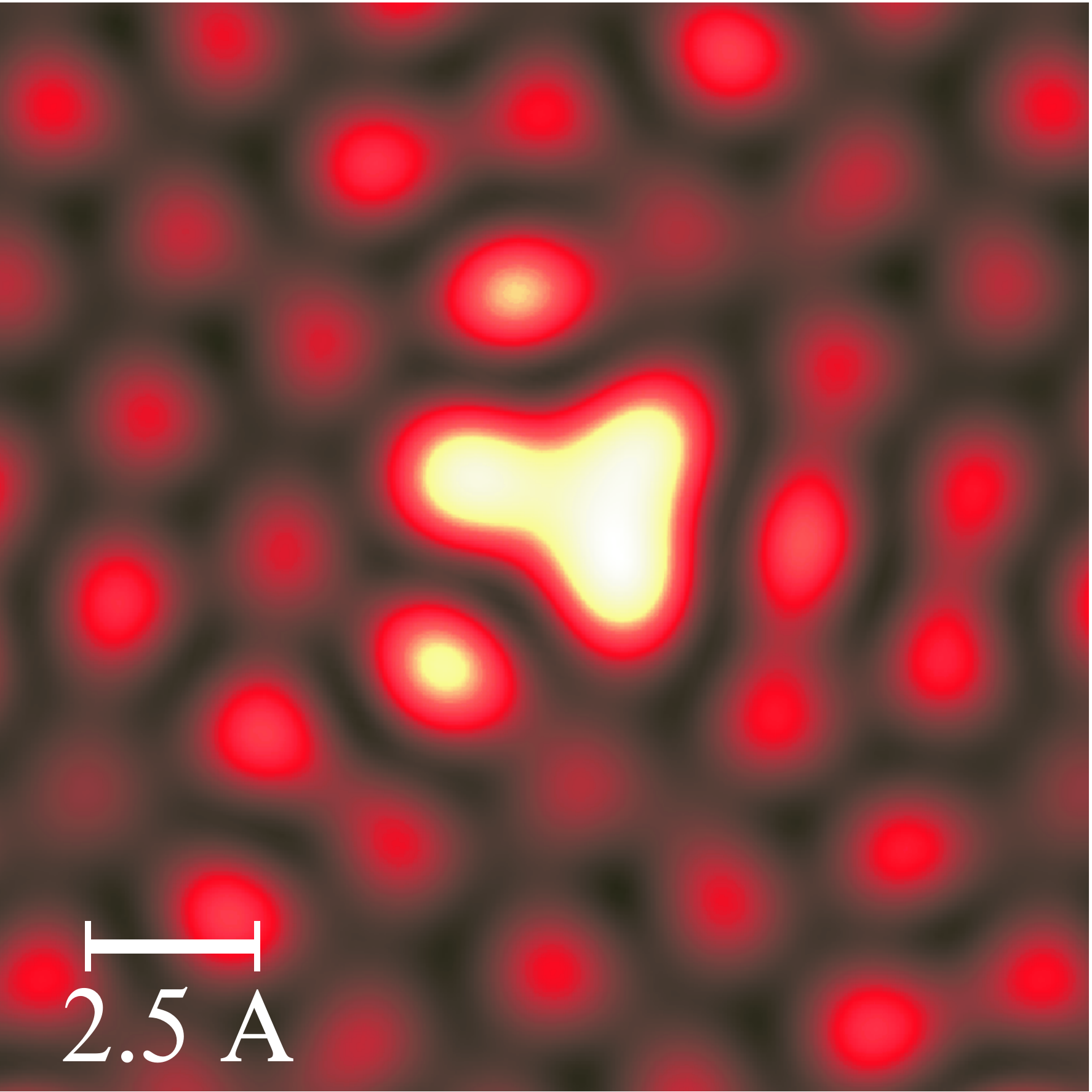}}
\quad
\subfigure[]{\label{tBLG-VacA-STM-Bot}\includegraphics[width=0.25\textwidth]{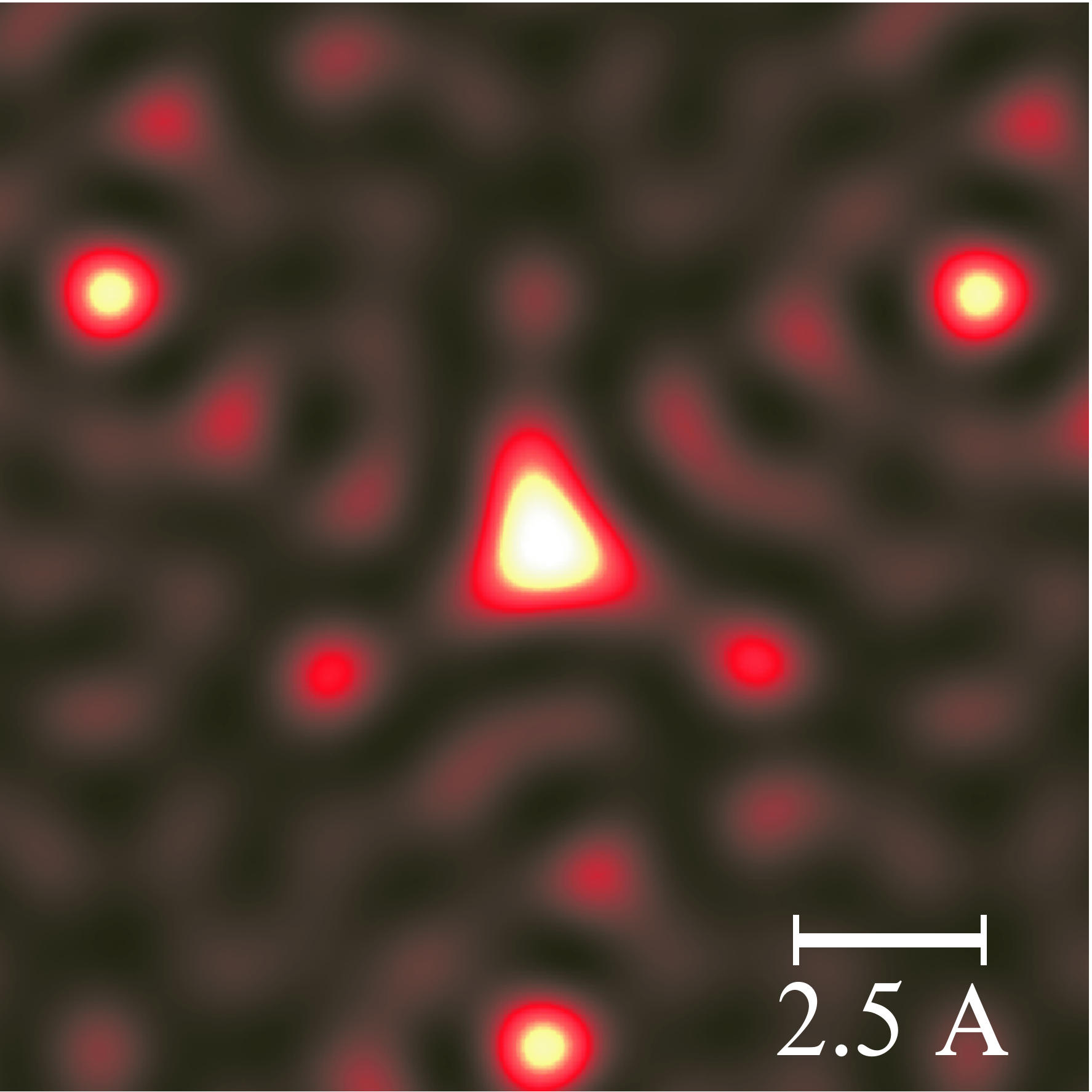}}
\quad
\subfigure[]{\label{tBLG-VacB-STM-Bot}\includegraphics[width=0.25\textwidth]{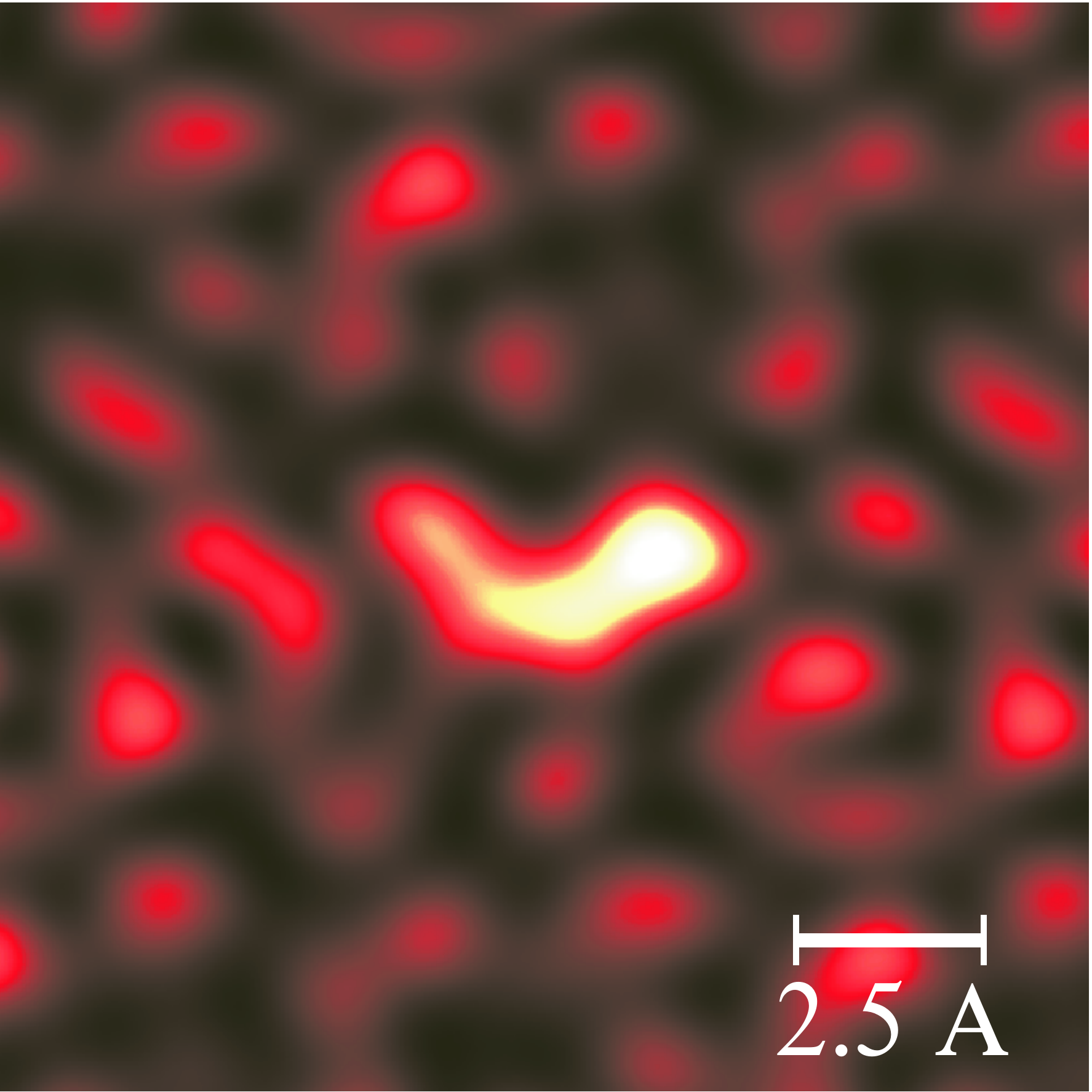}}
\quad
\subfigure[]{\label{tBLG-VacC-STM-Bot}\includegraphics[width=0.25\textwidth]{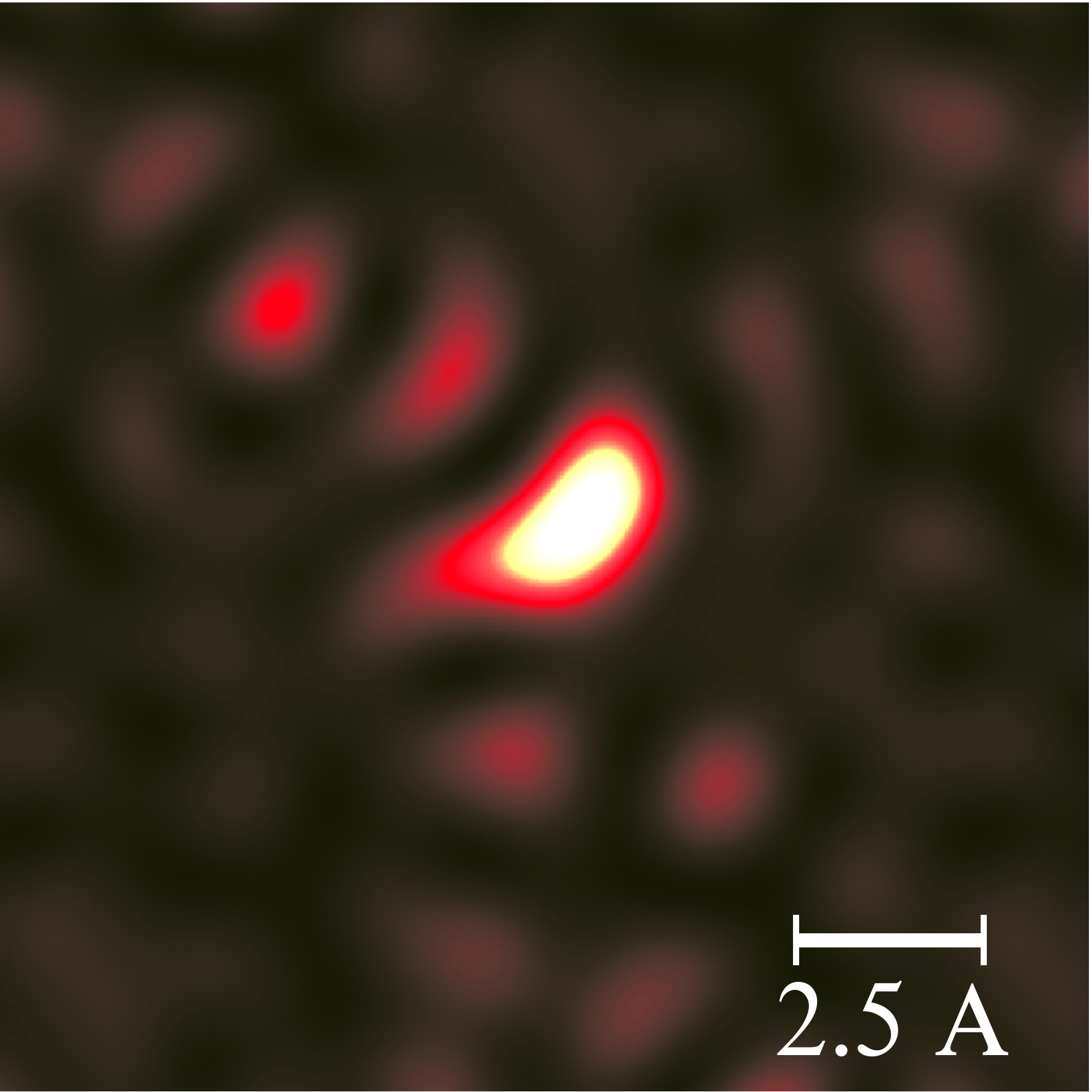}}
\caption{ Atomic structures for (a) $\alpha$-vacancy, (b) $\beta$-vacancy, and (c) $\gamma$-vacancy. Lower layer atoms are black, and the upper defective layer atoms are light gray. The C atoms near the vacancy-defect are colored red for visual ease. Simulated STM images ``taken" from the defective side for $\alpha$-, $\beta$-, and $\gamma$ vacancies are shown in Figs. (d), (e), and (f), respectively and those ``taken" from the undefective side are shown in Figs. (g), (h), and (i), respectively. The images from the defective side are essentially indistinguishable, but the STM images from the undefective side show the effects of the relative orientation with respect to the other layer.}
\label{tBLG-Vac}
\end{figure*}

\subsubsection{Structural Properties}
In all the three types of monovacancy considered for TBLG, both layers remain almost planar.  
The nearest-neighbor atoms of the vacancy site are displaced noticeably from their positions in the undefective structure.
The removal of a carbon atom results in the under-coordination of its three  
nearest neighbor carbon atoms, resulting in the creation of a lone-pair of electrons over each carbon atom. 
The three-fold symmetry about the vacancy site is broken by a Jahn-Teller type distortion, 
resulting in an effective bonding between two of the three nearest neighbor carbon atoms. 
We find that two of the atoms are attracted toward each other and the distance between 
these atoms is reduced significantly, showing a signature of bonding. 
A similar phenomenon has earlier been reported for a monovacancy in a single layer of graphene \cite{Lehtinen-PRL2004} and for an AB-stacked bilayer of graphene.\cite{Choi-JPCM2008} 
We observe such a phenomenon in all the three vacancy positions: $\alpha$, $\beta$ and $\gamma$-vacancy [see Figs.~\ref{tBLG-VacA}, \ref{tBLG-VacB} and \ref{tBLG-VacC}]. 
For the choice of defective supercell $S_2$, this bond distance is found to be 2.05 \AA,  2.03 \AA, and 2.04 \AA, for $\alpha$, $\beta$ and $\gamma$-vacancies, respectively. 
For the case of the $\alpha$-vacancy, we have also considered different supercells $S_1$ and $S_3$, corresponding to a defect density of 7.2\% and 0.8\%, respectively. 
In these cases, this shorter bond distance is found to be 2.29 \AA\ and 1.86 \AA, respectively.

\subsubsection{Simulated STM images}
Next, we compute simulated STM images of the monovacancy defects in TBLG (See Fig.~\ref{tBLG-Vac}), making use of a bias voltage of 0.4 eV above the Fermi energy. The STM images of the three type of monovacancies ($\alpha$,
$\beta$ and $\gamma$) are basically indistinguishable, if the STM image is taken from the defective side of the bilayer.  Interestingly, however if the STM image is taken from the {\it undefective} side of the bilayer, one can clearly distinguish between the three types of monovacancies, as their STM images show markedly different signatures;
see  the three lowest panels in Fig.~\ref{tBLG-Vac}.

\subsubsection{Electronic Structure and Magnetic Properties}

Creating a monovacancy in TBLG results in the appearance of a magnetic moment; its origin is similar to that in single layer graphene.\cite{Nanda-NJP2012,Yazyev-RepProgPhys2010} 
In a sheet of pristine graphene, each carbon atom is bonded to three of its neighbors by three $sp^2$ hybridized $\sigma$-bonds and one $\pi$-bond, each bond sharing two electrons. 
When a monovacancy is created, the missing carbon atom takes away its share of four electrons, and there are three unsatisfied $sp^2$ 
$\sigma$-electrons (one electron localized on each of the three nearest neighbor carbon atoms) and one $\pi$-electron, in the vicinity of the monovacancy. 
Thus an electron in each of these states will try to maximize its spin, giving rise to a net magnetic moment of 4 $\mu_B$. 
As already mentioned, of the three C atoms surrounding the vacancy, two atoms rebond, lowering the energy via a Jahn-Teller type distortion, 
and as a result the spins of the $\sigma$-electrons involved in this bonding orient antiparallel to each other, due to Pauli's exclusion principle. 
Thus the monovacancy is expected to have a net magnetic moment of 2 $\mu_B$, arising from one $\sigma$ state and one $\pi$ state. This is why creating a monovacancy in a graphene sheet results in the appearance of a magnetic moment. However, as the $\pi$-electrons have a somewhat itinerant character, and their bands cross the Fermi level, there is fractional occupation of this state so that the magnetic moment becomes less than 2  $\mu_B$.
In our case of a monovacacy in TBLG, in supercells of different sizes $S_1$, $S_2$ and $S_3$, we find a net magnetic moment of 1.81 $\mu_B$, 1.25 $\mu_B$ and 1.85 $\mu_B$ per defect, respectively. 
These values are in good agreement with earlier reported values for monovacancy defects in SLG (magnetic moments of ranging from 1 $\mu_B$ to 2 $\mu_B$ depending on the defect concentration \cite{Lehtinen-PRL2004, Nanda-NJP2012}) and in AB-BLG (magnetic moment of 1.3 $\mu_B$ \cite{Choi-JPCM2008}). 
In Table~\ref{tab-Vac-TBLG}, we have reported the values, for each of the cases considered by us in the supercell $S_2$, for the net magnetic moment per defect defined as $M_{net} = \int d^3r (\rho^\uparrow({\bf r}) - \rho^\downarrow({\bf r}))$ and the absolute magnetic moment per defect defined as $M_{abs} = \int d^3r |(\rho^\uparrow({\bf r}) - \rho^\downarrow({\bf r}))|$. Here $\rho^\uparrow({\bf r})$ and $\rho^\downarrow({\bf r})$ are the up-spin and down-spin charge densities, respectively, and the integral is carried out over the supercell $S_2$. All the three exchange-correlation functionals give $M_{net} = 1.25\ \mu_B$ for all systems. Interestingly however, the values obtained for $M_{abs}$ vary with the functionals used, being $\sim1.46\ \mu_B$ for LDA and $\sim1.63\ \mu_B$ for PBE and DFT-D2. No significant changes are seen on going from SLG to AB-BLG or TBLG.

\begingroup
\squeezetable
\begin{table}[tb]
\begin{center}
\centering
\begin{tabular}{  p {1.4cm}  p{1.3cm}  p{1cm} p{1cm}  p{1cm} p{1cm} c}
\hline 
\hline 
\multicolumn{7}{l}{Monovacancy defect: Formation energies $\Delta{E}_{\rm vac}$(eV)} \\ 
\hline 
\hline
Method       & SLG             & \multicolumn{2}{c}{AB-BLG}           & \multicolumn{3}{c}{TBLG}                              \\
(XC)            &                      & $\alpha$-Vac      & $\beta$-Vac      & $\alpha$-Vac    & $\beta$-Vac      & $\gamma$-Vac       \\
\hline
LDA       & 8.09               & 8.06            & 8.03           & 8.07          & 8.08           & 8.07             \\
PBE       & 7.72               & 7.71            & 7.71          & 7.72          & 7.72           & 7.72             \\
DFT-D2    & 7.77               & 7.76            & 7.72           & 7.77          & 7.79           & 7.77             \\
Expt.     & 7.0$\pm$0.5\cite{Thrower-StatSol1978}              &                   &                  &                 &                  &                    \\
\hline
\hline
\multicolumn{7}{l}{Monovacancy defect: Magnetic Moments ($\mu_{\rm B}$) } \\ 
\hline 
\hline
Method      & SLG             & \multicolumn{2}{c}{AB-BLG}           & \multicolumn{3}{c}{TBLG}                              \\
(XC)           &                      & $\alpha$-Vac      & $\beta$-Vac      & $\alpha$-Vac    & $\beta$-Vac      & $\gamma$-Vac       \\
\hline
LDA       &  1.25          & 1.25              & 1.25             & 1.25            & 1.25             & 1.25               \\
             &  (1.46)        & (1.46)            & (1.47)           & (1.46)          & (1.46)           & (1.46)             \\
PBE      &  1.25          & 1.25              & 1.25             & 1.25            & 1.25             & 1.25               \\
             &  (1.63)        & (1.63)            & (1.64)           & (1.63)          & (1.63)           & (1.64)             \\
DFT-D2  &  1.25          & 1.25              & 1.25             & 1.25            & 1.25             & 1.25               \\
             &  (1.63)        & (1.64)            & (1.65)           & (1.64)          & (1.63)           & (1.63)             \\
\hline
\hline
\end{tabular}
\caption{ Calculated values of monovacancy formation energy and net magnetic moments ($M_{net}$), for single layer of graphene, AB-BLG, and TBLG. The values in the parentheses denote the absolute magnetization per defect ($M_{abs}$).}
\label{tab-Vac-TBLG}
\end{center}
\end{table}
\endgroup


\begin{figure}[]
\centering
\vspace{0.7cm}
\subfigure[\ $\alpha$-vac TBLG]{\label{Bands-VacA-full}\includegraphics[width=8.5cm]{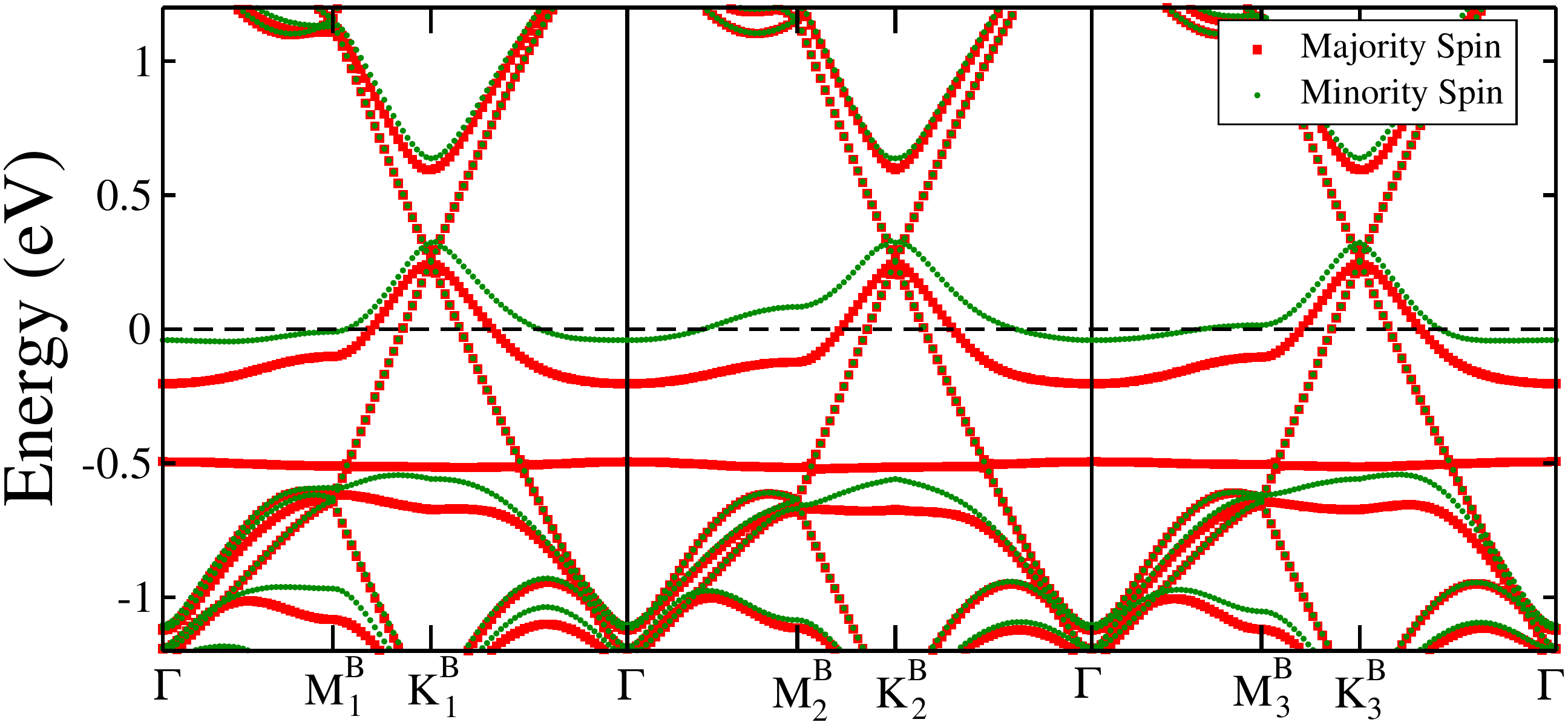}}
\hspace{0.5cm}
\subfigure[\ $\beta$-vac TBLG]{\label{Bands-VacB-full}\includegraphics[width=8.5cm]{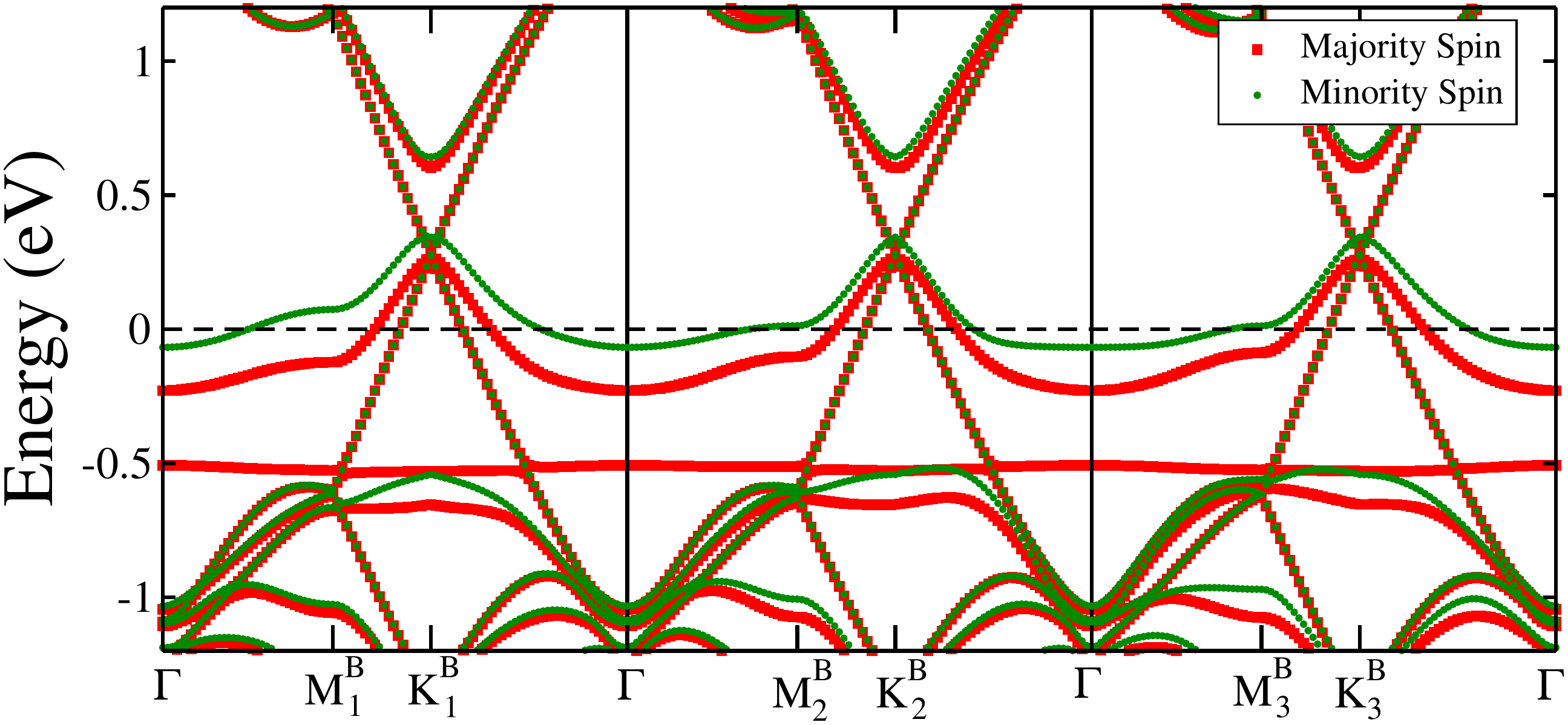}}
\hspace{0.5cm}
\subfigure[\ $\gamma$-vac TBLG]{\label{Bands-VacC-full}\includegraphics[width=8.5cm]{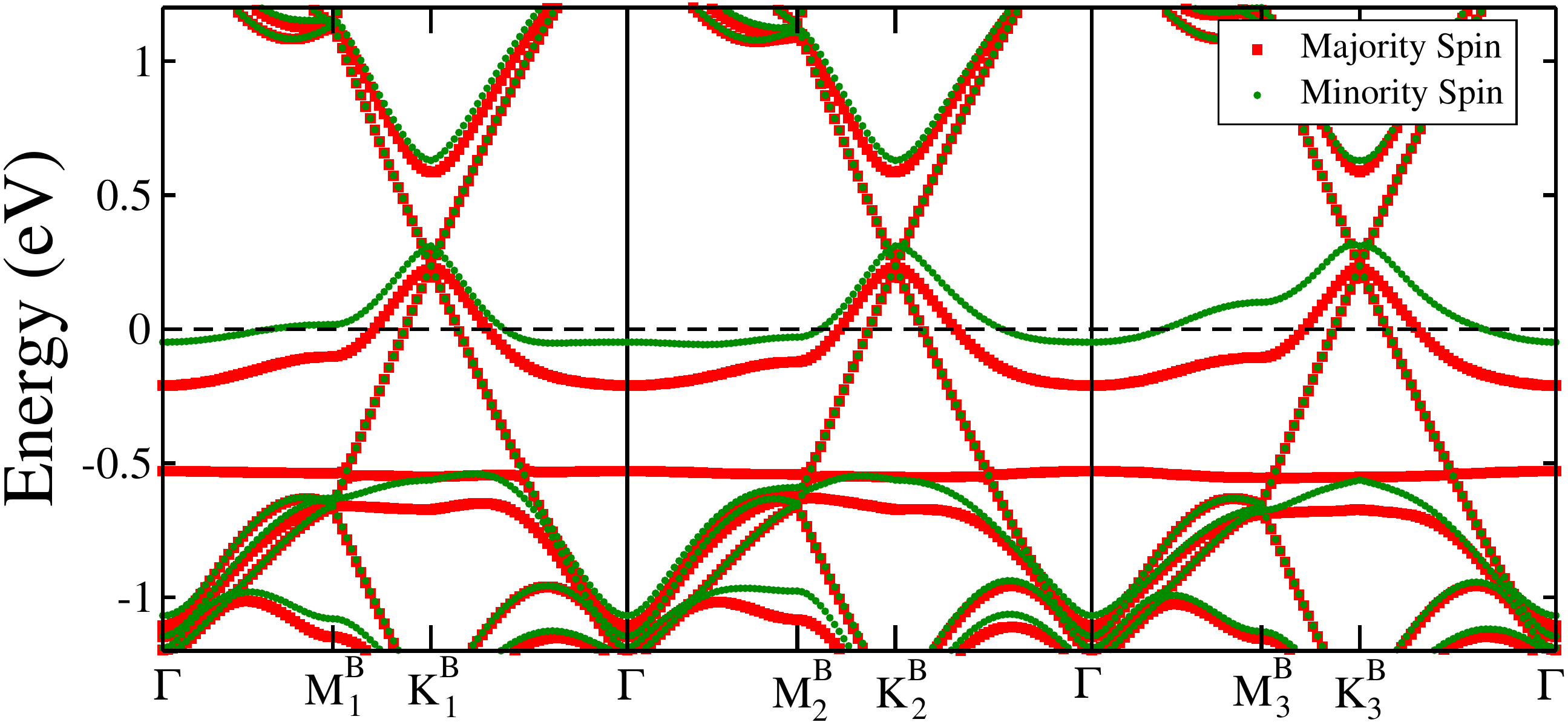}}
\caption{(color online) Electronic band Structure for (a) $\alpha$-, (b) $\beta$-, and (c) $\gamma$- monovacancies in TBLG, plotted along high-symmetry directions in the Brillouin zone shown in Fig.~\ref{Fig:BZ-S2}(a). Majority spin channel is shown in red and minority in green. The dashed line denotes the position of the Fermi energy $E_F$. }
\label{TBLG-vac-Bands} 
\end{figure}

\begin{figure*}[]
\centering
\vspace{0.7cm}
\subfigure[\ SLG]{\label{Bands-SLG}\includegraphics[width=5cm]{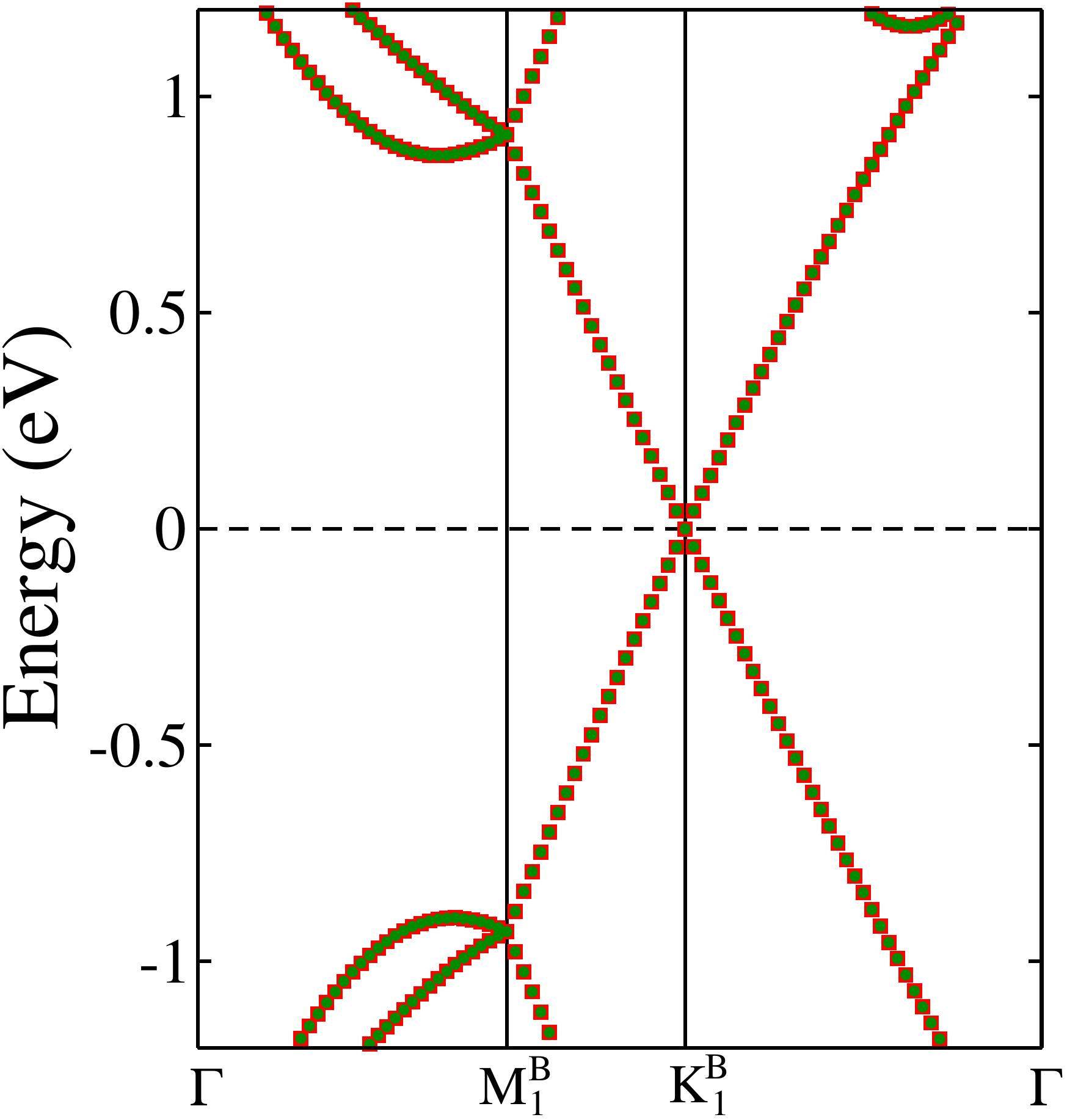}}
\hspace{0.5cm}
\subfigure[\ $\alpha$-vac SLG]{\label{Bands-SLGVacA}\includegraphics[width=5cm]{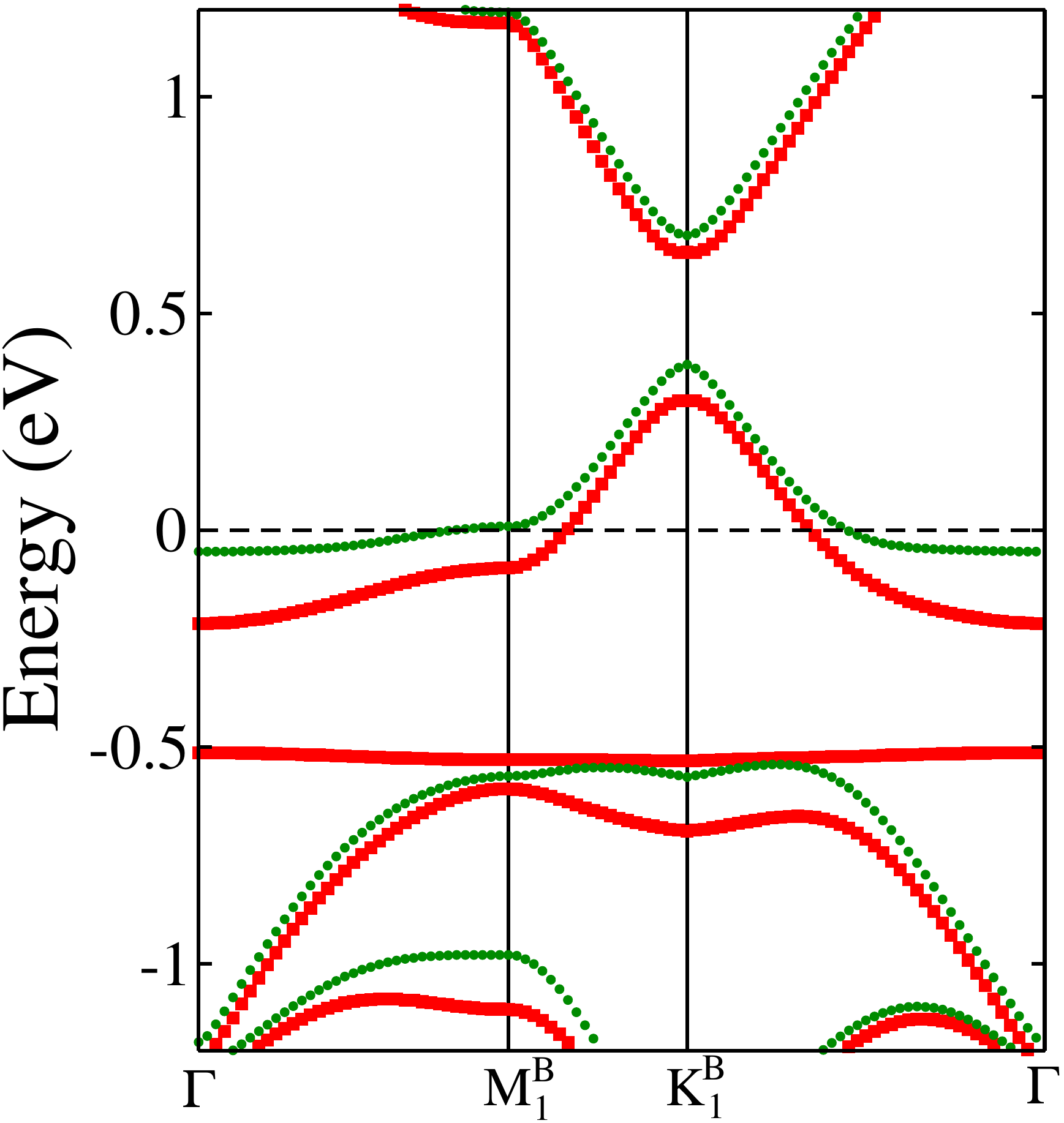}}
\hspace{0.5cm}
\subfigure[\ $\alpha$-vac TBLG]{\label{Bands-tBLGVacA}\includegraphics[width=5cm]{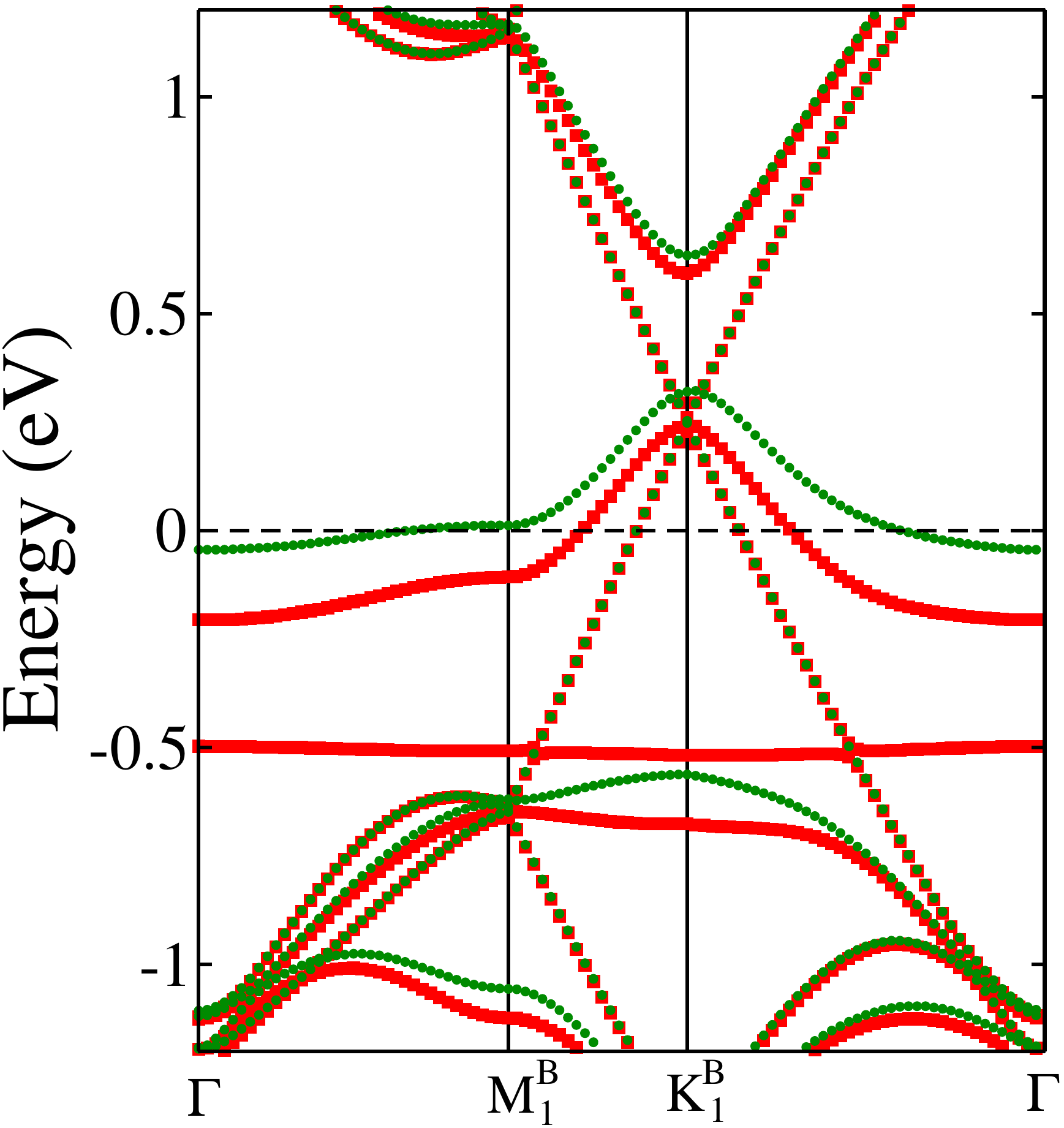}}
\hspace{0.5cm}
\caption{(color online) Electronic band Structure for (a) Single layer of graphene, (b) monovacancy in a single layer of graphene, and (c) monovacancy of type $\alpha$ in TBLG, plotted along high-symmetry directions in the Brillouin zone shown in Fig.~\ref{Fig:BZ-S2}(a). The dashed line denotes the position of the Fermi energy $E_F$. Red and green symbols indicate values for the majority and minority spins, respectively. }
\label{Bands-and-DOS} 
\end{figure*}

The effect of twist on the electronic structure of bilayer
graphene with a monovacancy is manifested most clearly when one examines the band structure; these results are shown in Fig.~\ref{TBLG-vac-Bands}. The presence of the vacancy in the twisted supercell results in an IBZ which is half of the entire BZ, similar to the case of the SW defect. As there is no shift in the position in k-space of the Dirac point, in the case of monovacancy defects, it suffices to plot the band structure along the conventionally used paths in the BZ; this is what we have done. We first compare the results obtained in the three types of monovacancy defects in TBLG [see Fig.~\ref{TBLG-vac-Bands}]. We see that
for a single type of monovacancy, the features of the four bands (two up-spin and two down-spin) that cross the Fermi level are largely similar in each third of the IBZ. Moreover, these four bands
are also quite similar for the three different types of monovacancies (differences in the band structure become more apparent as one moves further away from the Fermi level). Therefore, for simplicity, when
discussing the features of the band structure, and comparing it with the band structure of a monovacancy in SLG and pristine SLG, we will restrict ourselves to considering the band structure of the $\alpha$-monovacancy in TBLG, in one-third of its IBZ.

This is the comparison we carry out in Fig.~\ref{Bands-and-DOS}. We see that the band structure of TBLG with a monovacancy [Fig.~\ref{Bands-and-DOS}(c)] is clearly derived from the superposition of the band structures of a pristine SLG layer [Fig.~\ref{Bands-and-DOS}(a)] and SLG with a monovacancy [Fig.~\ref{Bands-and-DOS}(b)] with some important modifications. The bands derived from those of pristine SLG are shifted up in energy quite significantly, so that the Dirac crossing point $E_D$ now occurs $\sim0.25$ eV above $E_F$. In contrast, the bands arising from the layer with the monovacancy defect are shifted very slightly down in  energy, there is also a slight change in their shape, which has  a significant impact on the density of states because of the flat dispersion.

The shifting up/down of bands has three causes: (i) the fact that the layer with the monovacancy has four fewer valence electrons than the pristine layer (ii) a charge transfer between the layers, similar to
that observed in the case of the Stone-Wales defect (iii) a chemical interaction between the layers. The effect of (i) alone can be artificially examined by separating the two layers by a very large distance of 6.5 \AA, in that case, we find that the Dirac crossing energy $E_D$ arising from the pristine layer lies 0.32 eV above $E_F$; this shift can be attributed entirely to the requirement of equilibration of the Fermi energies of the two layers.  However, when the spacing between the two layers is reduced to its equilibrium value of 3.3 \AA, this upward shift is partially cancelled by a downward shift that arises due to the net combined effect of (ii) and (iii),
 resulting in the final upward shift of 0.25 eV observed. The initial shift of 0.32 eV can also be understood from computations of the work function $W$.  For SLG, we find that $W = 4.24$ eV, whereas for
 SLG with a monovacancy, $W = 4.56$ eV; both these values are computed making use of the same $S_2$ unit cell that is used for calculations of TBLG. When the two layers are brought together, with
 an interlayer separation of 6.5\AA, we again obtain $W = 4.56$ eV; this is because of the flat dispersion of the monovacancy states near $E_F$ and the resulting high value of the density of states, so that the states arising from the layer with the monovacancy do not shift perceptibly in energy, whereas the states arising from the pristine layer are shifted in energy by $4.56 - 4.24 = 0.32$ eV.
 
The charge transfer can be computed, as in the case of the Stone-Wales defect, by integrating the planar average of
the charge redistribution. The planar average is plotted as a function of the $z$ coordinate in Fig.~\ref{CT-TBLG-SW-VacA}(c) and is seen to be asymmetric. On integrating outward from the midpoint of the two graphene layers,
we find that the pristine layer has lost 0.02 electrons, and the layer with the monovacancy has gained 0.02 electrons [see Fig.~\ref{CT-TBLG-SW-VacA}(d)]. These values are in fairly good accordance with the values obtained from a Bader charge analysis, which yields a net charge on the pristine layer of 223.983 electrons, and on the defective layer of 220.016 electrons, i.e., 0.016 electrons have been transferred from the former to the latter.

\begin{figure}[]
\centering
\includegraphics[width=0.45\textwidth]{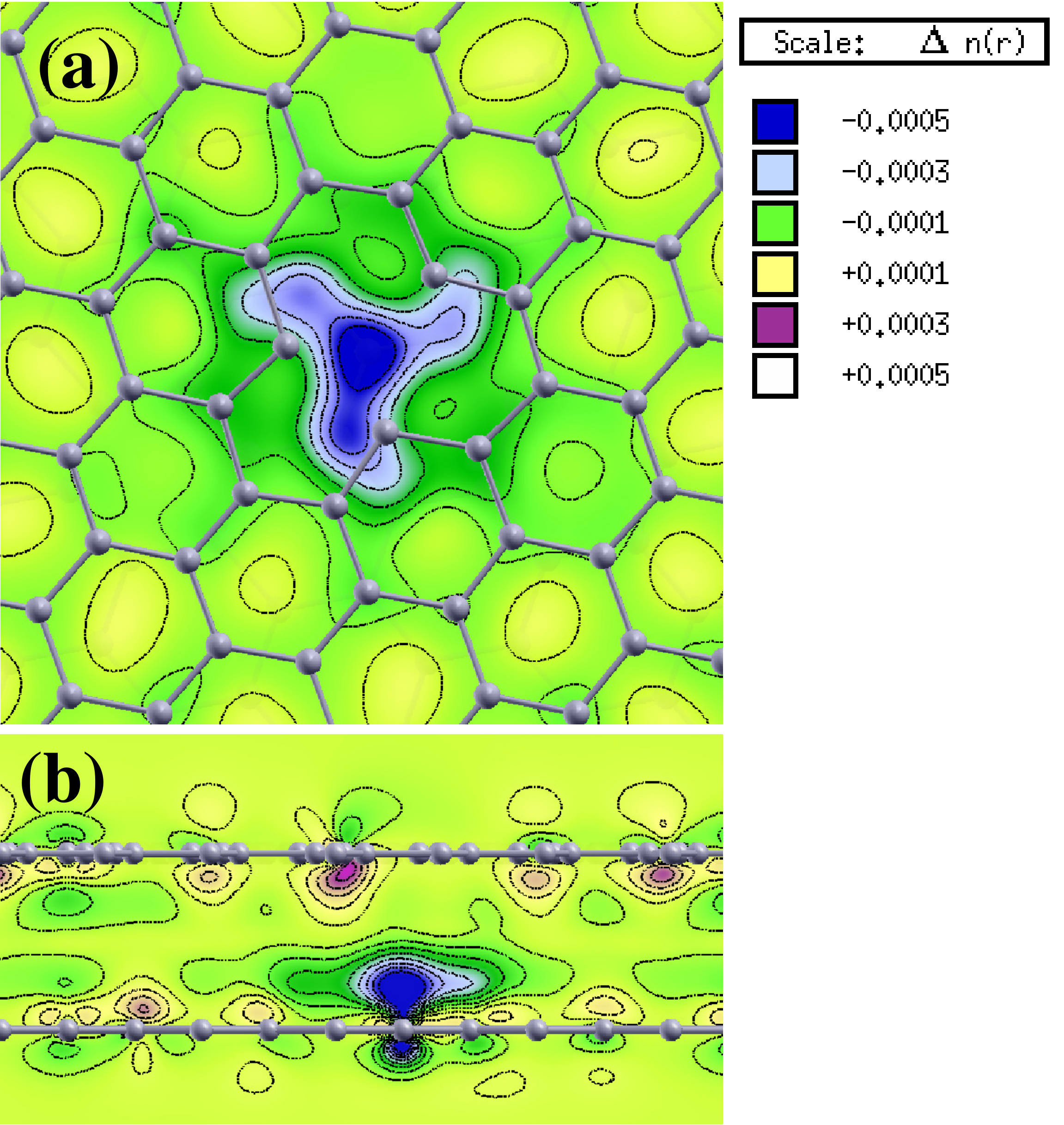}
\caption{\label{IsoSurf} (color online) Charge density redistribution in the vicinity of the undefective lower layer of TBLG due to the presence of an $\alpha$-monovacancy in the upper layer. (a) shows the contours of the depleted charge density in a plane about 0.8 \AA\ above the undefective layer. The atoms of the upper layer are shown as a reference. (b) shows a cross-sectional view of the redistributed charge in a plane perpendicular to the layers. }
\end{figure}

The redistribution in charge upon bringing the two twisted layers (one pristine, and one with a monovacancy) together can be seen clearly upon plotting a charge density difference map (See Fig.~\ref{IsoSurf}). To obtain this map, we subtract the individual charge densities of the defective SLG and the pristine SLG (maintaining their geometry as that in the combined system), from the charge density of the defective bilayer. A region of charge depletion is developed below the vacancy site, slightly above the undefective layer. We have plotted this for the $\alpha$ monovacancy, similar plots are obtained for the $\beta$ and $\gamma$ monovacancies. The charge redistribution in the presence of a monovacancy is again a manifestation of the interlayer coupling in twisted bilayer graphene. 

Once again, as in the
case of TBLG with a Stone-Wales defect, if one were to consider merely the simple model of the net charge transfer from the undefective to the defective layer, one would have expected $E_D$ to shift up further on reducing the separation from 6.55 \AA\ to
3.3 \AA. The fact that, in contrast, it shifted downward, is a manifestation of the chemical interaction between the layers, as already discussed in the case of the SW defect.

The magnetic defect state that arises due to the creation of the vacancy in a single layer of graphene (SLG) can be seen in the band structure as a spin-polarized flat band about 0.5 eV below the Fermi energy $E_F$, and can also be identified as a sharp peak in the density of states [See Figs.~\ref{Bands-and-DOS}(b) and (c)]. 
This corresponds to the unsatisfied $\sigma$-bond on one of the atoms near the vacancy.  
Further, the monovacancy acts as a scattering center for the Dirac-like $\pi$-electrons, thus resulting in the breakdown of the Dirac cone in the defective sheet, instead opening up a gap. The Fermi energy is lowered due to the removal of a carbon atom, and hence the Fermi level falls in the bands below this opened gap. The $\pi$-band also becomes spin polarized, as can be seen in the energy range -0.25 eV to 0.4 eV. The high density of states near the Fermi energy for one of the spin channels leads to a high spin polarization of $\pi$-electrons below $E_F$. 

\begin{figure}[]
\centering
\includegraphics[width=0.3\textwidth]{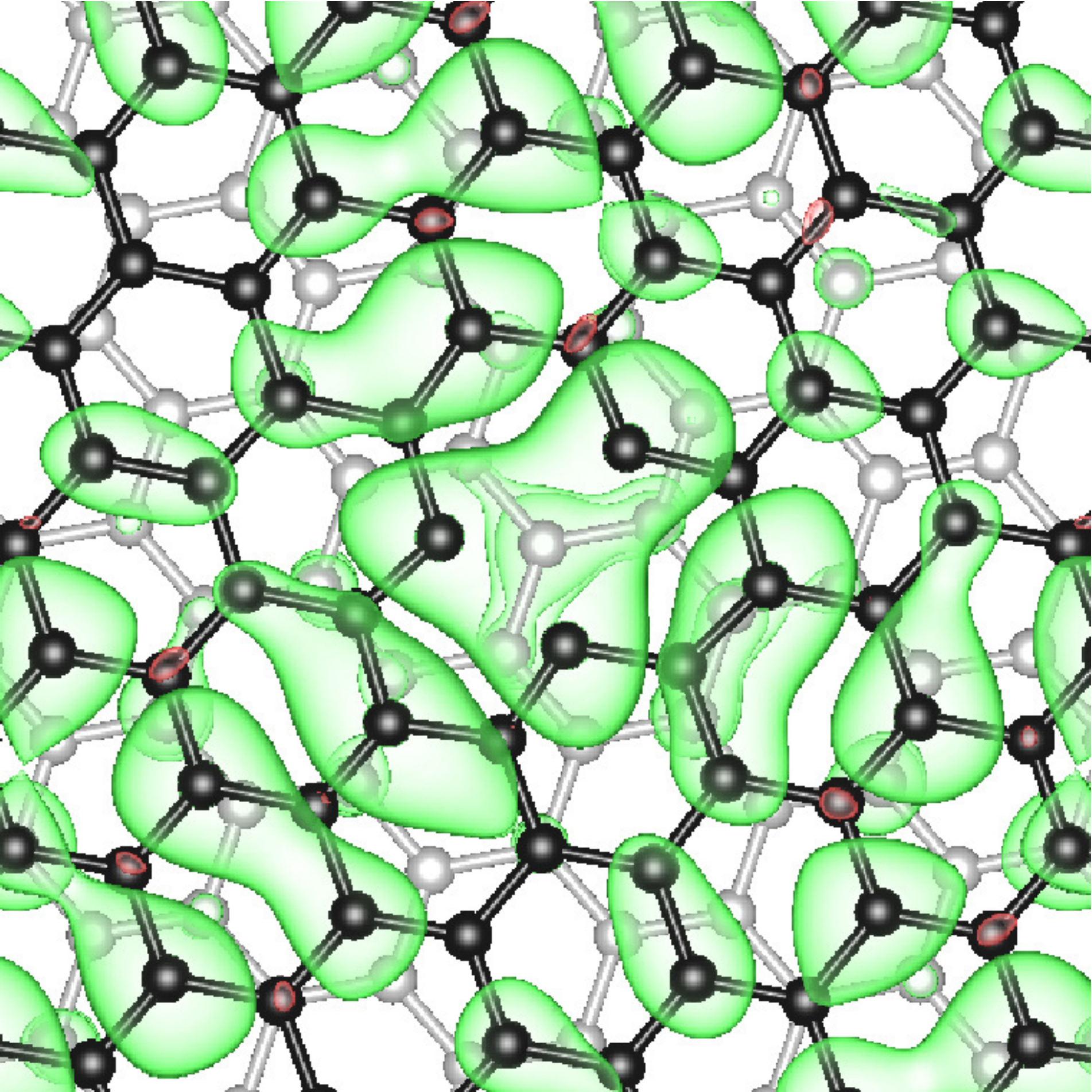}
\caption{(color online) Spatial extent of the spin polarized charge density $\rho_{SP}$ corresponding to the states in a energy window $\{E_F-0.05 {\rm eV}, E_F\}$ for the case of monovacancy in TBLG. The isosurfaces shown correspond to a density value of $\pm 2\times 10^{-5}$ e/bohr$^3$. Green lobes denote negative values, and red lobes (barely perceptible) denote positive values.}
\label{ILDOS-SLG-VacA} 
\end{figure}

We define the spin-polarized charge density $\rho_{SP}$ as the difference between the majority and minority spin charge densities. In Fig.~\ref{ILDOS-SLG-VacA}, we have plotted $\rho_{SP}$ for states that fall within an energy window of width 0.05 eV below $E_F$, for an $\alpha$ monovacancy in TBLG.  It is clearly evident that the spin polarization is not restricted to the immediate neighborhood of the monovacancy, but is
widespread in spatial extent. This may possibly lead to a high degree of spin polarization of a current that is passed through such a system. However, in a situation of randomly oriented monovacancy defects, the net spin-polarization of the charge density will possibly depend on the sublattice details of a particular distribution of the defects, similar to that seen in the case of SLG.\cite{Yazyev-RepProgPhys2010}

\section{Summary and Conclusions}

In this study, we have considered two types of point defects, viz., Stone-Wales defects and monovacancy defects, 
that can arise or can be induced in twisted bilayer graphene. We have studied the effect of these defects on the structural, electronic and magnetic properties of twisted bilayer graphene. We have compared these results with the same defects in single layer graphene and AB-stacked bilayer graphene.

The formation energy of these defects is very similar to those of the corresponding defects in single layer graphene and in AB-stacked graphene; 
this is to be expected because the coupling between the two layers
in twisted bilayer graphene is weak, and thus the effect on energetics of the second twisted layer is not
particularly marked. Similarly, the defect formation energy does not change appreciably when the defect is induced on various symmetry-inequivalent sites on TBLG.

However, signatures of interlayer coupling are manifested clearly when examining the band structure of TBLG with defects. When we consider TBLG consisting of one pristine layer and one with a SW defect, one gets the formation of two Dirac cones, shifted slightly in energy and momentum. The shift in energy of the two Dirac cones arises due to the combined effect of a transfer of electrons from the pristine to the defective layer, and a chemical interaction between the two layers; this mimics the effect of placing the system in an electric field. The band structure depends on the site at which the SW defect is created -- at some sites, the underlying double cone structure near the K points is distorted by the opening up of mini gaps -- it might however be difficult to control the site at which a defect is created in an actual experimental situation.

 We note here that the use of periodic boundary conditions in our plane wave calculations means that we actually have a periodic array of SW defects. Recent work has shown that it is possible to create multiple Dirac cones by artificially engineering a lateral periodic potential by creating a superlattice.\cite{Deshmukh-NL2013}

Creating a monovacancy in one layer of TBLG shifts up the Dirac cone of the other layer by $\sim0.25$ eV, due to a combination of the requirement of the equilbration of the Fermi energies, charge transfer and chemical interaction between the sheets. Both SW defects and monovacancies increase the density of states at $E_F$; this should be of relevance to transport properties. The monovacancy defect in TBLG leads to a magnetic defect state, very similar to the case of monovacancy defect in SLG and AB-BLG. A sharp magnetic defect state is the signature of the monovacancy, and arises from the dangling $\sigma$-bond on one of the nearest neighbor atoms of the vacant site. The defect has a magnetic moment of $\sim$ 1.2 -- 2.0 $\mu_{\rm B}$, depending on the defect density. The scattering of the $\pi$-states of the defective layer results in a highly spin-polarized density of states at the Fermi energy, both in the case of SLG and TBLG. The spin-polarized states in the vicinity of the Fermi energy are widespread in spatial extent, thus raising the  possibility of a large spin-polarization of the current in the defective layer. 

By comparing results obtained with the LDA, GGA (PBE) and DFT-D2 methods, we find that the PBE alone (i.e., without van der Waals interactions incorporated) is generally inadequate to describe bilayer graphene in general, and TBLG in particular.
In general, both LDA and DFT-D2 give very similar results, with a few notable exceptions: DFT-D2 gives a value for the exfoliation energy that is in closer agreement with experiment, and the two methods,
while giving similar results for the net magnetic moment $M_{net}$, give different values for the absolute magnetic moment $M_{abs}$ for TBLG with a monovacancy.

\end{document}